\documentclass[natbib,smallextended]{svjour3}       
\smartqed  
\usepackage{graphicx}
\usepackage{color,soul,amsmath,amssymb, lscape} 
\usepackage[colorlinks=false,citecolor=blue,urlcolor=blue,breaklinks]{hyperref}
\newcommand{\msano}{M_\odot~{\rm yr}^{-1}}
\newcommand{\e}[1]{\times 10^{#1}}

\newcommand{\mdot}{$\dot{M}$}
\newcommand{\jdot}{$\dot{J}$}
\newcommand{\ly}{Ly-$\alpha$}
\newcommand{\ha}{H-$\alpha$}

\newcommand{\fx}{$F_X$}
\newcommand{\bi}{\langle |B_I| \rangle}
\newcommand{\bv}{\langle |B_V| \rangle}
\newcommand{\om}{$\Omega_\star$}

\def\aj{{AJ}}                   
\def\araa{ARA\&A}             
\def\apj{{ApJ}}                 
\def\apjl{{ApJ}}                
\def\apjs{{ApJS}}               
\def\aap{{A\&A}}                
\def\aapr{{A\&A~Rev.}}          
\def\mnras{{MNRAS}}             
\def\pasj{{PASJ}}               
\def\solphys{{Sol.~Phys.}}      
\def\ssr{{Space~Sci.~Rev.}}     
\def\nat{{Nature}}              
\def\grl{{Geophys.~Res.~Lett.}} 
\def\jgr{{J.~Geophys.~Res.}}    
\def\prl{{Phys.~Rev.~Lett.}}     
\def\icarus{Icarus}

\journalname{Living Reviews in Solar Physics}

\begin{document}

\title{The evolution of the solar wind}
\author{Aline A.~Vidotto}

\institute{A.~A.~Vidotto \at
              School of Physics, Trinity College Dublin, The University of Dublin, Dublin-2, Ireland \\  \email{aline.vidotto@tcd.ie}   }

\date{Received: date / Accepted: date}

\maketitle

\begin{abstract}
How has the solar wind evolved to reach what it is today? In this review, I discuss the long-term evolution of the solar wind, including the evolution of observed properties that are intimately linked to the solar wind: rotation, magnetism and activity. Given that we cannot access data from the solar wind 4 billion years ago, this review relies on stellar data, in an effort to better place the Sun and the solar wind in a stellar context. I overview some clever detection methods of winds of solar-like stars, and derive from these an observed evolutionary sequence of solar wind mass-loss rates. I then link these observational properties (including, rotation, magnetism and activity) with stellar wind models. I conclude this review then by discussing implications of the evolution of the solar wind on the evolving Earth and other solar system planets. I argue that studying exoplanetary systems could open up new avenues for   progress to be made in our understanding of the evolution of the solar wind. 
\end{abstract}

\setcounter{tocdepth}{3}
\tableofcontents

\section{Introduction: placing the Sun in a stellar context} \label{intro}
It is fair to say that the Sun is the best studied star in the whole Universe: we can  measure its rotation, magnetic activity, composition, size, irradiation, and wind properties with accuracies like no other star in the Universe. However, all this information just tells us about how the Sun looks like \emph{now}. To understand the past, and future, evolution of the Sun, including its wind, magnetism, activity, rotation, and irradiation, astronomers rely on information from other ``suns'' in the Universe at different evolutionary stages. In a broad sense, these other ``suns'' belong to the group of solar-like stars.\footnote{Usually, stars similar to the Sun are classified into three groups: solar twins, which are a subset of so-called ``solar analogues'', which are in turn a subset of ``solar-like stars'' \citep{2010A&A...522A..98M}. Solar twins are stars that are (almost) identical to the Sun, with indistinguishable temperature, gravity, age and composition from solar \citep{2009A&A...508L..17R, 2014A&A...563A..52P}. Solar twins are  extremely rare. The second group, those of solar analogues, contains stars with spectral types between G0 and G5, within a factor of 2--3 in metallicity from the Sun \citep{2010A&A...522A..98M}. Finally, the largest group is formed by solar-type stars, which are stars with spectral types from late F to early K \citep{2009A&A...508L..17R}. The third group has the least stringent classification allowing not only for different compositions, but also for different ages.  When dealing with the evolution of the solar wind, I will draw stars from the group of solar-like stars. These are the definitions used in this review, but note that they can differ in the literature.}

Stellar evolution models predict that the Sun has evolved, since its pre-main sequence phase, from a spectral type K to its current G2 classification. It will leave the main sequence phase in another $\sim 4$ Gyr, when it will become a red giant star, go through a planetary nebula phase, until it will finish its days as a white dwarf that will cool down indefinitely. The large variation in solar photospheric temperature, radius and luminosity during the Sun's evolution is also accompanied by variations in the properties of the solar wind -- this outflow of particles, embedded in the solar magnetic field,  that propagates throughout the solar system. Although it is unlikely that the solar wind has been able to remove a significant amount of mass from the Sun, the solar wind has been able to remove significant amounts of angular momentum via magnetic field stresses. For this reason, the solar wind has played  a fundamental role in the Sun's evolution, as it regulates solar rotation.

\begin{figure}[ht]
	\centering
	\includegraphics[width=.98\columnwidth]{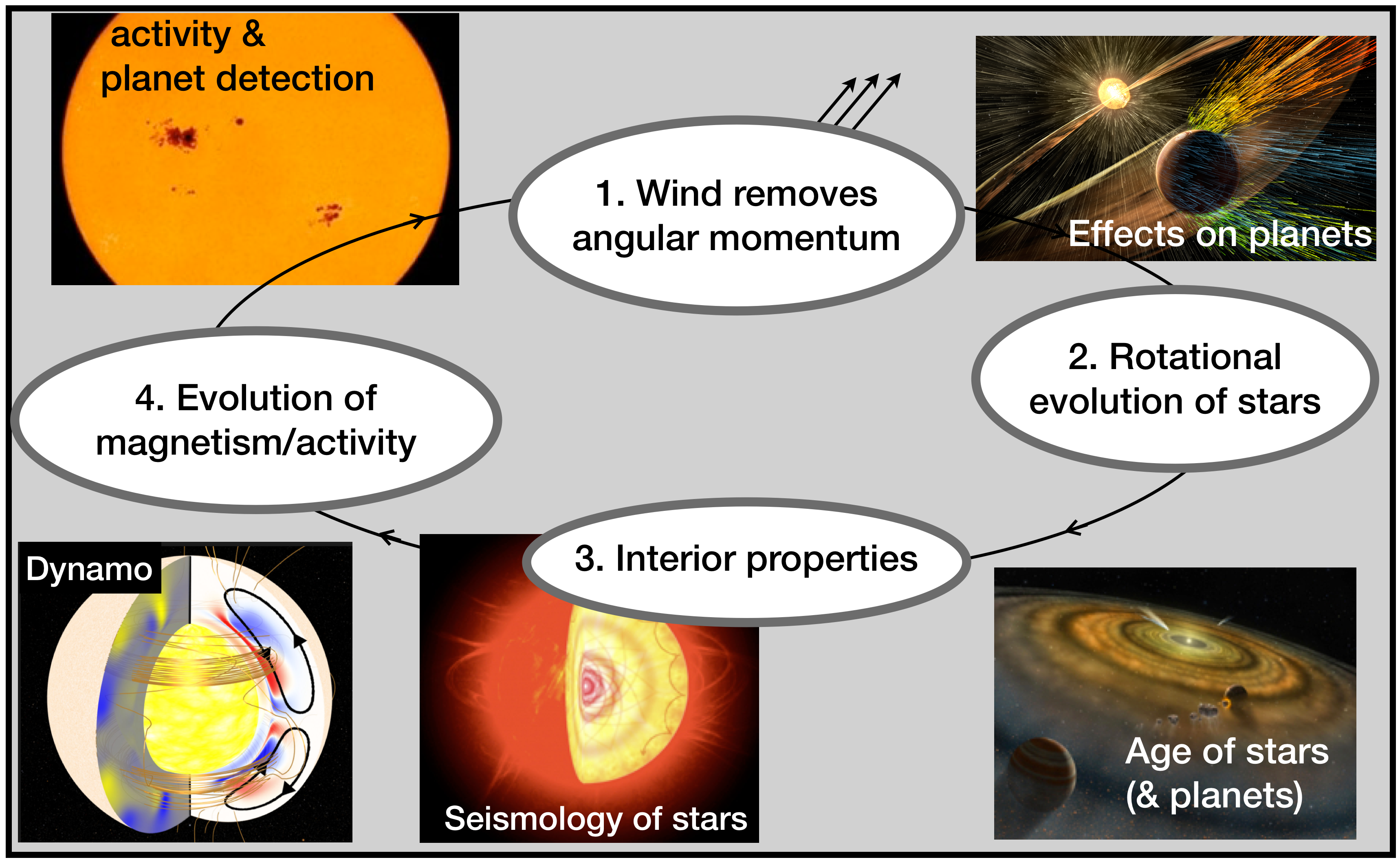}
	\caption{The big picture: evolution of winds of cool dwarf stars. As the star ages, its rotation and magnetism decrease, causing also a decrease in angular momentum removal. In the images, I highlight some of the areas to which wind-rotation-magnetism interplay is relevant. Image reproduced with permission from \citet{2016csss.confE.147V}.}
	\label{fig.big_picture}
\end{figure}

 In the sketch I show in Fig.~\ref{fig.big_picture}, I exemplify the ``life-cycle'' of the wind of an isolated solar-like star. I start with the text bubble \#1: winds of solar-like stars are magnetic in nature, and are thus able to carry away a significant amount of angular momentum from the star. Consequently, the star spins down as it evolves (\#2). Because of this variation in surface rotation, there is a redistribution of internal angular momentum transport, which changes the interior properties of the star (\#3). With a different internal structure, the dynamo that is operating inside the star changes, changing therefore the properties of the emerging magnetic fields (\#4). With a new surface magnetism, the stellar wind also changes (back to text bubble  \#1). This cycle repeats itself during the entire main-sequence lifetime of solar-like stars. Therefore, as an isolated solar-like star ages, its rotation decreases \citep{1967ApJ...150..551K,1972ApJ...171..565S} along with its chromospheric--corona activity \citep{2005ApJ...622..680R, 2008ApJ...687.1264M}, and magnetism \citep{2014MNRAS.441.2361V}. These three parameters are key ingredients in stellar wind theories. As a consequence of their decrease with age,  winds of main-sequence solar-like stars are also expected to decrease with time \citep{2004LRSP....1....2W,2019MNRAS.482.2853J}. 

\begin{figure}[ht]
	\centering
	\includegraphics[width=.98\columnwidth]{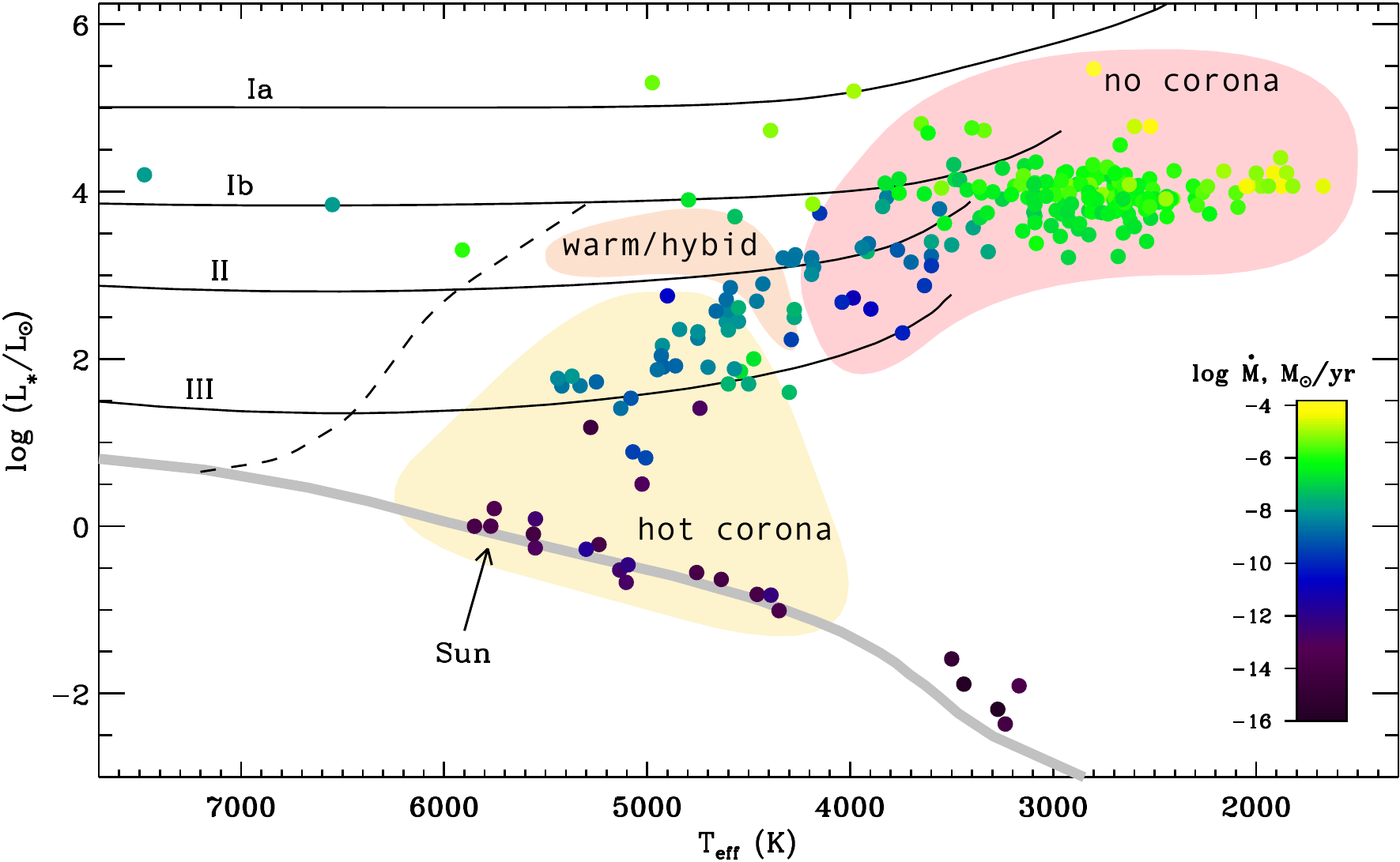}
	\caption{An overview of mass-loss rates (colour coded) in the cool-star HR diagram. Winds of {cool} stars evolve from hot ($\sim 10^6$~K) and tenuous (`hot corona') to cold  ($\sim 10^4$~K) and denser (`no corona'). Evolved low-mass stars that show weak or sporadic signatures of a hot corona are denoted `warm/hybrid'. The zero-age main sequence is shown by the grey line. Image reproduced with permission from \citet{2019ARA&A..57..157C}, copyright by Annual Reviews.}
	\label{fig.HR}
\end{figure}

Two physical quantities often quoted in stellar wind studies are the mass-loss rate \mdot\ and terminal velocity $u_\infty$, which is the speed winds asymptotically reach at large distances. These two quantities carry a lot information about stellar wind driving. Thus, they are key to inform wind models. Unfortunately, for solar-like stars,  these physical quantities are very  challenging to measure, because these winds are very tenuous and do not provide strong, detectable observational signatures, such as a P Cygni line profile, that is common in stars with denser winds. Figure \ref{fig.HR} shows an HR diagram, colour-coded by \mdot, where we see that estimated mass-loss rates in the lower main-sequence range between $\sim 10^{-16}$ to $\sim 10^{-12}\,\msano$. For comparison, the solar wind mass-loss rate, which is adopted throughout this review, is $\dot{M}_\odot = 2 \e{-14}\,\msano$. As these stars evolve off of the main sequence, their winds become more massive, and mass-loss rates increase by several orders of magnitudes. This increase in \mdot\ is caused by a change in wind driving, which likely has stronger contributions of thermal driving forces in the lower main-sequence phase, as these stars show signs of hot coronae  (seen in X-rays). As the low-mass stars evolve to become red giants in the post-main sequence phase, thermal driving becomes less important, resulting in cooler winds (moving from the yellow region in Figure \ref{fig.HR} to the top red region). The winds of evolved low-mass stars that do not show signs of coronae are likely driven by mechanical forces, such as for example pulsations and wave-driven mechanisms. In the lower main sequence, cool stars have hotter winds (on the order of $10^6$~K) and low mass-loss rates, while on the top right corner of the HR diagram cool stars have colder winds that can reach temperatures of $10^4$~K and maybe even lower, and high mass-loss rates. There is a relatively smooth transition between these two groups, with stars that show signs of weak/warm coronae belonging to an intermediate `hybrid' group, and they perhaps have a combination of wind driving mechanisms (thermal and mechanical, \citealt{lambda_and}). 

To derive an accurate picture of the solar wind evolution, a combined observational-theoretical approach is more suitable. Ideally, we would like to inform models by measuring winds of Sun-like stars at different evolutionary stages, thus sampling the physical properties of winds from young to old suns. In practice, however, this is not a simple task, as observing winds of cool dwarfs is currently very challenging. I start this review with an overview of observations of winds of solar-like stars (Sect.~\ref{sec.obs}). Using these results, I attempt to derive an evolutionary sequence of the solar wind mass-loss rate in Sect.~\ref{ref.Mdotevol}. One cannot talk about the solar wind evolution without talking about evolution of the key physical ingredients of the solar wind: rotation, magnetism and activity. Thus, in Sect.~\ref{sec.ingredients}, I introduce some key observations that have allowed us to infer the evolution of these three parameters. As we will see, data sampling  activity, rotation and even magnetism are more abundant in the literature than observational data sampling wind evolution. In Sect.~\ref{sec.models}, I present the results of some models investigating the solar wind evolution. In Sect.~\ref{sec.implications}, I discuss some implications of solar wind evolution on the past of the solar system, as well as on other extra-solar systems. I finish this review with a summary and a discussion of open questions in the field (Sect.~\ref{sec.open}).

From now on, I will refer to winds of isolated, main-sequence, Sun-like stars simply as winds or stellar winds. Otherwise, whenever I refer to winds of other types of stars, or at different evolutionary phases, I will specify (e.g., winds of red giants, or winds of pre-main sequence stars).

\section{How to detect winds of solar-like stars}\label{sec.obs}
 Although winds of low-mass stars are challenging to detect, astronomers have come up with some clever methods to infer the presence of a wind as well as to derive their physical properties. Table \ref{tab.wind_methods} summarises some proposed methods to detect winds of Sun-like stars, which I will discuss further. As I will show next, different methods  perform better for different systems, so a range of detection methods can help us probe the range of winds of Sun-like stars at different ages. 

\begin{table}
\caption{Some proposed methods to detect winds of Sun-like stars.} \label{tab.wind_methods}      
\begin{tabular}{lllll}
\hline\noalign{\smallskip}
Section number, Method & Requirement for detection & Key references  \\
\hline\noalign{\smallskip}
\ref{sec.astrosphere} Astrospheric Ly-$\alpha$ & partially neutral ISM, $\lesssim 10$pc  & \citet{2004LRSP....1....2W} \\
\ref{sec.radiowind} Radio free-free emission & denser winds and/or radio flares   & \protect\citet{1996ApJ...462L..91L, 1992AA...264L..31G} \\
\ref{sec.exodetection} Exoplanets as probes & evaporating planet  & \citet{2017MNRAS.470.4026V} \\
\ref{sec.promi}  Prominences in H-$\alpha$ & fast rotation  & \citet{2019MNRAS.482.2853J} \\ 
\ref{sec.cme} Detection of CME-dominated winds &fast CME associated with a flare & \citet{2018ApJ...856...39C}   \\
\ref{sec.pointsource} Propagation of radio emission & point source (planet?) within wind & \citet{2017AA...602A..39V}  \\
\ref{sec.xraydetection} X-ray emission from the stellar wind& hot coronal winds & \citet{1996ApJ...462L..91L, 2013MNRAS.436.2179L} \\
\ref{sec.chargeX} Charge-exchange  induced X-rays& partially neutral ISM &\citet{2001ApJ...546L..57W, 2002ApJ...578..503W}  \\
\ref{sec.whiteD} Accretion  onto white dwarfs& close binary with cool dwarf secondary& \citet{2006ApJ...652..636D}   \\
\hline\noalign{\smallskip}
\end{tabular}
\end{table}

\subsection{Detecting astrospheric absorption in Ly-$\alpha$}\label{sec.astrosphere}
\begin{figure}[ht]
	\centering
	\includegraphics[width=.98\columnwidth]{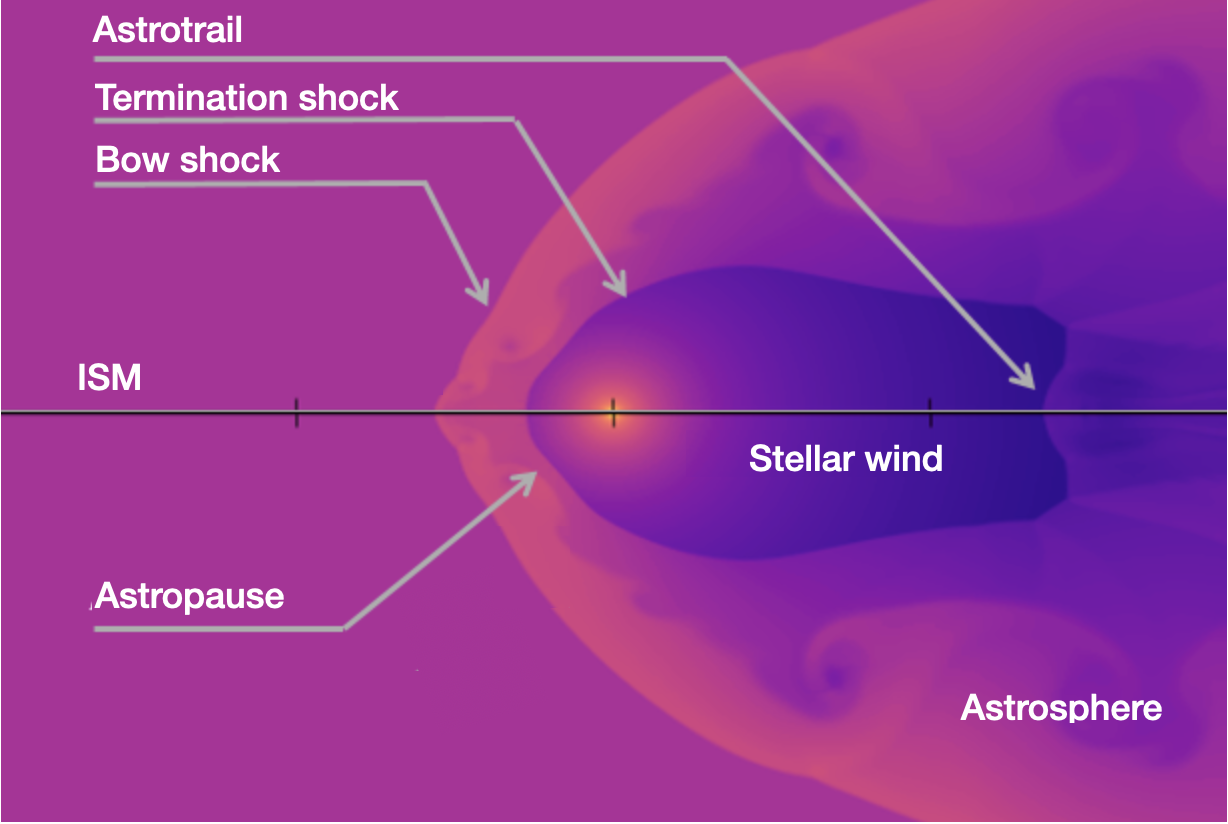}
	\caption{Sketch of the ISM (left) interacting with a stellar wind (right) and giving rise to an astrosphere, which is surrounded by a hydrogen wall and possibly a bow shock. The hydrogen wall, which is an enhancement of  hydrogen density, is located between the bow shock and the astropause. Figure adapted from \citet{2020PhDT.........2O}.}	\label{fig.ism}
\end{figure}

A stellar wind propagates from the star all the way to the interstellar medium (ISM), forming a similar structure as that observed around the solar wind, known as an ``astrosphere'', in analogy to the heliosphere around the Sun. Figure \ref{fig.ism} sketches an astrosphere, which has similar properties as the heliosphere. The inner part of the astrosphere is surrounded by the termination shock, where the stellar wind decelerates from supersonic to subsonic velocities.  Further out, the astropause separates the stellar wind and the ISM flow. Surrounding the astrosphere, if the relative motion between the stellar wind and the ISM is supersonic, a bow shock  forms. The enhancement in hydrogen density, known as the hydrogen wall, occurs between the astropause and the bow shock. Therefore, after  \ly\ photons  are emitted from the star, they have a long journey until they reach us. They first cross the stellar wind and the hydrogen wall around the astrosphere. Then comes the long journey through the ISM itself. Finally, they cross the hydrogen wall around the heliosphere and traverse the interplanetary space (solar wind) until they reach us and we can observe them (in space, above the Earth's atmosphere). During each stage of this journey, neutral hydrogen present in these different regions absorb the \ly\ spectral line at different velocities. As a consequence, the original \ly\ line is severely altered.  The hydrogen wall, in particular, is key for quantifying the stellar wind.  In this region,  ISM and stellar wind materials mix together, which can lead to charge-exchange between the ionised stellar wind material and the neutral component of the ISM. As a result,  stellar wind particles are neutralised, but they still retain their high velocity  and temperature. This hot neutral hydrogen, which causes substantial absorption in the   \ly\ line, provides the signature required to indirectly quantify stellar winds.
 
By modelling these different absorption components, one can then infer the column density of neutral hydrogen in the wall surrounding an astrosphere \citep{1998ApJ...492..788W}. With assistance from hydrodynamical models of the astrosphere-ISM interaction, it is then possible to estimate the wind ram pressure at the site of the interaction \citep{2001ApJ...547L..49W}. Given that the wind ram pressure is linked to the wind mass-loss rate as $P_{\rm ram} \propto \dot{M} u_\infty$, it is then possible to derive \mdot\ if  $u_\infty$ is known. In the case of astrospheric measurements, a unique $u_\infty = 400$~km/s has been assumed for all observed stars, allowing the mass-loss rate of the observed star to be inferred.  For more details, I refer the reader to the method review in \citet{2004LRSP....1....2W}. 

The ``astrosphere method'' has been the most successful method for measuring mass-loss rates of winds of solar-like stars, with nearly 20 measurements to date \citep{2018JPhCS1100a2028W}, from stars ranging from spectral types M to F on the main sequence, and a few more evolved cool stars as well. The reason this method has not provided measurements for a larger sample of stars is that it has a sweet spot for optimal performance. Firstly, the ISM needs to be neutral, or at least partially neutral, for charge-exchange to take place, and thus the astrospheric absorption signature to occur in the \ly\ line. This means that once the ISM becomes fully ionised, beyond 10--15~pc,  the method becomes inapplicable \citep{2008ApJ...673..283R}. Secondly,  if the stars are beyond $\sim10$~pc from us, the ISM column density can be too large and the ISM absorption obscures the astrosphere absorption. For these  reasons, \ly\ observations of nearly all stars beyond 10~pc have turned out to be non-detections \citep{2005ApJS..159..118W}. 

\begin{figure}
\centering
 \includegraphics[height=0.37\textheight]{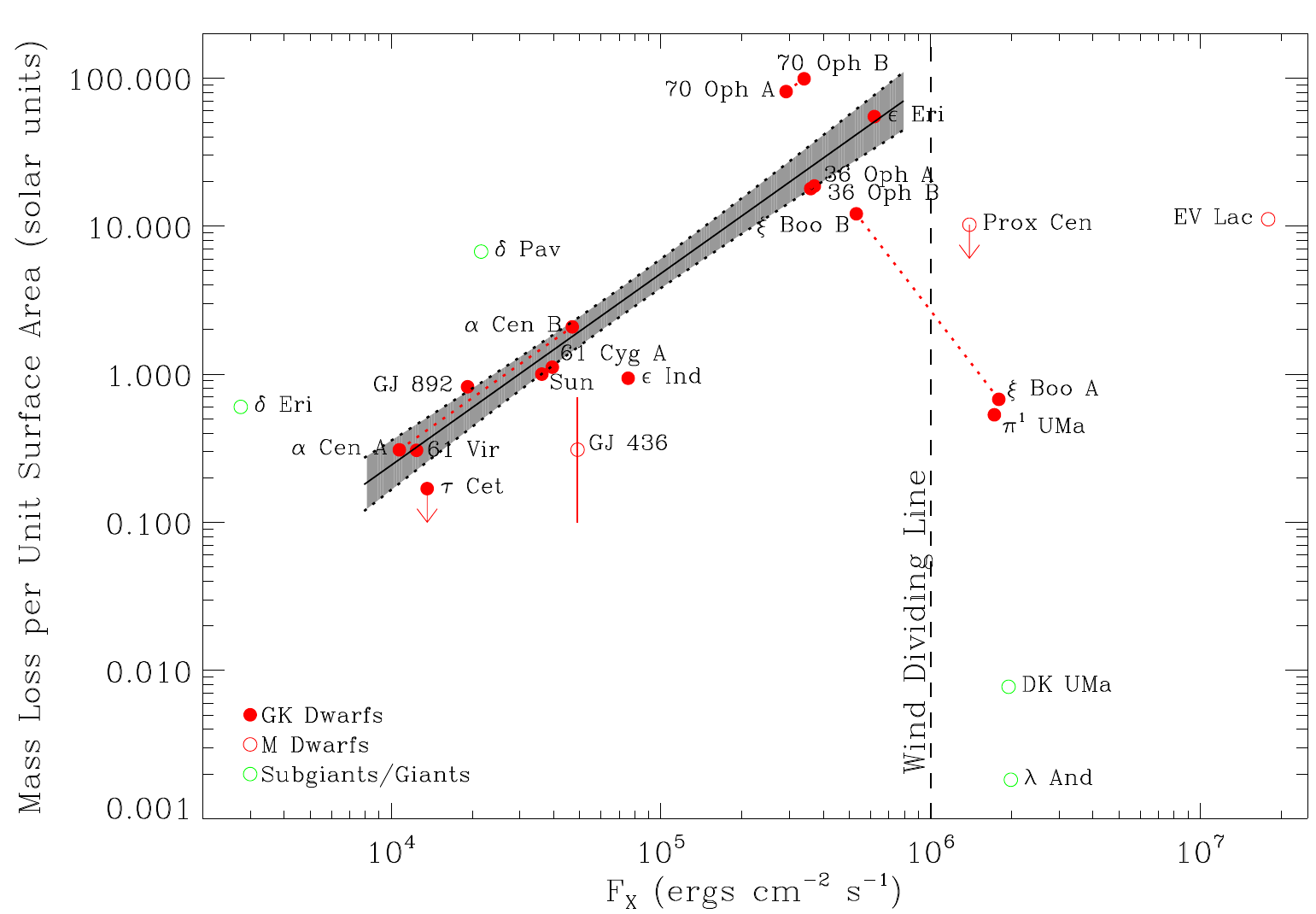}
 \caption{Summary of mass-loss rates for low-mass stars derived from the astrosphere method. For GK dwarfs (red filled circles), the mass-loss rate per unit surface area varies as a function of X-ray flux as $\propto F_X^{1.34}$ (shaded line). It has been suggested that active (and overall younger) stars with $F_X \gtrsim 10^{6}\mathrm{\ erg\ s^{-1}\ cm^{-2}}$ would have reduced mass-loss rates, thus giving rise to a `wind dividing line'. Some new mass-loss rates measurements indicate that mass-loss rates of young Suns could actually remain large (cf.\ Sect.~\ref{ref.Mdotevol}). Figure from \citet{2018JPhCS1100a2028W}.} \label{fig.detection_wood}       
\end{figure}

Figure \ref{fig.detection_wood} summarises the detections of winds of solar-like stars (filled red circles) using the astrosphere method. This figure shows that the mass-loss rate per unit surface area ($\dot{M}/R_\star^2$) varies as a function of X-ray flux $F_X$ as $\dot{M}/R_\star^2 \propto F_X^{1.34}$ (shaded line). Because X-ray flux is a measure of stellar activity, $ F_X$ can be used as a rough proxy for age -- stars with relatively large X-ray fluxes tend to be younger than stars with lower X-ray fluxes. For stars with  $F_X \gtrsim 10^{6}\mathrm{\ erg\ s^{-1}\ cm^{-2}}$, \citet{2004LRSP....1....2W} proposed the existence of a wind dividing line, beyond which the power-law fit  (shaded line) would cease to be valid. For solar-like stars, this X-ray flux of $10^{6}\mathrm{\ erg\ s^{-1}\ cm^{-2}}$ roughly corresponds to an age of 600~Myr. Would there be something happening at around this age that makes the wind mass-loss rate drop more than 2 orders of magnitude? It has been suggested that the magnetic topology of stars with $F_X \gtrsim 10^{6}\mathrm{\ erg\ s^{-1}\ cm^{-2}}$ suffers an abrupt change that could affect their  winds \citep{2005ApJ...628L.143W}, an idea that is backed-up by solar wind observations (see Sect.~\ref{ref.Mdotevol}  for the solar wind analogy). However, magnetic field reconstructions of young stars do not show abrupt changes in their field topology, but rather a smooth transition from a magnetic field with an important toroidal component at young ages (fast rotation) to a topology that is dominated by a poloidal field at older ages \citep{2008MNRAS.388...80P, 2018MNRAS.474.4956F, 2015MNRAS.453.4301S, 2016MNRAS.455L..52V}. 

$\pi^1$ UMa, a young solar-analogue star, is at the centre of the discussion on the existence of the wind dividing line. As can be seen in Figure \ref{fig.detection_wood}, the derived astrospheric mass-loss rate is quite low, and yet, the star has a high $F_X$  \citep{2014ApJ...781L..33W}.  $\pi^1$ UMa would, therefore, support the idea of the existence of a wind dividing line. However, an alternative explanation is that $\pi^1$ UMa might be in a strongly ionised ISM (the star is at a distance of 14~pc) and the measured \ly\ absorption could be taking place in a cloud between the star and the observer instead of in the hydrogen wall \citep{2014ApJ...781L..33W}. Indeed, spin down models (see Sect.~\ref{sec.jdot}) predict that solar analogues with $\pi^1$ UMa's rotation period and the age should have a mass loss rate that is about one order of magnitude higher than the present-day solar wind mass-loss rate   \citep{2015A&A...577A..28J}. One way to clarify this contradiction is to use multiple wind-detection methods applied to the same star. In fact, radio observations to detect the wind of $\pi^1$ UMa have been conducted \citep{2017A&A...599A.127F}. In the next section, I will discuss this method in more details.

\subsection{Detecting free-free radio emission from winds}\label{sec.radiowind}
What is unfortunate about the astrosphere method is that the non-detections do not allow us to derive upper limits for mass-loss rates. Although a non-detection could indeed be due to very low wind mass-loss rates, it could also be due to a fully ionised ISM around the astrosphere or  an ISM that causes too much absorption, rendering the \ly\ astrospheric signature undetectable. On the other hand, the method that I will discuss now, based on  thermal radio emission of winds, is a method that can extract meaningful information even in cases of non-detection of stellar winds. The main current disadvantage of this method is that no wind of low-mass star has ever been detected and, thus, all it has provided so far are upper limits on mass-loss rates. With the advance of radio technologies and construction of more sensitive radio telescopes, this could  change in the very near future. 

The stellar wind is a thermal, ionised plasma and, as such, it emits continuum bremsstrahlung (free-free) radiation across the electromagnetic spectrum. In particular, the densest region in a stellar wind (its innermost region) can emit at radio wavelengths, thus providing a way to directly detect the wind in radio \citep{2002ARA&A..40..217G}. For large enough densities, the innermost regions can become optically thick to radio wavelengths, creating a radio photosphere \citep{2019MNRAS.483..873O, 2019MNRAS.485.4529K}, that, if detected, can allow us to quantify the mass-loss rate of the wind. In this case, the underlying non-thermal radio emission from the star cannot be seen. 
On the other hand, a low-density wind would be radio transparent, and radio (non-thermal) emission from the stellar surface can pass through the wind unattenuated. Non-thermal radio emission, such as radio flares, has been detected in M dwarfs and solar-like stars \citep[e.g.,][]{1996ApJ...462L..91L, 2017A&A...599A.127F}. These radio-transparent winds can nonetheless provide important upper limits to wind mass loss rates. Regardless of whether the wind is optically thin or thick, the wind itself produces radio emission and that may provide a direct detection method.

The idea that  winds of early-type (hot) stars could emit in radio was proposed in the seminal works of \citet{1975A&A....39....1P},  \citet{1975MNRAS.170...41W} and \citet{1975A&A....39..217O}, which initially evolved from the theory of continuum emission in HII regions. The main modification to the previous theory was that, contrary to a constant density approximation adopted in HII regions, stellar winds were assumed to have a spherically-symmetric density  profile of the form  $\rho \propto r^{-2}$, where $r$ is the radial coordinate (this implicitly assumes a constant wind speed, which, as I will show next, is incorrect in the innermost region of stellar winds). This  modification led to a change in the shape of the spectrum.  In the asymptotic limit of an optically thick plasma ($\tau_\nu \gg 1$ at low frequencies $\nu$), the  flux density of an HII region behaves as $S_\nu \propto \nu^{2}$, while for a stellar wind, it was then found that $S_\nu \propto \nu^{0.6}$. In the optically thin asymptotic limit ($\tau_\nu \ll 1$ at high $\nu$), the radio flux density of a stellar wind has the same frequency-dependence as that of an HII region: $S_\nu \propto \nu^{-0.1}$. Later-on, \citet{1986ApJ...304..713R} dropped the assumption of spherically symmetric winds with constant speeds and demonstrated that spectral indices could range between $-0.1$ and $2$. Indeed, in the low-frequency, optically thick regime, wind models predict a range of slopes -- \citet{2019MNRAS.483..873O} derive indices from 1.2 to 1.6 in 3D wind simulations.

Although currently there has been no detection of free-free emission originating in winds of solar-type stars, radio observations have provided upper  limits of mass-loss rates for a number of objects \citep{1992AA...264L..31G, 1996ApJ...462L..91L, 1996ApJ...460..976L, 2000GeoRL..27..501G, 2014ApJ...788..112V, 2017A&A...599A.127F, 2017AA...602A..39V}. Even for the pre-main sequence phase, when stellar winds are expected to have higher mass-loss rates, only upper limits have been derived (e.g., $3\e{-9}\,\msano$ for the weak-lined TTauri star V830 Tau, \citealt{2017AA...602A..39V}), indicating that this is indeed a tricky detection with current instrumentation.  For stars identified as good solar analogues, like $\pi^1$ UMa, $\beta$ Com, $\kappa$ Ceti, EK Dra and $\xi^1$ Ori, upper limits of mass-loss rates are as low as $3\e{-12}\,\msano$ and as high as $1.3\e{-10}\,\msano$ \citep{2000GeoRL..27..501G, 2017A&A...599A.127F}. Another complication is that, if the wind has a small mass-loss rate, its weak emission competes against the expected thermal free-free radio emission from the chromosphere and  gyromagnetic emission from active regions. For such weak detections,  the distinction between a wind signature and the thermal component  from close to the stellar surface requires extremely accurate radio  spectra \citep{1993ApJ...406..247D, 2014ApJ...788..112V}. 

\begin{figure}
\centering
\includegraphics[width=.75\textwidth]{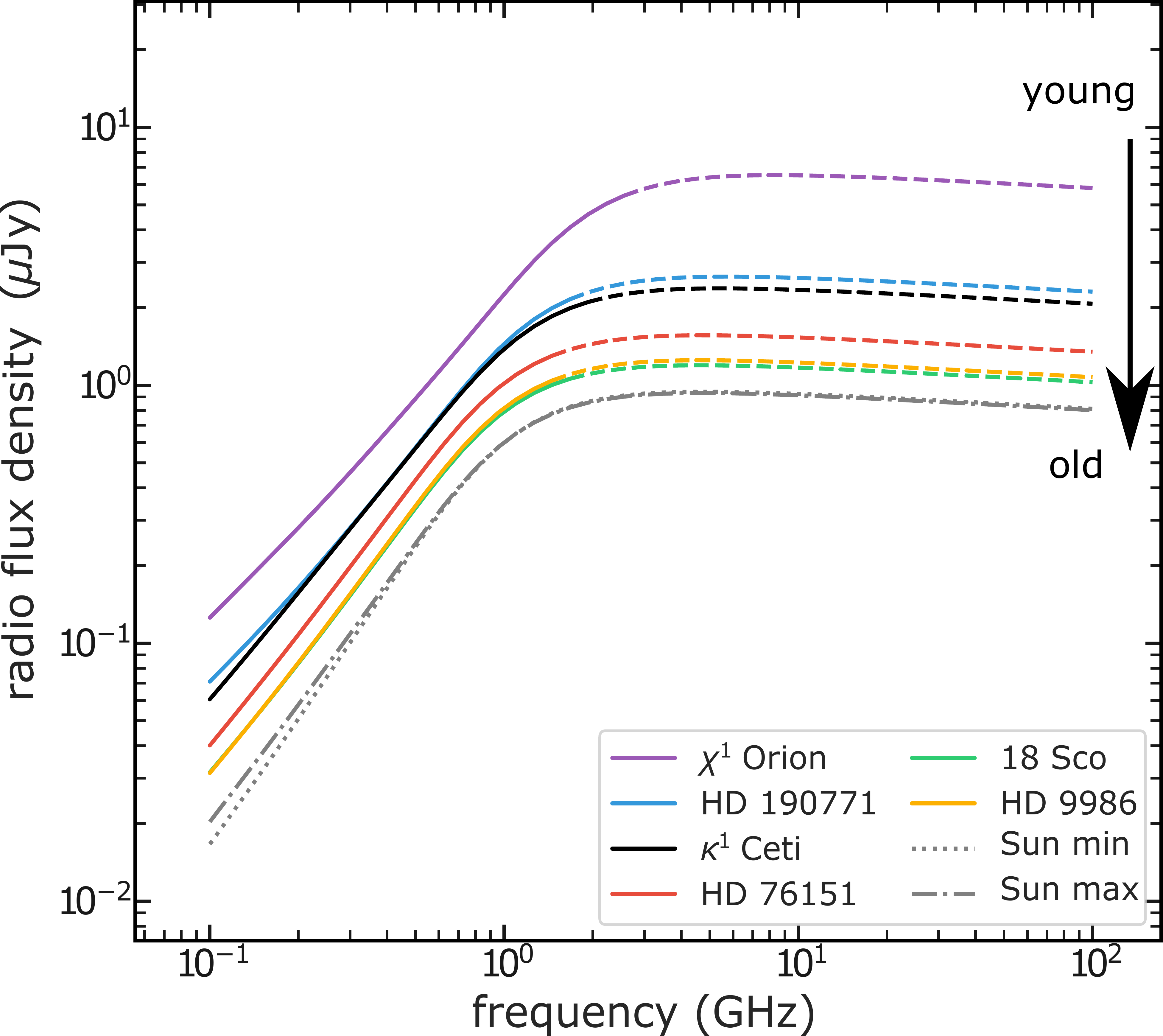} 
 \caption{Predicted evolution of the radio flux with rotation, used as proxy for stellar age, for a number of solar-like stars. The spectra of all these stars have been normalised to 10 pc. Young stars, close-by, are the targets which would present the strongest radio flux. Figure adapted from \citet{2019MNRAS.483..873O}.} \label{fig.radio_wind}       
\end{figure}

 In a number of simulations of winds of solar-like stars, \citet{2019MNRAS.483..873O} predicted that the radio flux density should increase for higher stellar rotation rates \om\ (Fig.~\ref{fig.radio_wind}). In particular, at $\nu = 1$ GHz, the radio flux density 
\begin{eqnarray}\label{eq.radio_prediction}
S_{\rm 1GHz} &\simeq& 0.7  \rm{\mu Jy} \, \left[ \frac{\Omega_\star}{\Omega_{\odot}} \right]^{0.7} \left[ \frac{10\, \rm pc}{d} \right]^2 \\
&\simeq& 0.7  \rm{\mu Jy}  \, \left[ \frac{4.6 \, \rm Gyr}{t} \right]^{0.39} \left[ \frac{10\, \rm pc}{d} \right]^2 \nonumber 
\end{eqnarray}
where $d$ is the distance, $t$  the age, and  $\Omega_\odot$ is the present-day solar rotation rate. In the equation above, I used the fact that $\Omega_\star \propto t^{-0.56}$, which applies to ages $\gtrsim 600$~Myr \citep[][see Sect.~\ref{sec.rotevol}]{2011MNRAS.413.2218D}, i.e., younger stars rotate faster. This implies that an analogue of our present-day Sun, when placed a 10~pc, would emit a flux density of $\sim 0.7 \mu$Jy at 1GHz.\footnote{From Fig.~\ref{fig.radio_wind}, we see that the solar magnetic cycle has little influence in the predicted radio spectra, due to small variations in the solar wind properties in a cycle timescale.} This is below detection limits of current radio telescopes. In fact, the distance-square decay  in Eq.~(\ref{eq.radio_prediction}) plays a more important role than the increase in rotation rate (or decrease in age) of the stars. Young suns, such as $\kappa$ Ceti and $\pi^1$ UMa, are also the closest to us in the simulated sample of  \citet{2019MNRAS.483..873O}  -- for these objects, the predicted 1-GHz flux density is still only a few $\mu$Jy. At the time of writing (Spring 2020), there exist plans to upgrade the existing VLA system, which would increase instrument sensitivity by a factor of 10 \citep{2018arXiv180305345O}. The expected  sub-$\mu$Jy sensitivity level of the future Square Kilometre Array (SKA) means that winds of close-by young suns (such as $\kappa$ Ceti and $\pi^1$ UMa) are potentially detectable below 1 GHz (see also discussion in the previous paragraph).

One consequence of the free-free emission of stellar winds is the creation of a radio photosphere, which is  paraboloidal in shape \citep{2019MNRAS.485.4529K}. Radio emission produced from sources inside the photosphere might be absorbed by the stellar wind. An expected source of low-frequency radio emission are magnetised close-in exoplanets \citep{1999JGR...10414025F}. If a planet is embedded in the radio photosphere of their host star, then it is possible that most of the planetary radio emission gets absorbed and does not escape \citep{2017AA...602A..39V, 2020MNRAS.493.1492K}. Although this might be problematic for planet detection in radio, close-in exoplanets can be a useful tool to probe the inner regions of stellar winds, as I explain next.

\subsection{Using exoplanets to probe stellar winds}\label{sec.exodetection}
Close-in, giant planets form  the majority of the exoplanet population known nowadays. With orbital distances significantly smaller than Mercury's orbit (and as small as  $\lesssim 0.02$~au), these exoplanets are embedded in a stellar wind regime  that is unprecedented for solar system planets. For comparison, the innermost planet in the solar system is Mercury, with a semi-major axis of 0.4 au. Although the solar wind speed at Mercury is similar to the wind speed at Earth's orbit (roughly 400 km/s), the local solar wind density at Mercury is about $1/0.4^{2}$ times larger. This is a consequence of the $r^2$-decay of  density for winds at asymptotic terminal speeds. However, going even closer to the Sun, one reaches the acceleration zone of the solar wind, in which the wind speed is still increasing with distance. Due to mass conservation, densities increase nearly exponentially towards small heliocentric distances. 

Not only the local stellar wind densities around close-in planets are orders of magnitude larger than the $\sim 5~{\rm cm}^{-3}$ at Earth's orbit, the  magnetic field embedded in the stellar wind is also expected to be several orders of magnitude larger \citep{2015MNRAS.449.4117V}. Additionally, the close orbital distances imply that close-in planets have high Keplerian velocities ($v_{\rm kep} \propto r^{-1/2}$). This means that, in the planet's reference frame, the stellar wind particles could still arrive at large velocities,  even though the local stellar wind itself might still not have reached terminal speed \citep{2010ApJ...722L.168V, 2011MNRAS.411L..46V}. Altogether, this shows that close-in planets  face extreme wind environments \citep{2015MNRAS.449.4117V}. 

Although an extreme stellar wind environment could be detrimental for life formation \citep{2007AsBio...7..185L,2007AsBio...7..167K,2007AsBio...7...85S,2013A&A...557A..67V}, it is precisely the extreme wind conditions around close-in planets that amplify signatures of the wind-planet interaction, potentially leading to  detection of such signatures. Signatures of wind-planet interaction have been detected in the hot-Jupiters HD209458b, HD189733b and the warm-Neptune GJ436b, that orbit main-sequence stars of spectral types F8, K2 and M3, respectively. These planets  show strong atmospheric escape, which can only be interpreted when considering the interaction with the winds of their host stars \citep[e.g.,][]{2008Natur.451..970H, 2013A&A...551A..63B, 2016A&A...591A.121B, 2014MNRAS.438.1654V, 2014Sci...346..981K}. Given that the escaping atmospheres of these gas giants are hydrogen dominated, their outflows are detected in the Ly-$\alpha$ line during planetary transits, in a technique known as transmission spectroscopy or spectroscopic transit observations \citep[for a review on the topic see, e.g.,][]{2018haex.bookE.100K}. 

By modelling the stellar \ly\ line profile that is transmitted through the planetary atmosphere, one can derive the conditions of the stellar wind surrounding the exoplanet.
The stellar wind has two effects that leave their fingerprints in the \ly\ line. Firstly, the wind shapes the outflow of the planet, in a similar way as the ISM shapes the astrosphere. This causes an asymmetry in the planetary outflow, as the interaction occurs preferentially on one side of the planet, causing lightcurve asymmetries \citep{2018MNRAS.479.3115V, 2019ApJ...873...89M}. Secondly, the ionised wind exchanges charge with the neutral hydrogen escaping the planetary atmosphere, creating a population of high-velocity neutrals (previously, these were high velocity ions from the stellar wind that became neutralised during the charge-exchange process). This leads to a high-velocity, blue-shifted component of the stellar \ly\ line \citep[e.g.,][]{2008Natur.451..970H}. Therefore, modelling these fingerprints left in the \ly\ line, one can then obtain the local densities and speeds of the stellar wind \citep{2016A&A...591A.121B}, and, consequently, the wind mass-loss rates as well (\citealt{2014Sci...346..981K}, \citealt{2017MNRAS.470.4026V}). 

In the case of GJ436b, a warm Neptune orbiting an M dwarf, models of the spectroscopic transit in \ly\ predict local wind speeds of 85~km/s and proton densities of $2\e{3}~{\rm cm}^{-3}$ \citep{2016A&A...591A.121B}, which translates to wind mass-loss rates of $1.2 \e{-15} \,\msano$ \citep{2017MNRAS.470.4026V}. For the solar-like star HD209458, \citet{2014Sci...346..981K} were able to model the \ly\ line with a wind that resembles that of the present-day Sun: with the same mass-loss rate, but with a scaled-up density at the orbit of HD209458b. This results in local wind speeds of $400$~km/s and wind densities of $5\e{3}~{\rm cm}^{-3}$. For HD189733, the derived local stellar wind speed at the orbit of the planet of $190$~km/s and density of $3\e{3}~{\rm cm}^{-3}$ \citep{2013A&A...557A.124B}  yielded a mass-loss rate of approximately $4 \e{-15} \,\msano$. One caveat to keep in mind, though, is that the values derived above are likely not unique, and some degeneracy might exist \citep{2020MNRAS.494.1297M}. 

Although the wind properties can be derived as a by-product of a method to detect escape in exoplanets, there are disadvantages to conduct transmission spectroscopy in the \ly\ line. The biggest of which is that the line falls in the ultraviolet part of the electromagnetic spectrum, whose observations require space instrumentation, which are very expensive. Currently, only the Hubble Space Telescope is able to observe in the ultraviolet. For this reason, the recent detection of escaping atmospheres in lines that can be detected with ground-based instrumentation, such as H-$\alpha$ or the HeI triplet at 10830\AA\ \citep{2018NatAs...2..714Y, 2018Sci...362.1388N,2018Natur.557...68S, 2019A&A...623A..58A}, can present new opportunities for probing escaping atmospheres, and likely the interaction region with stellar winds \citep{2018ApJ...855L..11O, 2021MNRAS.501.4383V}.

The typical wind conditions around exoplanets are soon to become better understood, with experiments that are taking place, right now, in our own solar system. Until very recently, no in-situ measurements of the solar wind plasma at close distances to the Sun existed. Measurements of the first encounter of the NASA's Parker Solar Probe (PSP) were recently released, probing the solar wind at distances as close as $36\,R_\odot \simeq 0.17$~au \citep{2019Natur.576..237B}. PSP is set  to remain in the heliospheric equator and, after a series of Venus flybys, it will get closer and closer to the Sun, reaching a highly elliptical orbit with a perihelion of 0.046~au -- typical of hot Jupiters. A second complementary spacecraft,  ESA's Solar Orbiter, was launched in February 2020 and will also study the solar wind at close distances, down to 0.29 au, albeit at high heliospheric latitudes (polar regions). Together, these two spacecrafts will allow in situ measurements of the solar wind at unprecedented close heliospheric distances. They will provide more information about the physical mechanism that accelerates the solar wind and will provide information about the environment at the orbits of close-in exoplanets.

\subsection{Using prominences to probe stellar winds}\label{sec.promi}
The last method for detecting winds of solar-like stars that I would like to discuss involves using observations of slingshot prominences to derive wind mass-loss rates. Slingshot prominences occur on fast rotating stars, and are very extended \citep{2005MNRAS.361.1173J}, in contrast to solar prominences. They are detected as absorption features  that travel in the \ha\ stellar line profile as the star rotates. Their observed velocities indicate that slingshot prominences occur at or beyond the co-rotation radius, which are several radii above the stellar surface \citep{1989MNRAS.236...57C, 1989MNRAS.238..657C}. \citet{2005MNRAS.361.1173J} suggested that prominences are formed at the  top of long magnetic loops, which are filled with mass from the stellar wind. The material in the prominence  cools to a few $10^4$~K -- thus, if the material has enough optical depth, slingshot prominences are seen in \ha\ Doppler maps in absorption, when they pass in front of the stellar disc.

\begin{figure}
\centering
 \includegraphics[width=.98\textwidth]{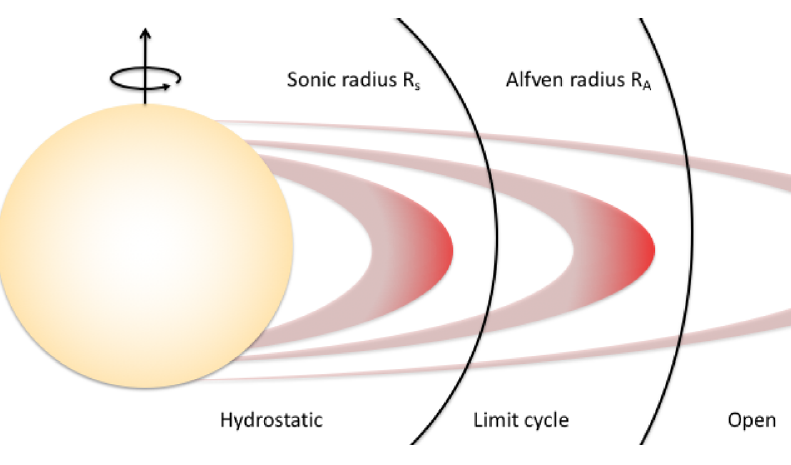}
 \caption{The formation of slingshot prominences occurs when in the  `limit-cycle regime'. In this case, the site of prominence formation, i.e., on loop-tops at the corotation radius, occurs above the sonic point (the star is ``unaware'' that the prominence has formed and thus keeps loading it with stellar wind material) and below the Alfv{\'e}n radius (beyond the Alfv{\'e}n radius all magnetic field lines will be open).  By observing slingshot prominences, one can estimate the rate at which mass is loaded into the loop tops and thus derive mass-loss rates of stellar winds. Figure from \citet{2019MNRAS.482.2853J}.} \label{fig.prominences}       
\end{figure}

The dynamical support and lifetime of slingshot prominences depend on the relative position between where they are formed (i.e., the co-rotation radius) and the Alfv{\'e}n and sonic surfaces of a stellar wind (Fig.~\ref{fig.prominences}). The Alfv{\'e}n and sonic surfaces are defined as the surface where the wind speed reaches the Alfv{\'e}n and sound speeds, respectively (more about this will be discussed in the modelling Sect.~\ref{sec.models}). Two main conditions should be kept in mind with regards to the existence of slingshot prominences.  Firstly, given that these prominences occur at loop tops, they can only exist when the co-rotation radius lies below the Alfv{\'e}n surface, as beyond the Alfv{\'e}n surface, magnetic field lines are open \citep{2019MNRAS.482.2853J}. \citet{2018MNRAS.475L..25V} demonstrated that for older solar-like stars, the co-rotation radius is above the Alfv{\'e}n radius, and these stars cannot support these types of prominences. Young, fast rotating stars, on the other hand, belong to the other group, in which prominences occur within the Alfv{\'e}n surface. 

The second important condition to consider is the relative location between a prominence (co-rotation point) and the sonic point. Information about what is happening in the wind cannot be passed back to the star for any event that takes place above the sonic point \citep{1998A&A...330L..13D, 2020MNRAS.494.2417V}. Thus, if the prominence, formed at the co-rotation radius, is formed above the sonic point, the star keeps loading the prominence with stellar wind material  and the loop top becomes denser and denser, until it eventually erupts, and the cycle starts again. This represents the `limit-cycle regime' proposed in \citet{2019MNRAS.482.2853J}. If the site of prominence formation at the co-rotation radius, on the other hand, occurs below the sonic point, the mass-loading from the surface gets readjusted -- although prominences could still be formed in the `hydrostatic regime', they would erupt on an occasional basis only.  

Altogether, these conditions imply that the formation of slingshot prominence occur in the `limit-cycle regime' (Figure \ref{fig.prominences}). Such a condition is more easily met in faster rotators, as these stars have smaller co-rotation radii and thus more easily to be formed in the sub-Alfv{\'e}nic regime. Their winds need to be relatively hotter, so that the sonic point occurs at lower heights. As I will discuss later on, we expect that active stars have hotter winds. These conditions are more easily met in fast and ultrafast rotators, in agreement with extended prominences seen in the ultrafast rotators such as Speedy Mic \citep{1993MNRAS.262..369J,  2006MNRAS.365..530D}, HQ Lup \citep{2000MNRAS.316..699D} and, more recently, in V530 Per \citep{2020A&A...643A..39C}, all of which with rotation periods $<0.4$~d. Slingshot prominences have been predicted to occur in stars that are rotating at less extreme rates as well, such as in the young Sun HIP 12545 \citep{2018MNRAS.475L..25V}, which shows a rotation period of nearly 5 days. 

From \ha\ observations, one can measure the prominence lifetime and the amount of mass contained in the prominence, giving the rate of mass that upflows into the prominence. To convert this mass-loading rate, which occurs in a  localised region of the stellar surface, to a wind mass-loss rate,  the prominence surface coverage is needed. \citet{2019MNRAS.482.2853J} estimated that about 1\% of the surface of a fast rotating star would be covered in prominences. These authors then derived mass-loss rates of 350, 4500, and 130 times the solar-wind mass loss rate for the young solar-type stars AB~Dor, LQ Lup and Speedy Mic, respectively, all of which rotate with a period shorter than half a day.\footnote{The sample in \citet{2019MNRAS.482.2853J} also contains a couple of active M dwarf stars, namely V374 Peg and HK Aqr, whose derived mass-loss rates are $4\e{-12}$ and $10^{-12} ~\msano$, respectively. These mass-loss rates are substantially larger than those derived for M dwarfs  with the astrosphere method, such as in the case of the active M dwarf EV Lac and the less active Proxima Cen ($2\e{-14}$ and $< 4\e{-15}\msano$, respectively \citealt{2004LRSP....1....2W}).}

Due to their youth, these three solar-type stars are  X-ray luminous. When placing them together in the diagram shown in Fig.~\ref{fig.detection_wood}, we notice that the trend of mass-loss rates with X-ray flux extends for a larger dynamical range in X-ray fluxes, compared to the work of \citet{2018JPhCS1100a2028W}. In Sect.~\ref{ref.Mdotevol}, I will discuss how the outcomes of these detection methods can be used to derive an observed evolutionary sequence for mass-loss rates. 

\subsection{Detecting coronal mass ejections (CMEs) through type II radio bursts}\label{sec.cme}
Although not all solar CMEs have a flare counterpart (and vice-versa), solar CMEs are often associated to flares. \citet{2011SoPh..268..195A} showed that stronger flares lead to  more massive CMEs and that the CME mass scales with the X-ray flare flux as $M_{\rm CME} \propto F_{\rm flare}^{0.7}$ or, in terms of the flare energy, as $M_{\rm CME} \propto E_{\rm flare}^{0.68}$ \citep{2012ApJ...760....9A}. If one were to extend these solar empirical relations to active stars, which show higher flare energies, one would expect that the mass contained in stellar CMEs could become substantially large. As a result, active stars, due to their higher flare rates, could have CME-dominate winds with mass-loss rates that could be substantially larger than solar. 

In the Sun, CMEs contribute, on average, to a mass-loss rate of  $\simeq 4 \e{-16}~\msano$ \citep{2010ApJ...722.1522V}, which is about only a few percent of the present-day solar wind mass-loss rate. In contrast are the predictions for young stars -- using flare--CME empirical scalings, \citet{2012ApJ...760....9A}, \citet{2013ApJ...764..170D} and \citet{2015ApJ...809...79O} estimated CME mass-loss rates for T Tauri stars in the range $\sim 10^{-12}$--$10^{-9}~\msano$, which are several orders of magnitude larger than the present-day solar wind. However, for  CME-dominated winds with rates $\gtrsim 10^{-10}~\msano$, \citet{2013ApJ...764..170D} argued that they would have kinetic energies that could amount to 10\% of the stellar bolometric luminosity, which is a rather substantial  energy budget associated to CMEs. These authors questioned that instead the solar flare--CME relation could not be extrapolated indefinitely to higher flare energies. For example, the relationship might flatten out towards higher flare energies or that observed stellar flares might not always be accompanied by a CME. This could happen if, for example, CMEs on active stars were more strongly confined, and so fewer of them would be produced for any given number of X-ray flares. 

This stronger confinement could be caused by the noticeable differences between the solar and stellar magnetic field characteristics. Indeed, more active stars seem to have more toroidal large-scale magnetic field topologies \citep{2008MNRAS.388...80P, 2015MNRAS.453.4301S, 2016MNRAS.455L..52V}, which could indeed result in more confined CMEs. Using numerical simulations, \citet{2018ApJ...862...93A} showed that a stronger overlying large-scale dipolar (poloidal) magnetic field of 75~G could prevent a typical solar CME from erupting and that only CMEs with magnetic energies $\gtrsim 30$ times larger than those in typical solar CMEs were able to escape. Those that escaped, however, were not  able to accelerate efficiently due to the strong overlying  magnetic field.

While \citet{2011SoPh..268..195A} and \citet{2013ApJ...764..170D} concentrated on flare energies in the X-ray band,  \citet{2015ApJ...809...79O} investigated empirical solar flare--CME relations using the bolometric energy from flares.  These authors provided energy partition calculations that can be used to relate the amount of radiated flare energy in one bandpass (e.g., in white light or X-rays) to the total bolometric energy of flares. White-light flares, for example, are frequently seen in high-cadence observations of missions such as Kepler, K2 and TESS. The long-term monitoring provided by these missions allows us to build more complete statistical studies of  flares from stars of different spectral types and ages \citep{2012Natur.485..478M, 2015EP&S...67...59M, 2014ApJ...797..122D}, which can then be used to study stellar CMEs. 

An alternative approach to estimate mass-loss rates of CME-dominated winds was proposed by \citet{2017ApJ...840..114C}. He used relations between solar magnetic energy flux and the kinetic energy flux of the solar wind/CME outflows to predict the evolution of CME mass-loss rates of solar-like stars. His models predict that both the wind and CME mass-loss rates are larger at younger ages, with CMEs dominating the mass loss process in the first 0.3~Gyr of the lifetime of a solar-like star. At these younger ages, \citet{2017ApJ...840..114C} estimated that  CME mass-loss rates are a factor of 10 to 100 higher than the quiescent (overlying) stellar wind. At ages $\gtrsim 1$~Gyr, the mass loss in the quiescent wind dominates that in CMEs. 

Although these models qualitatively agree  that CMEs can contribute significantly to the total mass-loss rates at younger ages, observationally confirming the presence and amount of stellar CMEs  is challenging and their detection has remained elusive \citep[see, e.g.,][for an updated census]{2020MNRAS.493.4570L}. Given that type II radio bursts are related to CMEs in the Sun (not all CMEs produce type II bursts though), one possible way to detect stellar CMEs is through observations of radio bursts \citep[see, e.g.,][and references therein]{2016ApJ...830...24C}. Solar type II radio bursts are believed to originate in the shocked material created as the CME propagates outwards at super-Alfv{\'e}nic velocities \citep{2019SunGe..14...111}. The plasma emission from this shocked material is seen in the dynamic spectrum as a  burst drifting in time (related to the speed of the shock/CME) and frequency (related to the density of the corona). Given that the coronal density decreases with height, type II radio bursts drifts towards lower frequencies as the CME propagates outwards -- starting (maximum) frequencies of $\sim 100$ MHz corresponds to electron densities of $\sim 10^8$~cm$^{-3}$ (see Equation \ref{eq.plasma} that will be discussed below).   

Not all CMEs should produce type II bursts though. For example, if  stellar CMEs are not sufficiently accelerated to become super-Alfv{\'e}nic (see, e.g., slowly accelerated CMEs as seen in the models from \citealt{2018ApJ...862...93A}), a shock wave would not be formed and thus a type II burst would not be seen. In spite of this, given that type II bursts are indicative of solar CMEs, by analogy, detecting stellar type II radio bursts could provide a lower limit on stellar CMEs  \citep{2016ApJ...830...24C}, which can be used to determine CME occurrence rates and characterise their properties. \citet{2016ApJ...830...24C} relates the shock speed $v_{\rm CME}$ to the frequency drift rate $\dot{f}$ and coronal scale height $H$ as
\begin{equation}\label{eq.crosley}
v_{\rm CME} = - \frac{2H\dot{f}}{f} ,
\end{equation}
where $f$ is the frequency. Although  $H$ has to be modelled, both $\dot{f}$ and $f$ are quantities that are directly obtained from type II radio burst observations. With this, one could derive the CME speed. One further step is required to convert the CME speed into CME mass. For that, it is assumed that energy equipartition between the bolometric radiated energy of the flare ($E_{\rm flare, bol}$) and the kinetic energy of its associated CME holds. Thus
\begin{equation}\label{eq.crosley2}
M_{\rm CME} = \frac{2E_{\rm flare, bol}}{v_{\rm CME}^2} = \frac{E_{\rm flare, bol}}{2 H^2 } \left(  \frac{f}{\dot{f}} \right)^2  ,
\end{equation}
where Eq.~(\ref{eq.crosley}) was used in the last equality of Equation (\ref{eq.crosley2}). $E_{\rm flare, bol}$ is not a directly observed quantity, but energy partition calculations  can be used to convert from observed flare energies at a given bandpass to $E_{\rm flare, bol}$ \citep{2015ApJ...809...79O}. Thus, while no type II burst has yet been observed associated with a stellar flare, if one were detected, the rate of frequency change would provide a very useful measure of CME speed (Equation \ref{eq.crosley}) and mass  (Eq.~(\ref{eq.crosley2}, cf.\ \citealt{2018ApJ...856...39C}). With an observationally derived flare frequency distribution, this could then be used to estimate the total mass loss in CME-dominated winds.
 
More recently, the prospects of using coronal dimming have been suggested as a means to identify stellar CMEs \citep{2016SoPh..291.1761H, 2020IAUS..354..426J}.  In the Sun, coronal dimming is seen in  certain  EUV coronal lines, tracing the post-CME evacuation of the corona. From the depth of the light curve, one can infer how much mass has been blown out, and from the slope of the light curve, one can determine the CME speed \citep{2016ApJ...830...20M}.  The suggestion of detecting the stellar equivalent of solar coronal dimming is still in its first steps and will be explored with future instrumentations \citep{2019SPIE11118E..08F}. Some points still need to be elucidated, such as whether dimming could  be detected in the case where CMEs are happening all the time, as could potentially be the case of young stars, or whether dimming could change the ``basal'' level of the stellar EUV emission to such an extent that one would not be able to disentangle particular CMEs. To answer these and other open questions, more detailed modelling studies are needed \citep[see, e.g.,][]{2020IAUS..354..426J}.

\subsection{Propagation/suppression of radio emission from a point source (planet) embedded in a stellar wind}\label{sec.pointsource}
\begin{figure}
\centering
\includegraphics[width=.49\textwidth]{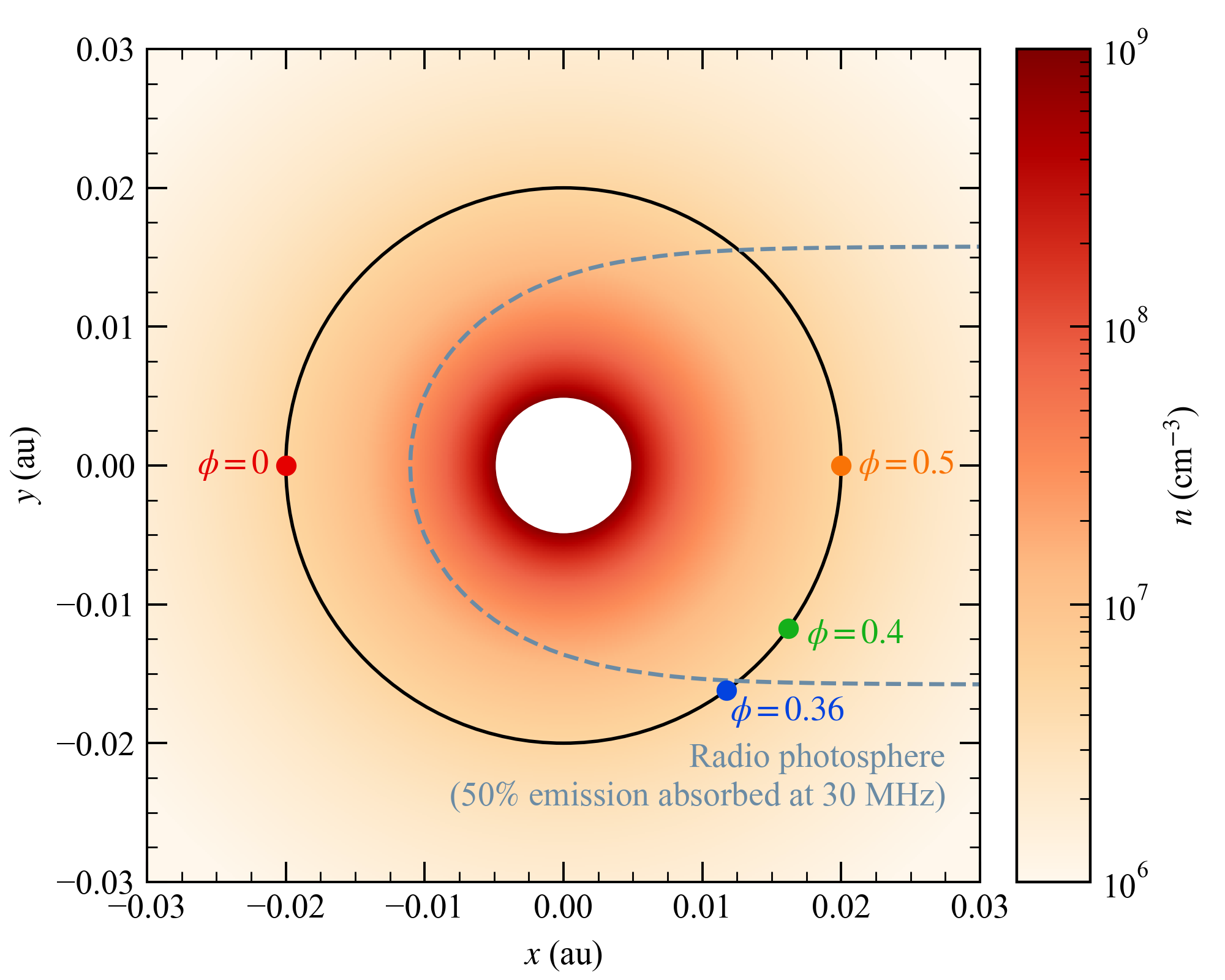} 
\includegraphics[width=.49\textwidth]{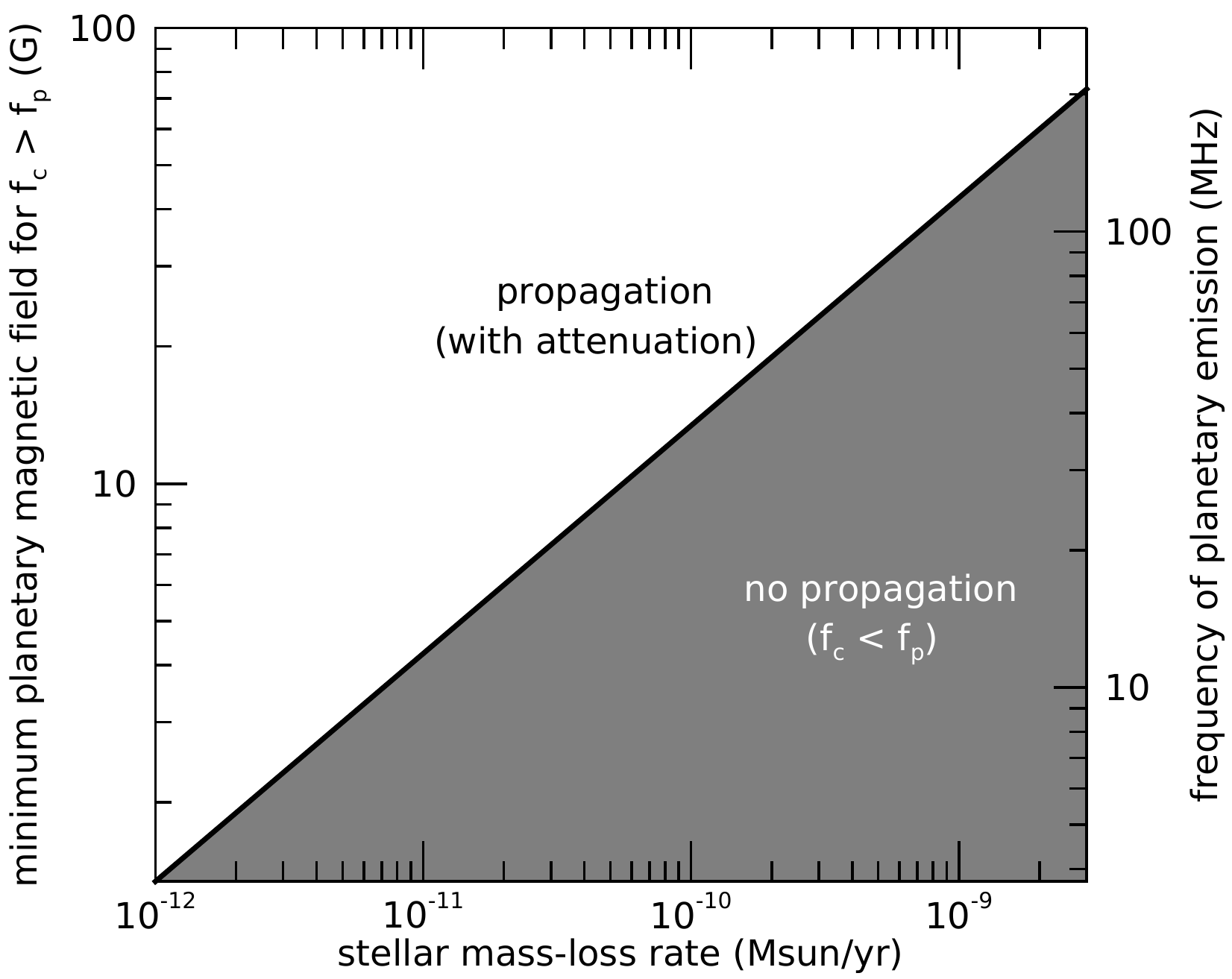} 
 \caption{Left: Number density of the wind of a solar-like star with a mass-loss rate of $2\e{-12}~\msano$. A planet is considered to orbit at 0.02~au, with the observer looking towards the system from the negative $x$-direction. The dashed line shows the radio photosphere where 50\% of a 30-MHz wind emission is produced. The hypothetical 30-MHz radio emission of this planet is increasingly more attenuated after the planet ingresses the radio photosphere and the attenuation peaks at orbital phase $\phi=0.5$. Its emission is least attenuated at $\phi=0$. Given that the position of the radio photosphere is linked to the stellar wind properties, monitoring of planetary radio emission could allow one to derive stellar wind properties. Figure from \citet{2020MNRAS.493.1492K}. Right: The situation on the left panel only occurs if the plasma emission $f_p$ is below the cyclotron frequency $f_c$ of the planetary emission (white area). If  the wind of the host star has a high mass-loss rate and the planet has a weak magnetic field, such that $f_p > f_c$, then the planetary radio emission cannot propagate through the wind of the host star (grey area). Detections of planetary radio emission can thus place an upper limit on the mass-loss rate of the star (Equation \ref{eq.bp3}). Note that the wind parameters used to produce this figure is based on  models of the weak-lined T Tauri star V830 Tau. Figure from \citet{2017AA...602A..39V}.} 
\label{fig.robert}       
\end{figure}

As discussed in Sect.~\ref{sec.radiowind}, a stellar wind, if sufficiently dense, can emit free-free emission at radio frequencies. We can define a boundary around the star wherein the bulk of the wind emission comes from. Here, we define this boundary, also known as the radio photosphere, as the isosurface where the frequency-dependent optical depth is $\tau_\nu$. The value of $\tau_\nu = 0.399$, for example,   delineates the region within which 50\% of the radiation is absorbed  by the stellar wind and 50\% of the emission escapes \citep{1975A&A....39....1P}.  The left panel in Figure \ref{fig.robert} illustrates the radio photosphere at $\nu = 30$~MHz (dashed line) and the density profile of the wind of  a sun-like star with a mass-loss rate of $2\e{-12}~\msano$ \citep{2020MNRAS.493.1492K}. In this figure, we are seeing a 2D cut of the radio photosphere, where the observer sees the system from the negative $x$ axis. 
In three dimensions, the radio photosphere resembles a `tea cup'  \citep[see Fig.~5 in][]{2019MNRAS.485.4529K}, which means that in the plane of the sky, this isocontour would have an approximately circular shape (the shape is only perfectly circular for a spherically symmetric wind though).

We can  now imagine a situation in which a  point-source radio emitter is embedded in the wind, which is optically thick at radio frequencies. 
The point source could be, for example, an exoplanet. Exoplanets, if they are magnetised, are believed to emit at radio frequencies,  analogously to the magnetised planets in the solar system \citep{1999JGR...10414025F}. Their emission is cyclotronic and thus takes place at cyclotron frequencies
\begin{equation}\label{eq.cyclotron}
f_c = \frac{e B_p}{2 \pi m_e c} = 2.8 \left( \frac{B_p}{1 \rm G} \right) {\rm MHz}
\end{equation}
where $B_p$ is the planet's magnetic field,  $c$ the speed of light, and $e$  and $m_e$ are the electron charge and mass. We explore here two different scenarios. In the first scenario, we investigate whether the emission of the point source could \emph{propagate} though the wind of the host star, while, in the second scenario, we assume the emission can propagate, but it is \emph{attenuated}.

The planetary radio emission can only propagate in the stellar wind plasma if the cyclotron frequency of emission $f_c$ is larger than the stellar wind plasma frequency  $f_p$  everywhere along the propagation path, where
\begin{equation}\label{eq.plasma}
f_p = \left( \frac{n_e e^2}{\pi m_e}\right) ^{1/2} = 9 \left( \frac{n_e}{1~ {\rm cm}^{-3}} \right)^{1/2} {\rm kHz} . 
\end{equation}
The   local electron density $n_e$ of a fully ionised hydrogen stellar wind is $n_e=n/2$, where $n$ is the total particle density. The condition $f_c > f_p$ is met when
\begin{equation}\label{eq.bp}
{B_p} > \left( \frac{n_e}{10^5 ~\textrm{cm}^{-3}} \right)^{1/2} \textrm{G ~~~~or ~~~~} n_e < 10^5 \left( \frac{1 \rm G}{B_p} \right)^2  ~\textrm{cm}^{-3}. 
\end{equation}
The density is related to the mass-loss rate of a stellar wind by the mass continuity equation. Therefore, the condition expressed in Equation (\ref{eq.bp}) can be translated into
\begin{equation}\label{eq.bp2}
{B_p} \gtrsim  \left( \frac{\dot{M}}{10^{-12}~\msano} \right)^{1/2} \left( \frac{400~ {\rm km/s}}{u (r_{\rm orb})}  \right)^{1/2} \left( \frac{0.05~\rm au}{r_{\rm orb}}  \right)  \textrm{G}, 
\end{equation}
or
\begin{equation}\label{eq.bp3}
\left( \frac{\dot{M}}{10^{-12}~\msano} \right)
\lesssim \left( \frac{B_p}{\textrm{G}} \right)^2 \left( \frac{u (r_{\rm orb})}{400~ {\rm km/s}}  \right)\left( \frac{r_{\rm orb}}{0.05~\rm au}  \right)^2 , 
\end{equation}
where we assumed a mass-loss rate of a steady, spherically symmetric,  fully ionised hydrogen wind: $\dot{M} = 4\pi r_{\rm orb}^2 m_p n_e(r_{\rm orb})  u(r_{\rm orb})$, with $u(r_{\rm orb})$ being the wind speed at the orbital distance $r_{\rm orb}$.  The right panel in Figure \ref{fig.robert} shows the region of parameter space where radio emission of the exoplanet V830 Tau b could propagate through the wind of the host star \citep{2017AA...602A..39V}. As we can see, the propagation depends on the combination of planetary and stellar wind characteristics. As such, if one can detect radio emission from the planet, the planetary magnetic field can be derived from the frequency of the emission (Equation \ref{eq.cyclotron}), and an upper limit of the wind mass-loss rates can be estimated  (Equation \ref{eq.bp3}).

Assuming the planetary radio emission can propagate through the wind of the host star, there is still the question of by how much  the planetary emission is attenuated by the wind. To answer this question, we turn our attention back to the left panel in  Figure \ref{fig.robert} and more specifically to the location of the planet through its orbit with respect to the radio photosphere (dashed curve). The planetary radio emission is least attenuated (i.e., least absorbed by the wind of the host star) when the planet is between the observer and the star at orbital phase $\phi=0$. As the planet ingresses in the region where the wind is optically thick, the emission from the planet will get increasingly more attenuated, due to an increase in optical depth. In the case shown in the left panel of Figure \ref{fig.robert}, the reduction in radio emission from the planet starts at orbital phase $\phi =0.36$, it is maximum when the planet is the furthest from the observer (at $\phi=0.5$) and decreases back again until the planet emerges from the radio photosphere (at $\phi=0.64$). Given that the size of the radio photosphere depends on the physical properties of the stellar wind, if radio emission is detected from the planet for a certain fraction of the orbit, one can then use this information to further constrain the stellar wind properties \citep{2020MNRAS.493.1492K}. 

\subsection{Constraining stellar winds  (upper limits) though X-ray emission}\label{sec.xraydetection}
As I will discuss further on, in the Sun, the bulk of the X-ray emission originates from active regions, whose magnetic loops can confine hot plasma. Nevertheless, a hot stellar wind, optically thin in X-rays, can also contribute to a fraction of the stellar X-ray emission. Assuming that the total stellar luminosity is $L_X = L_X ^{\rm AR} + L_X^{\rm wind}  $, with  the superscript `AR' denoting the contribution from active regions, we have that  $ L_X^{\rm wind}  < L_X $. Therefore, we can use the observed stellar luminosity $L_X$ to infer the upper limit of the wind contribution $L_X^{\rm wind}$. As I will show next, this can then place a constraint (upper limit) in the wind mass-loss rate. 

The X-ray emission of the wind can be estimated as
\begin{equation}
L_X^{\rm wind} =  \int_V \left( \int_{\nu_1}^{\nu_2} \epsilon_\nu d \nu \right) dV 
\end{equation}
where $\epsilon_\nu $ is the X-ray emissivity integrated over the X-ray frequency range $[\nu_1, \nu_2]$ and $V$ is the volume of the emitting wind. Considering the wind X-ray emission is caused by free-free radiation, the emissivity of an optically thin fully ionised hydrogen plasma is given by \citep{1986rpa..book.....R}
\begin{equation}
\epsilon_\nu = 6.8\e{-38} \frac{n_e n_i}{T^{1/2}} g_{\rm ff} \exp \left( -\frac{h \nu}{k_B T}  \right) \textrm{erg~cm}^{-3}~\textrm{s}^{-1}~\textrm{Hz}^{-1} 
\end{equation}
where $g_{\rm ff}$ is the Gaunt factor of the order of unit \citep{1961ApJS....6..167K}, $h$ is the Planck constant and $k_B$ is the Boltzmann constant. Note that, at temperatures from below 1 MK up to several MK, lines dominate the X-ray spectrum. Therefore, when assuming that the emissivity is entirely due to free-free radiation, we are providing a lower limit on the emission, given the probable temperature  range of coronal winds.  Let us first estimate the wind emissivity in the X-ray range
\begin{equation}
\epsilon_X =  \int_{\nu_1}^{\nu_2} \epsilon_\nu d \nu = 6.8\e{-38} \frac{n_e n_i}{T^{1/2}} \int_{\nu_1}^{\nu_2} \exp \left( -\frac{h \nu}{k_B T}  \right) d \nu
\end{equation}
where we assumed $g_{\rm ff}\simeq 1$. Solving the integral analytically, one gets
\begin{equation}
\epsilon_X = 1.4\e{-27} {n_e n_i T^{1/2}} \left[\exp \left( -\frac{h \nu_1}{k_B T} \right)  - \exp \left( -\frac{h \nu_2}{k_B T} \right) \right].
\end{equation}
With this, we get that the wind contribution to the X-ray luminosity is
\begin{equation}\label{eq.lxwind1}
L_X^{\rm wind}  = \int_V \epsilon_X  dV = 1.4\e{-27} {T^{1/2}} \left[e^{-{h \nu_1}/{k_B T}}  - e^{-{h \nu_2}/{k_B T}} \right] \textrm{EM},
\end{equation}
where we assumed the wind is isothermal and the emission measure is
\begin{equation}\label{eq.EM}
\textrm{EM} = \int_V n_e n_i  dV =  \frac{1}{4} \int_V n^2  dV ,
\end{equation}
where we used the fact that for a fully ionised hydrogen wind, we have  $n_i = n_e = n/2$.  

Equation (\ref{eq.lxwind1}) shows that the  X-ray emission coming from the wind  depends on the density profile $n$ of the wind. In general, the relationship between density and wind mass-loss rate is not straightforward, as $\dot{M}$ depends also on the velocity structure $u$ of the wind ($\dot{M} \propto u r^2 n$). However, for non-magnetised, thermally-driven winds at a given temperature, it can be shown that the mass-loss rate scales linearly with $n$, as the velocity structure only depends on the temperature of the wind (see Sect.~\ref{sec.thermalwinds} and in particular Equation \ref{eq.poly_mom}). Therefore, Equation (\ref{eq.lxwind1}) indicates that isothermal winds with higher densities and thus higher mass-loss rates would have higher $L_X^{\rm wind}$.

To compute $L_X^{\rm wind}$, we can use stellar wind models to predict how the density of the wind varies with $r$ and plug this in Eq.~(\ref{eq.EM}). Here, however, we proceed with a simplified approximation. For an isothermal wind, we can approximate the density structure below the sonic point by a hydrostatic density structure
\begin{equation}
n = n_0 \exp \left[ - \frac{R_\star}{H_\star} \left( 1 - \frac{R_\star}{r}\right) \right],
\end{equation}
where $n_0$ is the wind base density, $H_\star = k_B T / (g_\star m_p/2)$ is the scale height of a fully ionised hydrogen wind and $g_\star = G M_\star/R_\star^2$ is the stellar gravity at the surface. Thus
\begin{eqnarray}
\textrm{EM} = \frac{1}{4} \int_V n^2 dV = \frac{1}{4} n_0^2 \int_V  \exp \left[  \frac{2R_\star}{H_\star} \left(  \frac{R_\star}{r} -1\right) \right]dV \nonumber \\
 \simeq \pi R_\star^3 n_0^2 \int_1^{x_{\max}} \exp \left[ \frac{2R_\star}{H_\star} \left( \frac{1}{x} -1\right)  \right] x^2 dx , \label{eq.integral}
\end{eqnarray}
where $x = r/R_\star$. The limits of the integral above should be from the stellar surface $x=1$ to the observer $x_{\max} \to \infty$. However, given the validity of the hydrostatic density structure, the value of $ x_{\max}$ should not be larger than the sonic point. Numerically, we can see that the integrand decays quite fast with $x$, having its maximum of $1$ at $x=1$. For the sake of simplicity, we will take $x_{\max} \simeq 2 R_\star$. For isothermal winds of solar-like stars, this is actually not a bad assumption, as the bulk of the emission for winds with temperatures ranging from $1.0$ to $2.5$ MK occurs within $2 R_\star$. 

For the present-day solar wind, with a temperature of $\simeq 1.5 \e{6}$~K and base density of $n_0 = 10^8$~cm$^{-3}$, solving the integral in Equation (\ref{eq.integral}) numerically gives $\simeq 0.1$ and thus $\textrm{EM}  \simeq 10^{48}$ cm$^{-3}$.  Substituting this in Equation (\ref{eq.lxwind1}), for an X-ray range from 0.2 keV to 10 keV, our estimated X-ray emission of the solar wind is quite low, on the order of $10^{-10} L_\odot$. The total X-ray luminosity of the Sun varies during the solar cycle from about $7.0\e{-8}$ to $1.2 \times 10^{-6}L_\odot$ \citep{2000ApJ...528..537P}, which is 700 to 12000 times larger than our estimated X-ray wind emission. 

As we can see, the present day solar wind indeed contributes to a small fraction of the total solar X-ray luminosity. Although we know the base density of the solar wind, this parameter cannot be easily derived for winds of solar-like stars. We can then ask ourselves, if the Sun were instead a distance star, what would be the base density and, thus mass-loss rates, we would derive from its X-ray luminosity? Given that  $L_X^{\rm wind} \propto n_0^2$, we would obtain a density value that is larger by a factor of $\sqrt{700}$ to $\sqrt{12000}$ the $n_0$ value we adopted before. For the same wind temperature of $\simeq 1.5 \e{6}$~K, this implies that our estimated wind mass-loss rates would be larger by a factor of $\sim 26$ to $110$  and  the derived maximum mass-loss rate of the Sun-as-a-star would have been $\lesssim 5 \times 10^{-13}$ to $2\times 10^{-12}~\msano$. The Sun-as-a-star experiment illustrates that stellar X-ray emission can only provide upper limits for wind mass-loss rates.

There has not been many examples in the literature where the technique presented here has been used to measure mass-loss rates in cool dwarf stars. \citet{1996ApJ...462L..91L} used soft X-ray observations of the M dwarf star YZ CMi to derive an upper limit of its mass-loss rate. \citet{2013MNRAS.436.2179L} used the X-ray observations of the K dwarf HD 189733 to constrain the base density $n_0$, which is a free parameter in their models. 

\subsection{Detecting charge-exchange  induced X-ray emission}\label{sec.chargeX}
The interaction between an ionised stellar wind with a neutral ISM can lead to charge-exchange, in which the ISM neutral atom transfers charge to a solar wind ion during a collision. In this process, a highly-charged ion, particularly oxygen, is excited to a high excitation state, which is then followed by a single or a cascade of radiative decays. These decays lead to emission in the X-ray range. Contrary to the charge-exchange process discussed in the context of \ly\ observations (Sect.~\ref{sec.astrosphere}) that only occurs at the site of the interaction between wind and ISM, the  charge-exchange  induced X-ray  emission should take place throughout the stellar wind, as  neutral ISM atoms penetrate in the astrosphere. This gives rise to an X-ray `halo'. 

This method was first idealised in \citet{2001ApJ...546L..57W}, using the solar wind as an example. Considering that low-mass stars have winds that are embedded in a partially neutral ISM, \citet{2001ApJ...546L..57W} further suggested that  stars with mass-loss rates not much greater than solar could also produce  charge-exchange  induced X-ray  emission. This emission has a distinct profile with a steep rise closer to the star and then a slow decay (see  Figure 1 of \citealt{2001ApJ...546L..57W} for a Sun-as-a-star wind emission). In a follow-up study, the same authors observed Proxima Centauri \citep{2002ApJ...578..503W}, an M dwarf star, but unfortunately the signature of charge-exchange induced X-ray emission was not detected. With their non-detection, they were able to place an upper limit for the mass-loss rate of Proxima Centauri of about 14 times that of the present-day Sun. 

\subsection{Accretion onto white dwarfs as probe of the wind of secondary companion}\label{sec.whiteD}
The last method I would like to present here consists of observing signatures of mass accretion in close binary systems, in which the primary is a white dwarf. The assumption is that material from the wind of the secondary is accreted onto the white dwarf and thus, if one can measure the accretion rate, the wind mass-loss rate of the secondary can be inferred. To the best of my knowledge, this method has not yet been used in systems with solar-like stars, and estimates of mass-loss rates so far exist for some M dwarf stars that are members of eclipsing binary systems \citep{2006ApJ...652..636D,2012MNRAS.420.3281P}.

For the interacting stars where only mass-accretion rates onto the white dwarf companions have been reported \citep{2012MNRAS.420.3281P}, these accretion rate values can be used as lower limits for the mass-loss rates of the M dwarf stars. These are considered lower limits of wind  mass-loss rates, because not necessarily all the mass lost in the stellar wind of the secondary will be accreted into the primary white dwarf. \citet{2006ApJ...652..636D} finds that M dwarf mass-loss rates are about 15 to 100 times larger than white dwarf mass-accretion rates.

As I mentioned before, this method has been used to model mass-loss rates in M dwarfs only (see Table~1 in \citealt{2017MNRAS.470.4026V} for a compilation of such values). At the moment, it is unclear if the coronae and winds of M dwarf stars in these binary systems are similar to those of isolated/non-interacting M dwarf stars.

\bigskip

\noindent From the nine methods presented in Sect.~\ref{sec.obs}, the first four are some of the key methods of wind detection, with the other proposed methods mostly used as case studies and applied on individual stars. 

\section{An observed evolutionary sequence for mass-loss rates?}\label{ref.Mdotevol}
\begin{figure}
\centering
 \includegraphics[width=.8\textwidth]{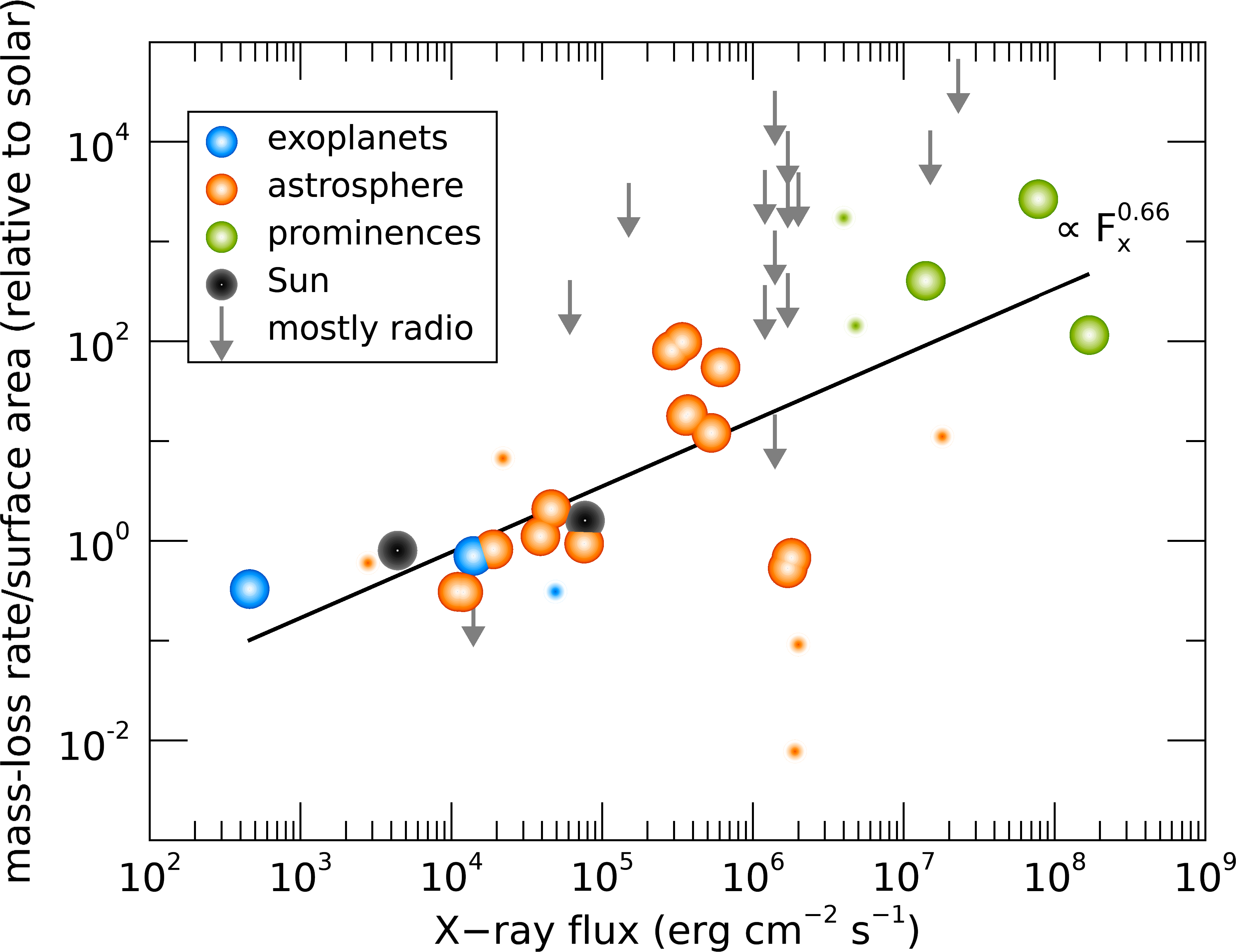}
 \caption{Summary of derived mass-loss rates for low-mass stars combining results from the different methods discussed in Sect.~\ref{sec.obs}. The $y$-axis is given in solar  values, i.e., $\dot{M}_\odot/R_\odot^2$, with $\dot{M}_\odot = 2\e{-14}~\msano$. Colour indicates the method used in the derivation: {blue} for exoplanets, {orange} for astrospheres, {green} for prominences, {black} for the Sun at minimum/maximum of its sunspot cycle. {Grey arrows indicate upper limits, which are mostly derived from radio observations}. The solid line is a power-law fit through the {larger} circles.  The smaller symbols are either evolved stars or M dwarfs, which were not included in the fit and neither were the stars for which only upper limits exist (arrows). The values used in this plot were compiled from the following works: \citet{1993ApJ...406..247D, 1996ApJ...460..976L, 2000GeoRL..27..501G, 2001ApJ...547L..49W,2002ApJ...574..412W, 2005ApJ...628L.143W, 2014ApJ...781L..33W, 2010ApJ...717.1279W, 2018JPhCS1100a2028W, 2002ApJ...578..503W,  2013A&A...551A..63B, 2014Sci...346..981K,  2017A&A...599A.127F, 2017MNRAS.470.4026V,  2017AA...602A..39V, 2019MNRAS.482.2853J, 2019ApJ...885L..30F, lambda_and}. 
 } \label{fig.detection_summary}       
\end{figure}

Figure \ref{fig.detection_summary} summarises the stellar wind measurements derived using the methods discussed here. Colour indicates the method used in the derivation: blue for exoplanets, orange for astrospheres, green for prominences, and the Sun is indicated in black (values at minimum and maximum of cycle are provided). Note that for the radio observations, only upper limits have been extracted and these are represented by the arrows pointing down (a few upper limits obtained from other methods are also included). The larger symbols are the stars that are relevant for this review -- they are main-sequence, solar-like stars, while the smaller symbols are either evolved stars or M dwarfs. The solid line is a fit through the larger circles. Figure \ref{fig.detection_summary} shows that, overall,  mass-loss rates increase with X-ray flux \fx\ and, given that young stars have higher \fx , this implies that young stars have overall higher mass-loss rates. The recent stellar wind measurements from \citet[][green circles]{2019MNRAS.482.2853J}, when contextualised with other measurements, suggest that mass-loss rate continues to increase with \fx\ and that a wind dividing line (cf.~Sect.~\ref{sec.astrosphere}) is no longer required.

Quantifying the wind of the same star through the use of different methods is an ideal approach to verify and validate measurements. Multiple measurements already exists for some stars in Fig.~\ref{fig.detection_summary}. One example is $\pi^1$ UMa -- in Fig.~\ref{fig.detection_summary}, the astrospheric measurement of $\pi^1$ UMa can be identified at $F_X = 1.7 \times 10^6\mathrm{\ erg\ cm^{-2}\ s^{-1}}$ and $\dot{M}/R_\star^2 = 0.53 ~\dot{M}_\odot/R_\odot^2$  \citep{2004LRSP....1....2W}, while radio observations derived an upper limit of  $\dot{M}/R_\star^2 \lesssim 270 ~\dot{M}_\odot/R_\odot^2$  \citep{2017A&A...599A.127F}. The spin down model of \citet{2015A&A...577A..28J}  predicts a ten times larger than solar mass-loss rate for  $\pi^1$ UMa, i.e., at $\dot{M}/R_\star^2 \sim 11  ~\dot{M}_\odot/R_\odot^2$.  Although the upper limit provided by the radio observation does not contradict the astrospheric measurement nor the spin down modelling, the astrospheric measurement for this star is in contradiction with spin-down tracks that are statistically observed for this type of star. As I discussed in Sect.~\ref{sec.astrosphere}, it is possible that  $\pi^1$ UMa is in a high-ionisation region, which might have affected the result obtained in the astrospheric method (see \citealt{2014ApJ...781L..33W}). 
Another star in Fig.~\ref{fig.detection_summary} with multiple wind measurements is the M dwarf Proxima Centauri ($F_X =1.4 \times 10^6\mathrm{\ erg\ cm^{-2}\ s^{-1}}$): the astrospheric method \citep{2001ApJ...547L..49W}, the charge-exchange  induced X-ray  emission \citep{2002ApJ...578..503W} and the free-free radio emission \citep{1996ApJ...460..976L} all produced upper limits of $\lesssim 4\e{-15}$, $\lesssim 2.8\e{-13}$ and $\lesssim 7\e{-12}~\msano$, respectively. Unfortunately, in this case, the three measurements are less stringent, as they are all upper limits. 

Figure \ref{fig.detection_summary} also shows that there is a significant spread in the observations. \citet{2016MNRAS.455L..52V} noted that the more active stars can show a significant variation (cyclic or not) of stellar magnetism in timescales of the order of years. Such variations could increase the spread in stellar wind properties, that is worsened when the X-ray flux and mass-loss rates are not contemporaneously derived. Note for example in Fig.~\ref{fig.detection_summary} the case of the present-day Sun, shown in black, where there is a variation of about one order of magnitude in \fx\ during the solar cycle. 

\begin{figure}
\centering
 \includegraphics[width=.99\textwidth]{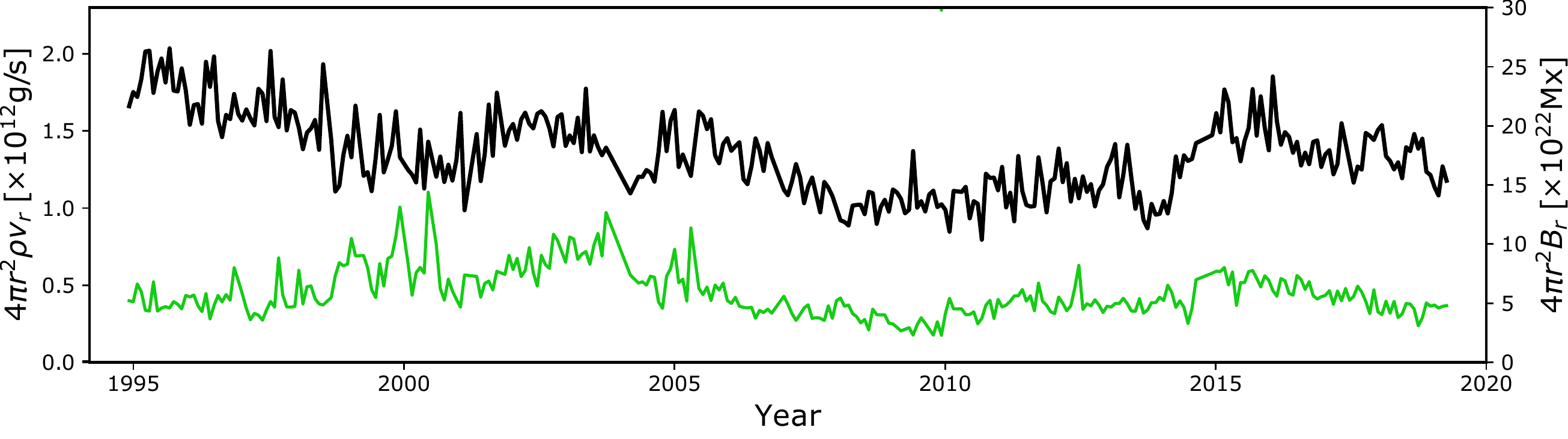}
 \caption{The mass-loss rate of the solar wind  (solid black line) has a small variation during the solar cycle. The green line is the estimated open magnetic flux of the solar wind. Data shows cycles 23 and 24. Figure adapted from \citet{2019ApJ...885L..30F}.} \label{fig.Mdot_cycle}      
\end{figure}

\begin{figure}[!b]
\centering
 \includegraphics[width=.98\textwidth]{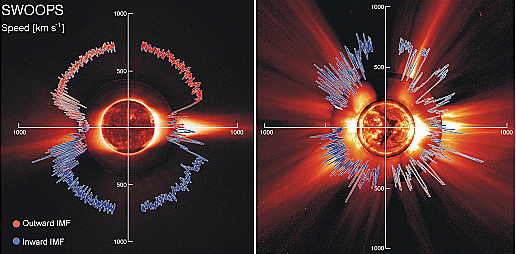}
 \caption{Solar wind variations during its magnetic cycle. Left: At cycle minimum (cycle 22), the solar magnetic field resembles an aligned dipole and the solar wind shows a bimodal velocity distribution, with faster streams emerging from high-latitude coronal holes and slower streams remaining in the equatorial plane. Right: At cycle maximum (cycle 23), the solar magnetic field geometry is more complicated, which is reflected in the solar wind velocity distribution. The background image shows a zoom-in of the solar corona extending out to a few solar radii, while the polar plot shows the solar wind speed measured by Ulysses at several au from the Sun. Colour indicates the magnetic field polarity (red for outward, blue for inward). Figures  from \citet{2003GeoRL..30.1517M}.} \label{fig.ulysses}      
\end{figure}

The mass-loss rate of the Sun, however, does not seem to vary significantly during the cycle \citep{2011MNRAS.417.2592C}, with \citet{2019ApJ...885L..30F} estimating that the solar wind mass-loss rate varies in the  range $\sim [1.6, 3.2]\times 10^{-14}\,\msano$ as shown by the black line in Fig.~\ref{fig.Mdot_cycle} (see also \citealt{1998ASPC..154..131W}).  However, other wind properties, like density and velocity and their latitudinal distribution vary more significantly during the cycle (Fig.~\ref{fig.ulysses}). When the Sun is at minimum activity, its large-scale magnetic field is dominated by a dipolar field, whose axis is roughly aligned with the rotation axis \citep{2003JGRA..108.1035S, 2012ApJ...757...96D, 2018MNRAS.480..477V}. In this case, the solar wind velocity distribution is organised into a faster stream ($\sim 700$--$800$~km/s) emerging from high latitude coronal holes and a slower stream ($\sim 400$ km/s) around the equatorial plane \citep{1998GeoRL..25....1M}. Conversely, at solar maximum, the large-scale magnetic field geometry is more complex, with a dipolar component vanishing and giving rise to higher-order fields, in particular, even-mode components become more important, such as the quadrupole component \citep{2012ApJ...757...96D, 2018MNRAS.480..477V}. The solar wind speed in this case has a more complex distribution, as shown in the right panel of Fig.~\ref{fig.ulysses}. The solar case illustrates the dependence of wind properties on stellar magnetism and later on in this review I will come back to the evolution of stellar magnetism.

I would like now to come back to the fits in Figures~\ref{fig.detection_wood} and \ref{fig.detection_summary}, which  represent the fits to the observations of mass-loss rates for solar-like stars. Using measurements derived from the astrosphere method (Fig.~\ref{fig.detection_wood}), \citet{2005ApJ...628L.143W} found that $\dot{M}/R_*^2 \propto F_X^{1.34\pm 0.18}$, with the fit being valid for $F_X \lesssim 10^6\mathrm{\ erg\ cm^{-2}\ s^{-1}}$. On the other hand, the solid line in Fig.~\ref{fig.detection_summary} fits through all  the solar-like star measurements (larger symbols) derived from several of the methods discussed in Sect.~\ref{sec.obs}. Note that this fit shows no break within the range of $F_X$ shown in the $x$-axis. For those stars, I find a less steep dependence with $F_X$, with a significant (20\%) 1-$\sigma$ uncertainty in the slope of the fit\footnote{Using simple stellar wind models, \citet{2017MNRAS.466.1542S} predicted a very similar relationship: $\dot{M} \propto F_X^{0.79}$.}
\begin{equation}\label{eq.mdot.fit}
\frac{\dot{M}}{R_\star^2} =10^{-2.75\pm 0.68} \left[ \frac{F_X}{{\rm erg\, cm}^{-2}{\rm s}^{-1}}\right]^{0.66\pm0.12}  \,\, \frac{M_\odot}{{\rm yr}~R_{\odot}^{2}}  \, .
\end{equation}  

\begin{figure} 
\centering
 \includegraphics[width=.8\textwidth]{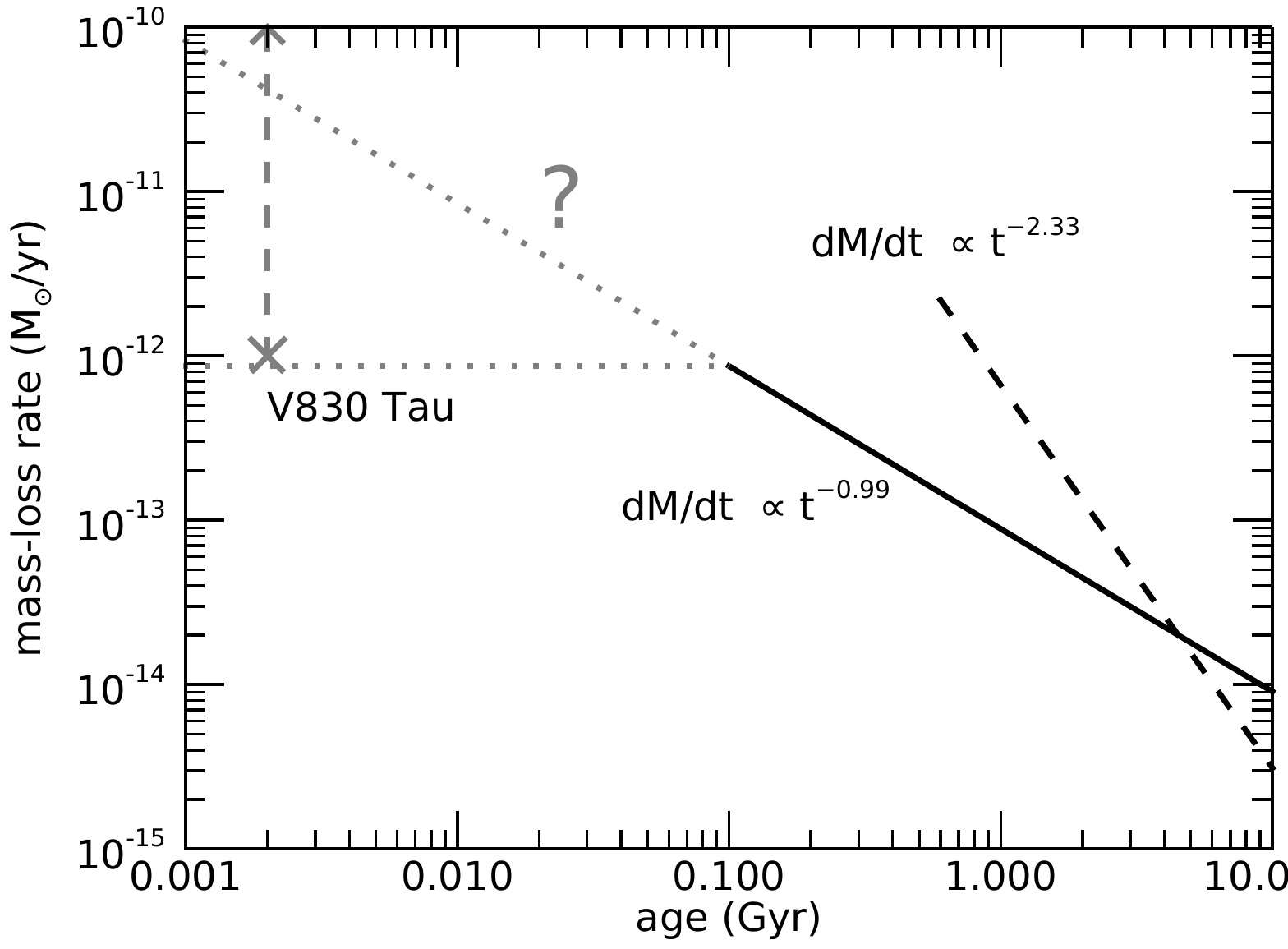}
 \caption{An evolutionary mass loss sequence. Predictions on how the solar wind mass-loss rate would have been in the past. These power-laws are discussed in the text and here they are normalised to match  the present-day solar wind mass-loss rates. Note that the power-laws carry large uncertainties in their slope, which are not illustrated in this figure. The proxy for the young Sun at 2 Myr, V830 Tau, is shown on the left of the plot. The question mark here emphasises that, from observational studies, we are not sure how the solar wind has evolved from the end of the pre-main sequence/beginning of the main sequence, until today. The mass-loss rate saturation discussed in \citet{2015A&A...577A..28J} is shown schematically by the horizontal line.} \label{fig.sequence}       
\end{figure}

I can now derive an approximate evolutionary sequence for the solar wind from this.  I am only focusing on  stars older than $\sim 600~$Myr, after which their rotational evolution has converged (see Sect.~\ref{sec.rotevol}). These are also stars that are no longer in the saturated regime, i.e., those for which X-ray increases with rotation (see Sect.~\ref{sec.activity}). There are several relations in the literature that link X-ray flux or luminosity with age or rotation rate  \citep{2005ApJ...622..680R, 2007LRSP....4....3G, 2014ApJ...794..144R}.  Here, I  adopt the relation from  \citet{2007LRSP....4....3G}, who found that  $L_X \propto t^{-1.5\pm 0.3}$, for solar-like stars in the non-saturated regime. From this, Eq.~(\ref{eq.mdot.fit}) becomes 
\begin{equation} 
\dot{M}   \propto t ^ {-0.99}\, .
\end{equation}
 \citet{2005ApJ...628L.143W} did this exercise, albeit using a different X-ray--age relationship. Using their fit, the  line presented in Fig.~\ref{fig.detection_wood}, \citet{2005ApJ...628L.143W} derived that $\dot{M} \propto t^{-2.33}$, which is steeper than our derivation. These two power-laws are shown in Fig.~\ref{fig.sequence}. Both of these  power-laws indicate that the solar wind had higher mass loss rate in the past. The dispersion is however quite significant, due to all uncertainties in the involved power-laws. For example, the uncertainty in the slope led \citet{2005ApJ...628L.143W} to estimate a solar-wind mass-loss rate in the range   $[5,30]\e{-12}\,\msano$ for when the Sun was 600~Myr-old. 

Regardless of which of the two fits we use ($\dot{M} \propto t^{-2.33}$ or $\dot{M} \propto t^{-0.99}$) for main-sequence stars, there is still a great unknown on how the solar wind has evolved from early stages in the main sequence (or final stages of the pre-main sequence) until the age of $\sim 600$ Myr. Can we extrapolate the $t^{-0.99}$ dependence to early times in the main sequence?  It has also been suggested that mass-loss rates increase with rotation (towards lower ages) but then saturate. This suggestion was discussed in  \citet{2015A&A...577A..28J} and is necessary to avoid the most rapidly rotating stars to spin down too fast. I will come back to this in Sect.~\ref{sec.jdot}. In Fig.~\ref{fig.sequence}, I represent the mass-loss rate saturation only schematically. Figure~\ref{fig.sequence} also shows the mass-loss rate derived for the weak-lined TTauri star V830 Tau, which is considered to be a 2 Myr-old `baby Sun'. With a mass of about $1M_\odot$, this star still has an inflated radius of $2R_\odot$. The upper limit for its mass-loss rate is estimated to be $< 3\e{-9}\,\msano$, with likely values ranging between $[1\e{-12}, {1\e{-10}}]\,\msano$ \citep{2017AA...602A..39V}. 

Summarising the discussion presented in this Section, it  seems  natural that the solar wind has evolved from a high mass-loss rate at early ages until today. However, observations so far do not give us very stringent limits on how precisely this evolution took place.

\section{Observed evolution of the main ingredients of the solar wind}\label{sec.ingredients}
The primary ingredient of the solar wind driving is the stellar magnetism, which can be  measured at the surface of solar-like stars via Zeeman broadening of spectral lines or through polarimetric techniques. Indirectly, surface magnetism can manifest itself through activity proxies, such as through brightness variations due to spots and plages as they come in and out of view in a rotation cycle,  emission detected in the core of certain chromospheric  lines (e.g., CaII H\&K lines), coronal high energy radiation in X-rays and ultraviolet, among others. All these magnetic proxies evolve with stellar rotation, the clock that tracks the passage of time in solar-like stars. 

This clock is ultimately regulated by stellar winds, which remove angular momentum from the star and thus force single solar-like stars to spin down with time. In this Section, I discuss the observational point of view of the evolution of stellar rotation, activity and magnetism as they play important roles in the theory of winds of solar-like stars, which will be discussed in Sect.~\ref{sec.models}. For a recent review on the theory of stellar magnetic field generation and links with stellar rotation, I point the reader to \citet{2017LRSP...14....4B}.

\subsection{Evolution of magnetism}\label{sec.evolB}
\begin{figure}
	\centering
\includegraphics[width=.99\columnwidth]{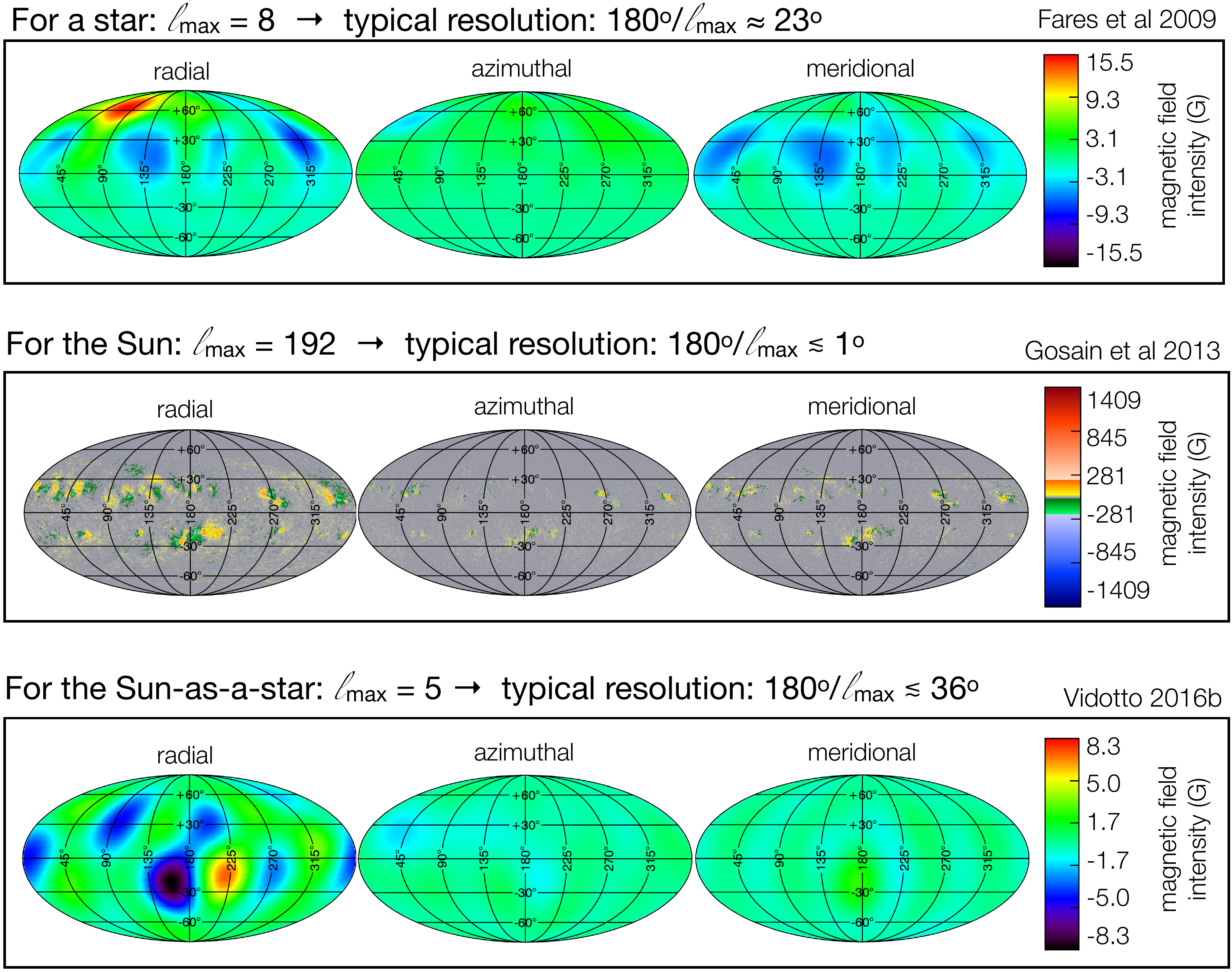}
	\caption{Top: The reconstructed stellar magnetic field using the ZDI technique for the F-type star $\tau$ Boo \citep{2009MNRAS.398.1383F}. Middle: Synoptic map of the Sun plotted with data from \citet{2013ApJ...772...52G}. The solar map, due to its increased resolution, shows a lot more structure in the surface magnetic field. Bottom: To compare the solar map with the stellar map, I filtered out the small-scale magnetic field structure of the Sun \citep{2016MNRAS.459.1533V}. At the top of each panel, I show the maximum harmonic order $\ell_{\max}$ for each map and its typical spatial resolution.}
	\label{fig.zdi_output}
\end{figure}

Stellar magnetism can be observationally probed with different techniques. A technique that has been particularly successful in imaging stellar magnetic fields is the Zeeman Doppler Imaging (ZDI, \citealt{1997A&A...326.1135D}; for a review see \citealt{2009ARA&A..47..333D}). Through a series of circularly polarised spectra (Stokes V) distributed over one or more rotational cycles, this technique has allowed the reconstruction of the large-scale surface fields (intensity and orientation) of more than one hundred cool stars to date \citep[e.g.,][]{2006MNRAS.370..629D, 2008MNRAS.390..567M, 2008MNRAS.388...80P, 2011MNRAS.413.1922M, 2016MNRAS.459.4325M, 2016MNRAS.457..580F}. The top panel in Fig.~\ref{fig.zdi_output} illustrates an output of this technique \citep{2009MNRAS.398.1383F}, in which the three components of the stellar magnetic field are reconstructed: radial, azimuthal (West-East) and meridional (North-South). For comparison, I show in the middle panel a synoptic map from the Sun, produced using SOLIS data  \citep{2013ApJ...772...52G}. The difference between both sets of maps is striking. Firstly, the solar magnetic field strength is an order of magnitude larger than the stellar map. Secondly, we see all this salt-and-pepper structure in solar maps that are not seen in the stellar maps. The reason for these differences lies ultimately in the resolution of the data. Due to its proximity,  synoptic magnetic maps of the Sun can be reconstructed down to much smaller scales than  stellar maps. 

If we decompose these maps using spherical harmonics, a stellar map would typically reach up to a maximum harmonic order\footnote{For fast rotating stars,  higher maximum harmonic orders can be achieved. See for example the case of the young solar-type star HD 141943 with  $\ell_{\max} \sim 30$ \citep{2011MNRAS.413.1922M}).}  $\ell_{\max} \sim 10$. For the solar map, on the other hand, the high resolution allows harmonics of order $\ell_{\max} = 192$ to be achieved. One can estimate how the maximum order is linked to angular resolution as $\Delta \theta \simeq 180^\circ / \ell_{\max}$, which means that the surface of the Sun can be mapped down to $\sim 1^\circ$, while for a star similar to the one shown in the top panel in Fig.~\ref{fig.zdi_output}, the resolution is on the order of $\sim 23^\circ$. 
Because of the lower resolution of ZDI maps, magnetic fields of opposite polarities that fall within an element of resolution cancel out. Given that the small-scale field (e.g., concentrated in spots and active regions) have high intensities and that these fields cancel out, the ZDI maps of older solar-like stars do not reach the large strength fields observed in the Sun and allow only the large-scale field to be reliably reconstructed \citep{2010MNRAS.404..101J, 2011MNRAS.410.2472A, 2014MNRAS.439.2122L}. 

In the bottom panel of Fig.~\ref{fig.zdi_output}, I show how the Sun would look like if it were observed as a star. This is done by filtering out the small-scale field of the solar synoptic map. In practice, I decompose the solar map shown in the middle panel using spherical harmonics as presented in \citet{2016MNRAS.459.1533V}. This allows me to calculate the spherical harmonics coefficients for each given harmonic order $\ell$. Note that the smaller scale structure is described by increasingly larger  values of $\ell$. Thus, to filter out the small-scale component, I only use the derived coefficients up to a maximum harmonic order of $\ell_{\max} =5$  for reconstructing the large-scale field. This method is commonly used in the literature to separate the large-scale field, i.e., the low harmonic orders  \citep[e.g.,][]{2012ApJ...757...96D,2013ApJ...768..162P,2018MNRAS.480..477V,2018MNRAS.478.4390L, 2019MNRAS.483.5246L}. With this, we can now more easily compare the large-scale field of the Sun (bottom panel of Fig.~\ref{fig.zdi_output}) to the ZDI map of a solar-like star (top panel). 

A thorough investigation of the limitations of ZDI was recently presented in \citet{2019MNRAS.483.5246L}, a study that incorporated simulated data in the ``ZDI machinery''. The authors then compare what the ZDI method derives against the controlled input data. This is done for simulated stars with two different inclination angles of the spin axis with respect to the plane-of-sky (20$^\circ$ and 60$^\circ$) and three rotational periods (17, 19 and 27 days). \citet{2019MNRAS.483.5246L} showed that for these  low-activity stars, ZDI recovers relatively well the large-scale field morphologies, but magnetic energies can be underestimated by up to one order of magnitude (roughly a factor of $\sqrt{10}\sim 3$ in field strengths). They also showed that the reconstruction of the field geometry is less reliable if the star is viewed more pole-on (lower inclination angles). For this reason, over the years, ZDI studies have favoured mapping the magnetic field of stars that have intermediate inclinations.  

One of  the major advantages  of ZDI maps is  that they allow the derivation of the three components of the stellar magnetic field topology. Even though this topology is only restricted to the large-scale field, it can still be useful in stellar wind modelling. As I will show in Sect.~\ref{sec.models}, surface maps have indeed been increasingly adopted in 3D simulations of stellar winds \citep[e.g.,][]{2012MNRAS.423.3285V, 2016A&A...594A..95A, 2020A&A...635A.178B} and the limited resolution of ZDI magnetograms has been demonstrated not to affect  stellar wind models \citep{2017MNRAS.465L..25J, 2020A&A...635A.178B}. The reason for this is the following. If you consider a multipolar field, the dipolar field ($\ell =1$) decays with $r^{-3}$; a quadrupolar field ($\ell =2$)  with $r^{-4}$; an octupolar field ($\ell =3$)  with $r^{-5}$; and so on. Generalising, a field with a harmonic order $\ell$ will decay with $r^{-(\ell +2)}$. This means that small-scale fields (i.e., with large $\ell$ values) have very short reach. Thus, it is the large-scale field that is embedded in winds of stars. Stellar winds flow through open magnetic field lines, which ultimately are formed by the stellar large-scale fields that were blown open by the wind outflow. The small-scale field can also affect stellar winds, through, for example, determining the ``micro-physics'' of wind acceleration, which  is key for understanding the heating of the solar lower atmosphere \citep{2005ApJ...632L..49S, 2007ApJS..171..520C, 2019ApJ...880L...2S}. 
 
While the ZDI is blind to the small-scale magnetic field, the Zeeman broadening technique is not. This technique uses Zeeman-induced line broadening of unpolarized light (Stokes I) to derive  the average unsigned surface magnetic field $\bi$ \citep{2007ApJ...664..975J, 2019A&A...626A..86S}. When we compare the unsigned average field strength $\bv$ from ZDI maps with $\bi$, we see that $\bv$ is only a fraction ($\sim 10\%$) of $\bi$ \citep{2009A&A...496..787R, 2010MNRAS.407.2269M}. Note that here I use the subscripts $I$ and $V$ to differentiate between Stokes I and Stokes V derived fields. For a review on this technique, see \citet{2012LRSP....9....1R}.

\begin{figure}
	\centering
\includegraphics[width=0.99\columnwidth]{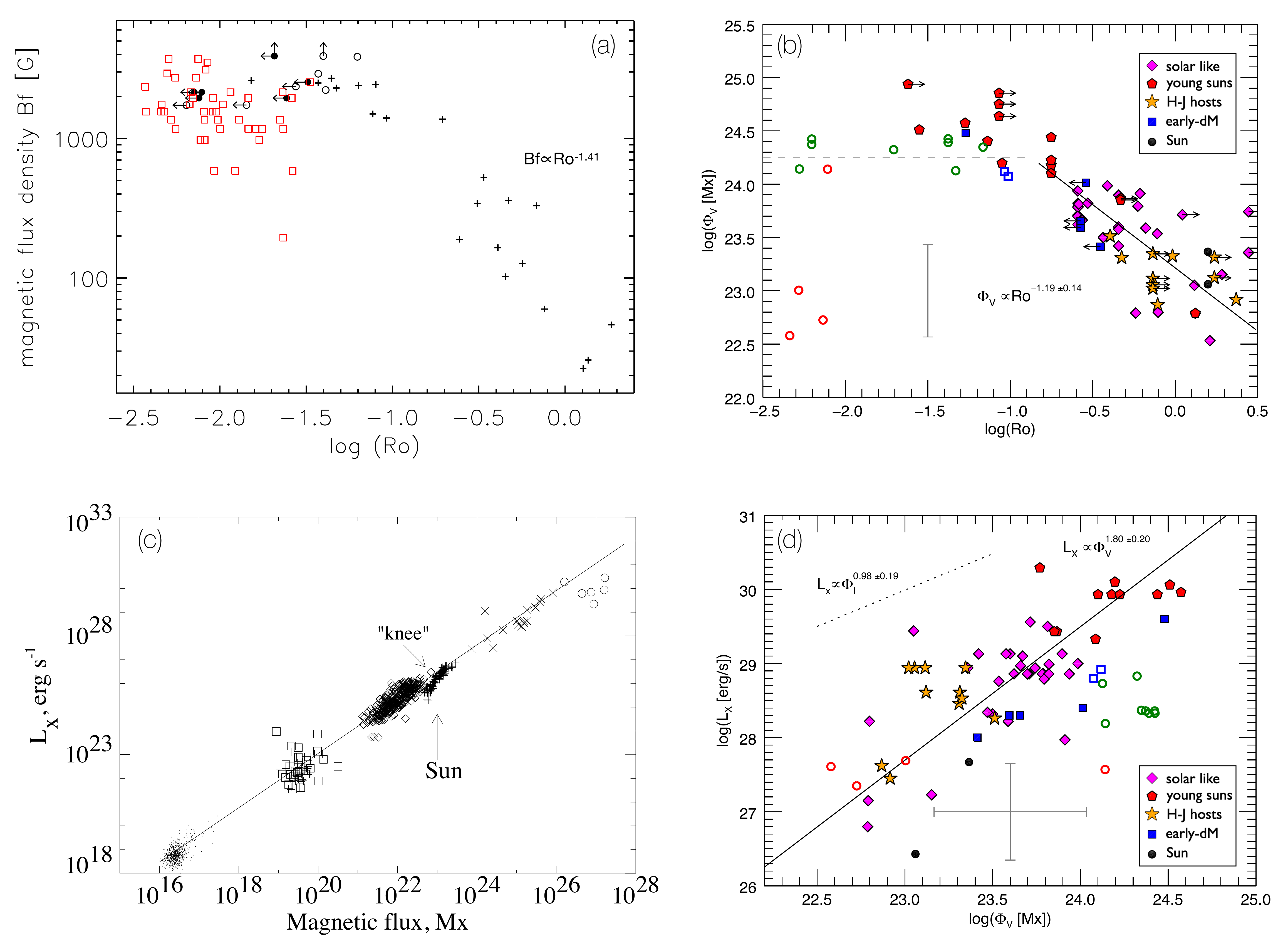}
	\caption{Magnetic flux relations for low-mass stars. The columns on the left show measurements of Zeeman broadening fields and the right columns present measurements of ZDI fields. The top row shows magnetic field as a function of Rossby number and the bottom row shows X-ray luminosity as a function of magnetic flux. (a): Figure from \citet{2012LRSP....9....1R}, where crosses are Sun-like stars, circles are M-type of spectral class M6 and earlier, red squares are late M dwarfs. I performed a rough computation of the power-law dependence with Ro for the unsaturated stars: $\bi \propto {\rm Ro}^{-1.41\pm0.22}$. (b) and (d): Figures from \citet{2014MNRAS.441.2361V}, where different symbols correspond to different ZDI surveys (open circles are the late and mid M dwarfs, not considered in the fits). (c): Figure from \citet{2003ApJ...598.1387P}, where dots are for the quiet Sun, squares for X-ray bright points, diamonds for solar active regions; pluses for solar-disc averages, crosses for G, K, and M dwarfs, and circles are for T Tauri stars. Solid lines shown in the last three panels are power-law fits.}
	\label{fig.mag_flux}
\end{figure}

$\bi$ is the product of the intensity-weighted surface filling factor of active regions $f$ and the mean unsigned field strength in these regions, which is roughly assumed to be the same as the equipartition field $B_{\rm eq}$:  $\bi = f B_{\rm eq}$ \citep{2011ApJ...741...54C, 2019ApJ...876..118S}.  The equipartition field is found by balancing the thermal and magnetic pressure at the photosphere of the star. Recent results  \citep{2011ApJ...741...54C, 2019ApJ...876..118S, 2020A&A...635A.142K} suggest that $B_{\rm eq}$ itself does not change significantly from star to star, but the filling factor $f$ increases  for fast rotators (younger stars). Eventually, the surface of the star becomes covered in active regions and the filling factor saturates for the very fast rotating stars at $f\sim 1$. This leads to a saturation in the magnetic flux of fast rotating stars \citep[Fig.~\ref{fig.mag_flux}a, ][]{2012LRSP....9....1R}. 

Figure \ref{fig.mag_flux} shows a compilation of a few trends derived empirically considering measurements from these two aforementioned techniques: Zeeman broadening (left panels) and ZDI (right panels). The top panels show how magnetism varies with Rossby number\footnote{The use of Ro instead of $P_{\rm rot}$ is commonly found in the literature, as it allows comparison across different spectral types and thus reduces the spread commonly noticed in trends involving $P_{\rm rot}$. One issue to be aware is that the Rossby number is model dependent, as $\tau_c$ is not an observable. }. The Rossby number Ro is defined as the ratio between rotation period $P_{\rm rot}$ and convective turnover time $\tau_c$
\begin{equation}
{\rm Ro} = \frac{P_{\rm rot}}{\tau_c}
\end{equation}
Three panels of this Figure show magnetic fluxes, defined as
\begin{equation}\label{eq.fluxesB}
\phi_I  = 4\pi R_\star^2 \bi  ~~~~~~ {\rm and} ~~~~~~ \phi_V  = 4\pi R_\star^2 \bv
\end{equation}
where the unsigned average field strength from ZDI observations is given by
\begin{equation}
 \bv = \frac{1}{4\pi}\int  |B_{\rm ZDI}(\theta,\varphi)| \sin \theta d\theta d\varphi \, .
\end{equation}
Here,  $\theta$ and $\varphi$ are the surface colatitude and longitude, and $|B_{\rm ZDI}(\theta,\varphi)|$ is the unsigned surface field strength of the ZDI map at $(\theta, \varphi)$. Note that in Figure  \ref{fig.mag_flux}b and d, only the radial component was used to calculate $\phi_V$.

The trend we see in Fig.~\ref{fig.mag_flux}a is also seen in other tracers of stellar activity, such as in X-ray luminosity versus rotation. I will come back to this later on this Section. This trend consists of two parts: a flat, or saturated, part that is independent of stellar rotation, and a power-law that shows magnetism/activity decreasing with the increase of Ro (i.e., towards slow rotators). Saturation occurs in fast rotating stars and the break occurs roughly at Ro $\sim 0.1$.  Here, we are not focusing on the red squares in panel a nor on the open circles in panel b, as those represent mid to late-M dwarfs. Panels a and b in Fig.~\ref{fig.mag_flux} show that both the large-scale field $\bv$ and the total field $\bi$ behave similarly and are summarised as follows: 

\begin{itemize}
\item I computed a rough fit for Ro $\gtrsim 0.1$ for panel Fig.~\ref{fig.mag_flux}a and found a power-law slope of $-1.41\pm0.22$ (cf, with $-1.2$ from \citealt{2001ASPC..223..292S}), which is consistent with the slope found in ZDI studies of $-1.19 \pm 0.14$ (panel b).

\item The saturation in panel a occurs at about $\bi_{\rm sat} \sim 2400$~G, or, assuming an average stellar radius of $1\,R_\odot$ for the solar-like stars, a magnetic flux of $\phi_{I, {\rm sat}} \sim 1.4 \e{26}$~Mx. Saturation for ZDI measurements is currently not clear -- the indicative grey dashed line in panel b is at $\phi_{V, {\rm sat}} \sim 1.8 \e{24}$~Mx (see discussion in \citealt{2014MNRAS.441.2361V})
\end{itemize}

Another very interesting relation is between stellar X-ray luminosities $L_X$ and magnetic fluxes, as shown in the bottom panels of Fig.~\ref{fig.mag_flux}. Combining solar data and Zeeman broadening measurements, \citet{2003ApJ...598.1387P} reported a large dynamical range spanning about 12 orders of magnitude in $L_X$  and $\phi_I$. If we focus only on dwarf stars (crosses in Fig.~\ref{fig.mag_flux}c), the dynamical range is more modest and spans about 2 orders of magnitude, which is also similar to the dynamical range in $\phi_V$ (Fig.~\ref{fig.mag_flux}d). Comparing panels c and d, we see that in both cases, the large-scale field and the total field increase with X-ray luminosity. This is an indication that magnetic fields power coronal activity, in the form of X-rays. Considering only the dwarf stars, \citet{2003ApJ...598.1387P} found a power-law slope of $0.98\pm 0.19$ for Zeeman broadening measurements, while ZDI measurements show a slope of $1.80\pm 0.20 $ \citep{2014MNRAS.441.2361V}. Technically, these slopes are similar within $3\sigma$, but the differences in slopes could also be real. If this is the case, this could indicate  different `efficiencies' in producing large- and small-scale fields  \citep{2008MNRAS.390..545D, 2008MNRAS.390..567M}. One needs to keep in mind that the X-ray values and magnetic field values shown in each panel are not contemporaneous. For example, similar to the Sun, stars have cycles, which affect both field strengths and X-ray luminosities. Thus, non-contemporaneous X-ray and magnetic measurements would lead to increased scatter in the relations shown in Fig.~\ref{fig.mag_flux}. The scatter is seen in all four panels for stellar measurements (it is less evident in panel c, due to the large dynamical range of the plot).

\begin{figure}
	\centering
	\includegraphics[width=.8\columnwidth]{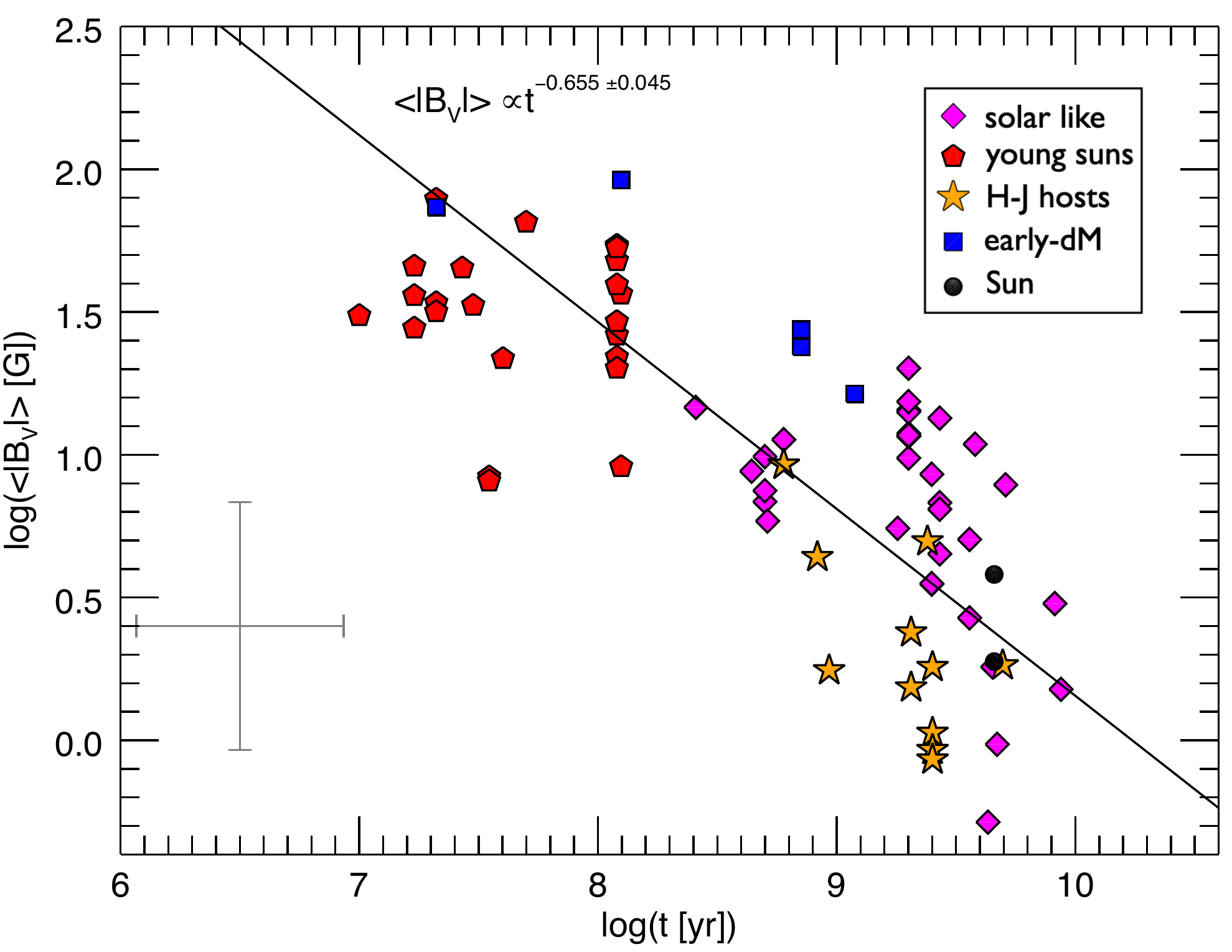}
	\caption{The average unsigned large-scale magnetic field decays with age as $t^{-0.655}$. This is a similar trend as seen in $\Omega_\star(t)$ (Equation \ref{eq.gyrochronology}). Different symbols correspond to different ZDI surveys. Note that the Sun is represented at minimum and maximum phases of its magnetic cycle. Typical error bars are indicated on the bottom left. Figure from \citet{2014MNRAS.441.2361V}.}
	\label{fig.B_age}
\end{figure}

Figure \ref{fig.mag_flux} shows that overall, $\bi$ and $\bv$ follow activity trends that are similar to those reported in chromospheric and coronal activity proxies \citep[e.g.,][]{1972ApJ...171..565S, 2011ApJ...743...48W}. An interesting question still remains: how does the magnetic field \emph{evolve} with age itself? 
Figure \ref{fig.B_age} shows that the average magnetic field intensity decays with age$^{-0.65 \pm 0.04}$  \citep{2014MNRAS.441.2361V}. This trend is valid over three orders of magnitude in field intensity (from G to kG fields) and four orders of magnitude in ages (from Myr to several Gyr). Although the trend is clear, it also presents a large scatter. Part of this scatter is due to short-term evolution of magnetism, including magnetic cycles. Additionally, in the case of  stars younger than $\sim 600$~Myr, part of the scatter is also caused by stars of similar ages and masses, but different rotation rates, showing different levels of magnetisation \citep[see Figure 7a in][where we see a factor of 10 in $ \bv$ for solar-mass stars at 120~Myr]{2016MNRAS.457..580F}. It is fascinating, nevertheless, to see that the relation in Figure \ref{fig.B_age} is in accordance to the observed decrease of $\Omega_\star$ with approximately the square-root of age. I will discuss age--rotation relations in Sect.~\ref{sec.rotevol}. 

\subsection{Evolution of stellar rotation}\label{sec.rotevol}
\begin{figure}
	\centering
\includegraphics[width=.99\columnwidth]{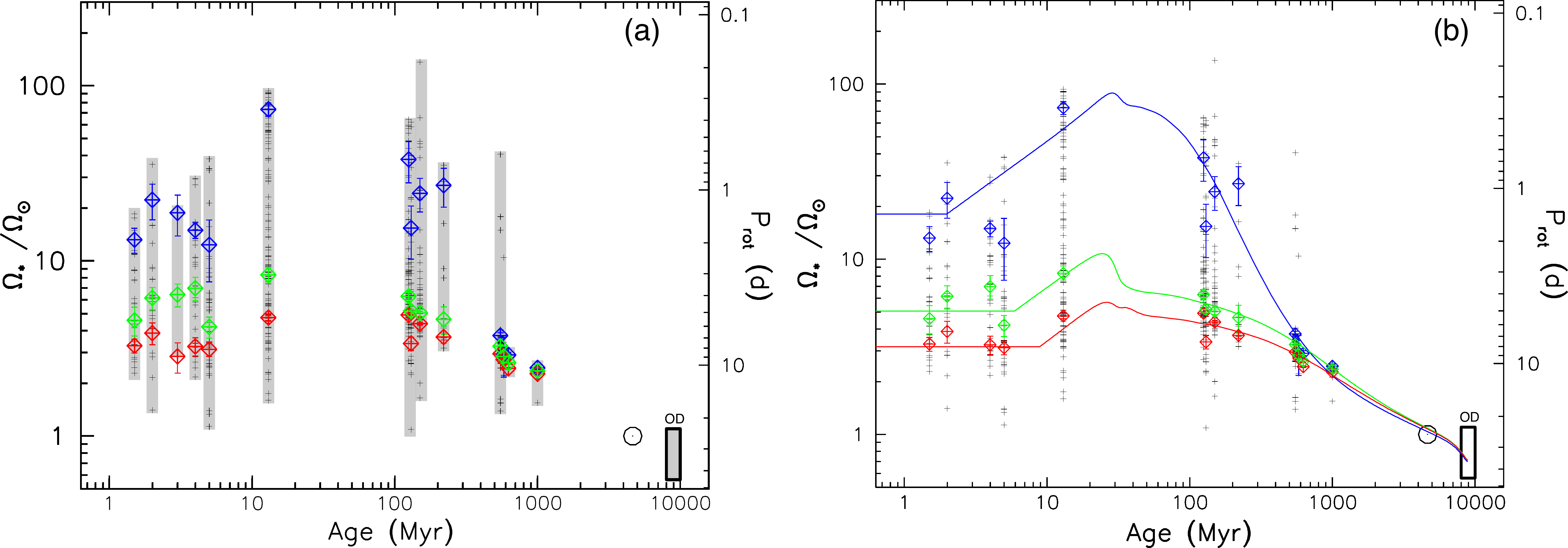}
	\caption{Rotational evolution from solar-mass stars. Pluses are observations of rotation rates of open cluster stars, circle is the present-day Sun and the rectangle shows the velocity dispersion of old disc field stars. The grey shaded bars in panel (a) highlight the velocity dispersion at each age. The red, green, and blue diamonds represent the 25th, 50th, and 90th rotational percentiles, respectively.  In panel (b), models are over-plotted to these observations. Image adapted and reproduced with permission from \citet{2015A&A...577A..98G}, copyright by ESO.}
	\label{fig.fig_gallet}
\end{figure}

Stellar clusters are laboratories for stellar evolution models. In particular, observations of rotation rates of stars in clusters at different ages allow us to draw an evolutionary history for stellar rotation.  Figure \ref{fig.fig_gallet} shows age--rotation diagrams from \citet{2015A&A...577A..98G}. The observations of rotation rates of stars with masses $\sim 1~M_\odot$ are shown as pluses, the Sun is depicted as an open circle and rotational velocity dispersion of old field stars in the Galactic disc is represented by the rectangle labeled `OD' (old disc). The right panel shows, in addition, model tracks as solid (describing the rotation of the envelope) and dotted (same, but for the core) lines. The observations shown in these diagrams are from stars in open clusters, which have  good age estimates. The variety of stellar rotation rates found in each cluster results in the  vertical alignment of the pluses. For each cluster, the blue, green and red diamonds represent the 25th, 50th, and 90th rotational percentiles, respectively, and the associated models are shown in curves with the same colours.

Two clear behaviours are seen in Fig.~\ref{fig.fig_gallet}. Firstly, it is noted that stars, even those with different rotation histories at the early stages of their lives, converge to a unique rotation-age relation approximately at the age of the Hyades open cluster ($\simeq 625$~Myr, \citealt{2011MNRAS.413.2218D}). From then on, observations show that their rotation rates evolve as
\begin{equation}\label{eq.gyrochronology}
\Omega_\star \propto t^{-b}, 
\end{equation}
where $b$ is known as the magnetic braking index. The equation above forms the basis of the gyrochronology dating method  \citep{2003ApJ...586..464B}, whereby a measurement of rotation rate allows one to infer the age of the star.  \citet{1972ApJ...171..565S} derived a braking index of $b\simeq0.5$, but more modern calibrations have suggested  small variations around the square-root. For example,  \citet{2009MNRAS.400..451C} and \citet{2011MNRAS.413.2218D} obtained $b=0.56$, \citet{2008ApJ...687.1264M} found $b=0.566 \pm 0.008$ and  \citet{2016A&A...592A.156D} found  $b= 0.62 \pm 0.36$.  The gyrochronology relation (Equation \ref{eq.gyrochronology}) can be very powerful for stars older than $\sim 600$~Myr, as rotation rates are relatively simple to measure and allows us to derive the {age} of stars. The problem of deriving ages using this method really starts for stars younger than $\sim 600$~Myr. 

 The second behaviour seen in  Fig.~\ref{fig.fig_gallet} is that younger clusters show  a wider dispersion of rotation rates, as opposed to older ones. Given that the dispersion of rotational velocities is already seen at very young, Myr-old open clusters, the dispersion is attributed to different `initial' conditions acquired at star formation. It is believed that a star born as, say, a fast rotator will remain as a fast rotator during its evolution, i.e., it will evolve along the upper envelope shown by the blue solid line in Fig.~\ref{fig.fig_gallet}b.  Broadly speaking, the rotational evolution models represent three phases. Starting from the young ages, we find the first phase, when the profile shows a constant \om . This is the disc-locking phase, in which the accretion disc regulates stellar rotation. Meanwhile, the star is contracting onto the main sequence. Once the disc is dispersed, disc locking ceases to exist and only contraction continues to take place and, mostly due to angular momentum conservation, the star then spins up. This is the only phase in which rotation increases and we see the peak in rotation happening at around 20--30~Myr. From there on, rotation decreases with age, which means that single stars on the main sequence will spin down with time. The main driver for this spin down, as we already discussed in the introduction, is the magnetised stellar wind, which carries away angular momentum. 

The particular model I present in Fig.~\ref{fig.fig_gallet}  \citep{2015A&A...577A..98G} includes more physical aspects, such as the evolution of the moment of inertia (mass, radius) and coupling between the rotations of the core and of the envelope. Other models exist in the literature -- they differ, for example, on which stellar evolution code is used,  different treatments for the core-envelope decoupling, or different prescriptions for the `stellar wind braking law', among others \citep[e.g., ][]{1988ApJ...333..236K, 1991ApJ...376..204M, 1995A&A...294..469K, 2011MNRAS.416..447S, 2012ApJ...746...43R, 2014ApJ...780..159E, 2015A&A...577A..28J, 2015ApJ...799L..23M, 2019A&A...631A..77A}.  The braking law describes how angular momentum loss rate \jdot\ depends on the stellar  properties, such as mass, radius, stellar magnetism, rotation, surface metallicity, and stellar wind mass-loss rate. The constructions of braking laws are  based on theoretical models, empirical scalings, or a combination of both. I refer the reader to \citet{2016A&A...587A.105A} for a study on how different braking laws can affect rotational evolution models and thus the goodness of fits to observed rotation rates. Reviews on the topic include the extensive lecture notes from \citet{2013EAS....62..227P} and \citet{2013EAS....62..143B}. I will come back to angular momentum losses in Sect.~\ref{sec.jdot}.

\subsection{Evolution of coronal activity}\label{sec.activity}
The third wind ``ingredient'' I would like to discuss is stellar activity. In particular, I would like to focus on coronal activity. The corona starts above the chromosphere (and above the transition region) and it is characterised by a high temperature ($\simeq 10^6$K) and low density plasma. The emission from the corona  falls in the X-ray and UV part of the electromagnetic spectrum. The solar wind is, in a sense, a continuation of the solar corona, expanding into the interplanetary space. However, in the cool-star community, it is common to separate the corona as the region of complex, \emph{closed} magnetic field lines and the wind as belonging to the region of \emph{open} field lines. For this reason, the corona is sometimes referred as the `closed corona' and it is the source of X-ray and UV emission, with the wind-bearing regions being X-ray dark (see Sect.~\ref{sec.xraydetection}). There is still no consensus at how precisely the corona and the wind are heated to such high temperatures and it is possible that a combination of different mechanisms heat the plasma in different regions and at different times, as discussed in the recent   review on solar coronal and wind heating by \citet{2019ARA&A..57..157C}.

\begin{figure}[ht]
	\centering
	\includegraphics[height=.6\columnwidth]{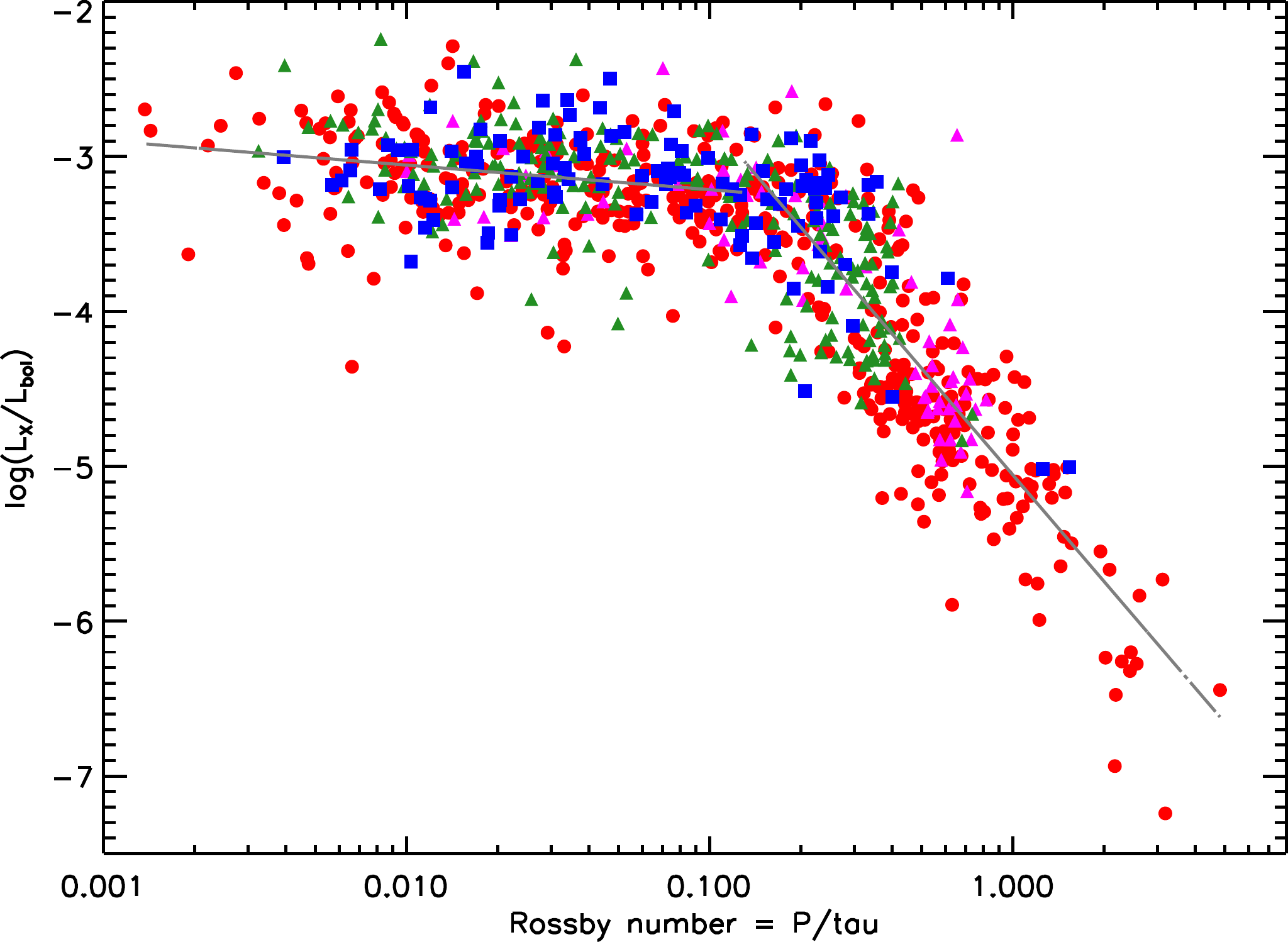}
	\caption{The X-ray-rotation activity relation shows how the normalised X-ray luminosity $R_X = {L_X}/{L_{\rm bol}}$ varies as a function of Rossby number (fast rotators on the left, slow rotators on the right of the $x$ axis). Figure from \citet{2014ApJ...794..144R}.}
	\label{fig.Xray_rot}
\end{figure}

Coronal activity is observed in other stars through X-ray and UV observations. Figure \ref{fig.Xray_rot} shows how the normalised X-ray luminosity $R_X = {L_X}/{L_{\rm bol}}$ evolves with Rossby number. Similar to what was seen in Figures \ref{fig.mag_flux}a and b, the emission saturates at $R_X\simeq 10^{-3}$ for fast rotating stars with Ro $\lesssim 0.1$ and a decrease in X-ray emission is seen with Ro$^{-2}$, or $\sim \Omega_\star^{2}$, in the unsaturated part \citep{2003A&A...397..147P, 2014ApJ...794..144R}. As discussed in Sect.~\ref{sec.rotevol}, rotation can be used as a proxy for age and thus this decay in X-ray can be attributed to age evolution.  Three factors contribute   to the large spread seen in Fig.~\ref{fig.Xray_rot}: inaccuracies in X-ray measurements, variability caused by stellar cycles and intrinsic differences between stars (some being more active than others for same mass, age and rotation, \citealt{2020arXiv200907695J}).

\begin{figure}[ht]
	\centering
	\includegraphics[height=.38\columnwidth]{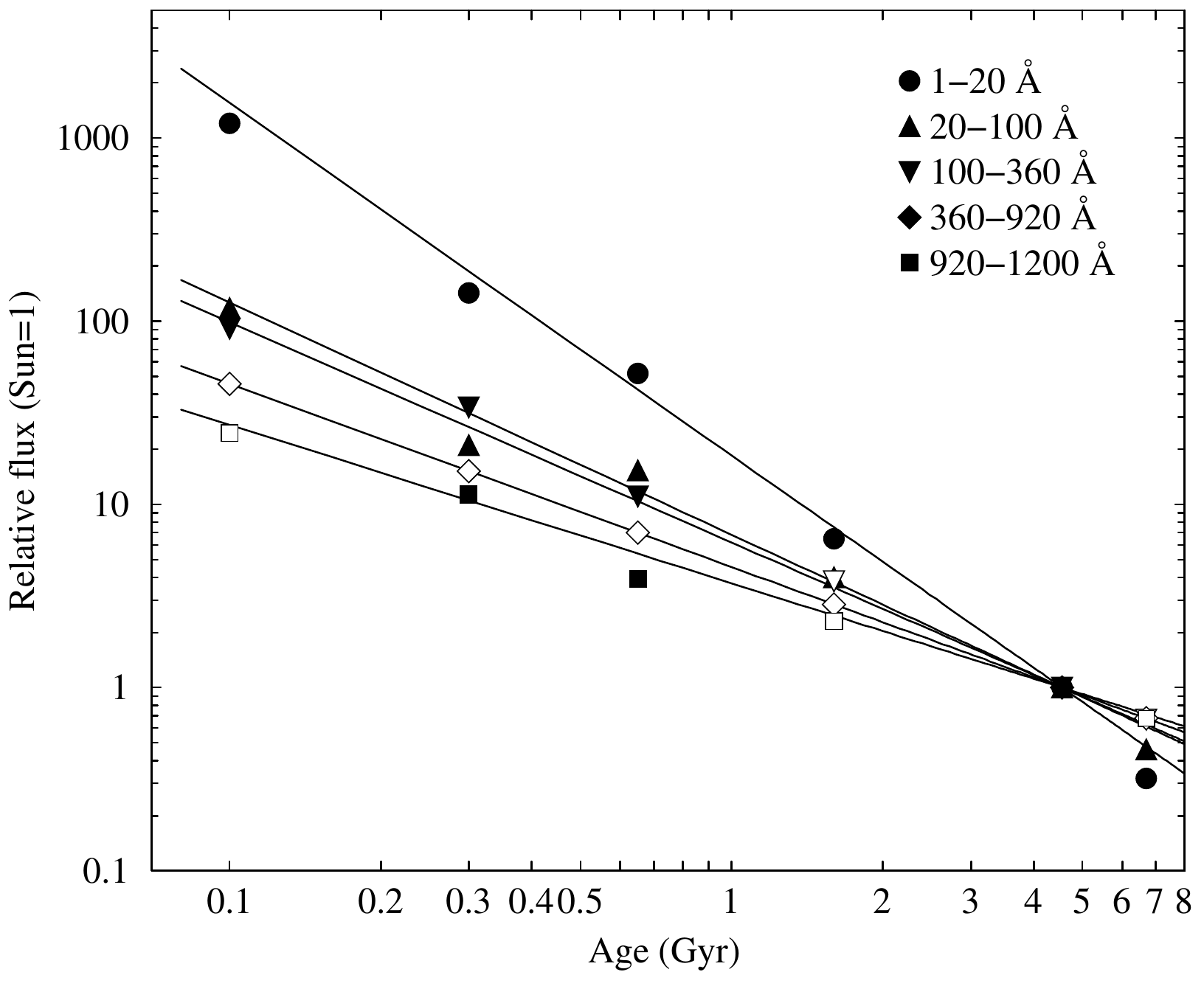}
	\includegraphics[height=.38\columnwidth]{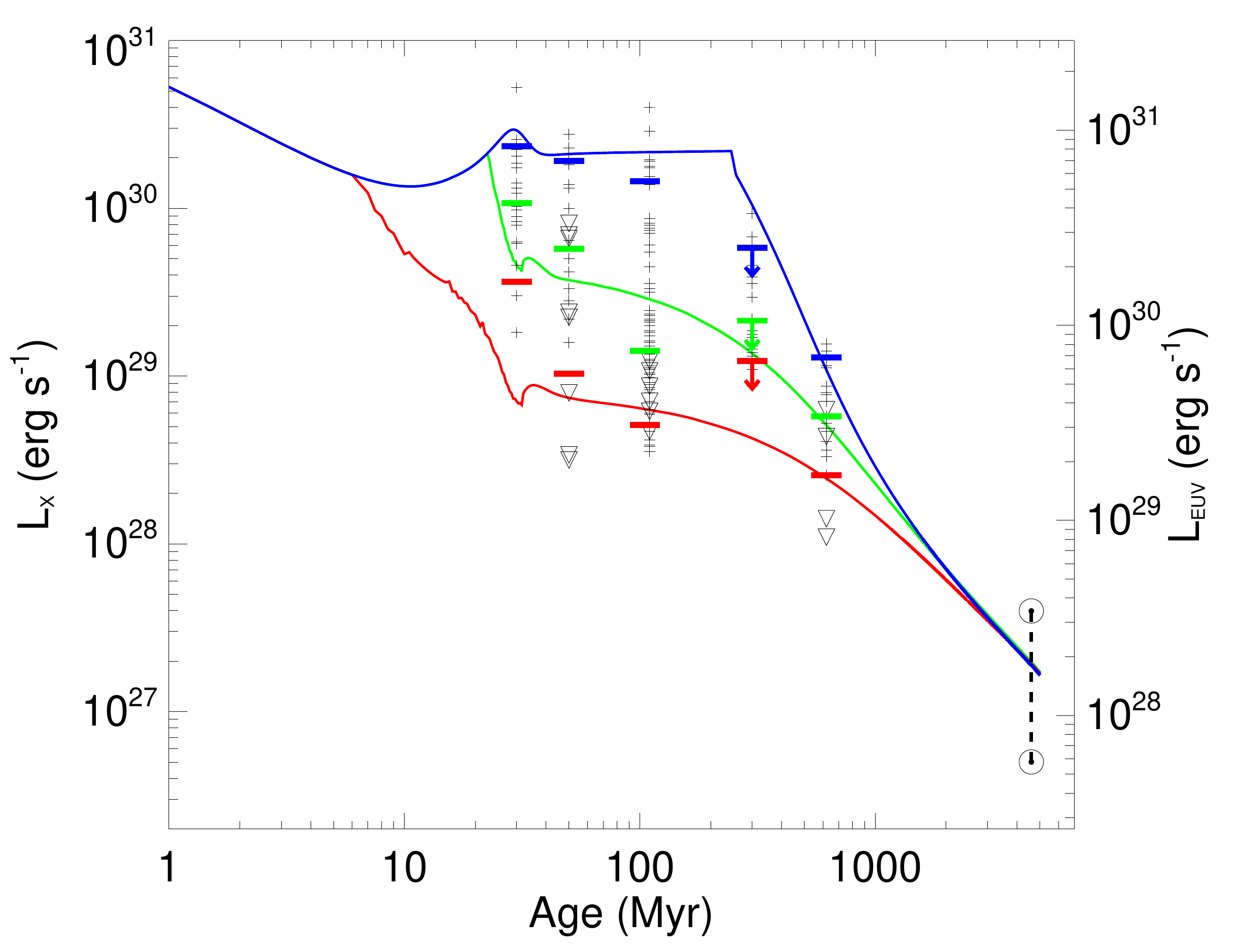}	
	\caption{Evolution of high-energy irradiation of solar type stars.  Left: The observed evolution of high-energy flux at a given orbital distance at different bands. Right: The  X-ray (left axis) and extreme ultraviolet (right axis) luminosities derived from rotational evolution tracks (e.g., Fig.~\ref{fig.fig_gallet}) as a function of stellar age. Images reproduced with permission from [left]  \citet{2005ApJ...622..680R}, copyright by AAS; and [right] \citet{2015A&A...577L...3T}, copyright by ESO.}
	\label{fig.Xray_evol}
\end{figure}

Figure \ref{fig.Xray_evol} shows the evolution of high-energy radiation of solar-type stars. The left panel divides the high-energy flux into 6 bands: the [1,\,20]\AA\ band corresponds to  hard X-rays, [20,\,100]\AA\  to the ROSAT band,  [100,\,360]\AA\ to the EUVE band and  [920, \,1180]\AA\ corresponds to FUSE (Far Ultraviolet Spectroscopic Explorer) bands. There is a gap between 360 and 920\AA\ for solar-like stars, so the fit for this band shown in the figure relies on interpolations from neighbour wavelength bands. The slopes in the power-law fits are shown in Table 5 of \citet{2005ApJ...622..680R} -- they range from the steepest value of $-1.92$ for the hard X-rays to $-0.85$ for the less energetic band. The fluxes are normalised to the present-day Sun values. We see that the contribution of the X-ray part of the spectrum  is quite strong at young ages, compared to present-day solar values. 

The EUV part of the spectrum (100--920\AA) is  relevant in studies of atmospheric escape in planets and exoplanets. \citet{2011A&A...532A...6S} estimated  that the EUV luminosity $L_{\rm EUV}$ is related to the X-ray luminosity (5--100\AA ) as
\begin{equation}\label{eq.sanz}
L_{\rm EUV}  = 10^{4.8}  L_X^{0.86}, 
\end{equation}
with  luminosities  given in erg~s$^{-1}$. Using the X-ray-rotation activity relation (similar to shown in Fig.~\ref{fig.Xray_rot}), the equation above and rotational evolution models (similar to shown in Fig.~\ref{fig.fig_gallet}), \citet{2015A&A...577L...3T} derived an evolutionary model to describe the evolution of $L_X$ and $L_{\rm EUV}$ that is shown in the right panel of Figure  \ref{fig.Xray_evol}. 
The spread in rotation rates observed for young stars implies a spread in XUV (X-ray plus UV) luminosities, such that slow rotators at a given age will have lower XUV luminosities than a fast-rotating star at the same age \citep{2015A&A...577L...3T}. 

More recently, \citet{2020arXiv200907695J} derived a relationship between the surface fluxes in X-ray and in the EUV bands (100--360\AA) and (360--920\AA): 
\begin{equation}
\log(F_{\rm EUV} ^{[100-360]}) = 2.04  + 0.681 \log(F_X) \,, 
\end{equation}
with 
\begin{equation}
\log(F_{\rm EUV} ^{[360-920]}) = -0.341 + 0.920 \log(F_{\rm EUV} ^{[100-360]}) \,, 
\end{equation}
with fluxes given in $\mathrm{erg\ cm^{-2}\ s^{-1}}$. The latter equation is constrained using the Sun only, as there is essentially no measurements in the band (360--920\AA) for stars other than the Sun. To compare to the `total' EUV band (100--920\AA) used in  Eq.~(\ref{eq.sanz}),  one should then sum $F_{\rm EUV} ^{[100-360]}$ and $F_{\rm EUV} ^{[360-920]}$, obtaining
\begin{equation}
F_{\rm EUV} = 10^{2.04} \left(  F_X^{0.681} + 0.31 F_X^{0.626} \right) .  \label{eq.johnstone}
\end{equation}
Comparing Eqs.~(\ref{eq.sanz}) and (\ref{eq.johnstone}), we notice a smaller exponent is the latter. This is because one relation uses flux and the other luminosity.  \citet{2020arXiv200907695J}  demonstrate that using surface fluxes reduce spread in empirical relations between X-ray, \ly\ and overall EUV emissions. 

\begin{figure}[ht]
	\centering
	\includegraphics[height=.48\columnwidth]{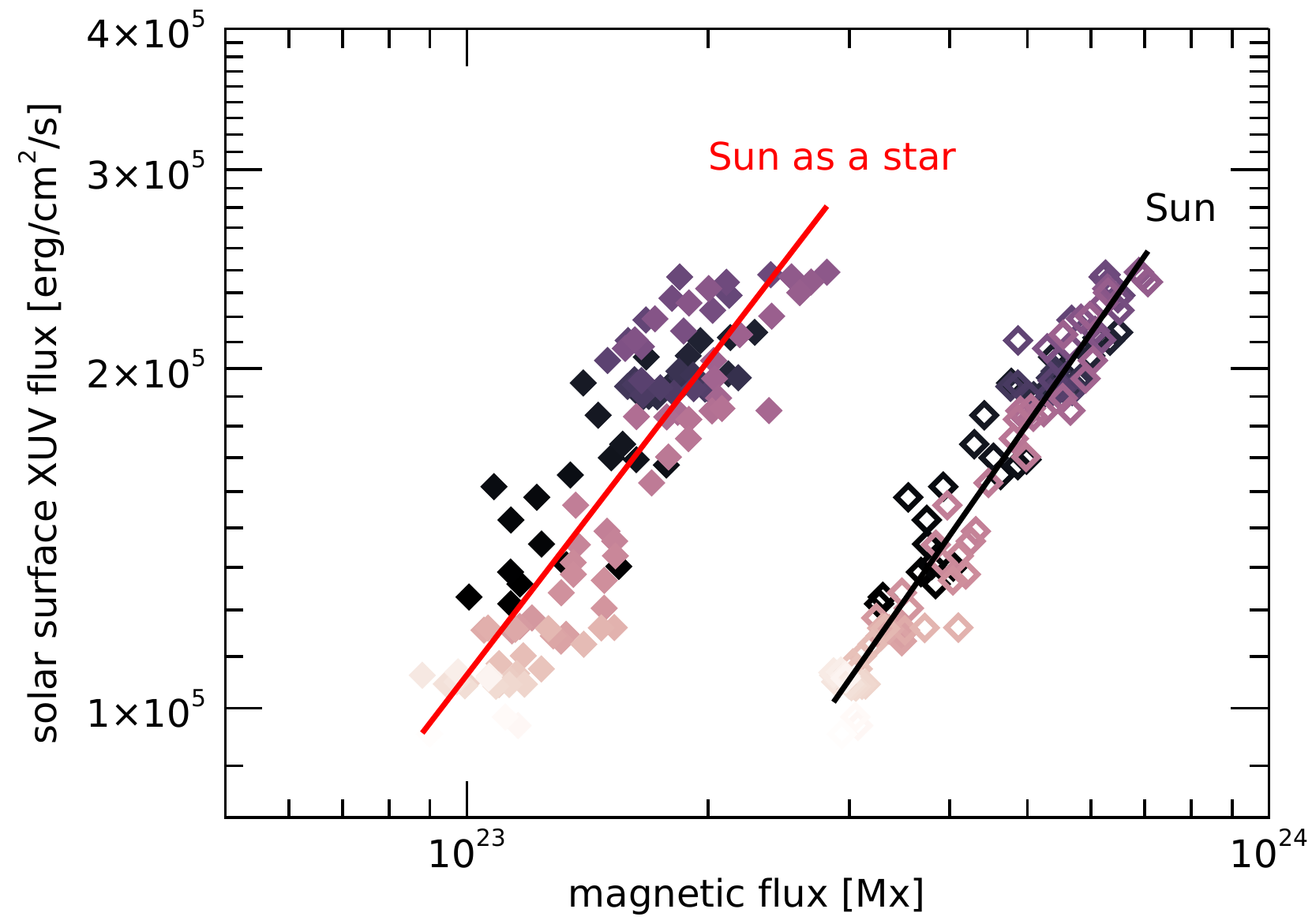}	
	\caption{Solar XUV flux as a function of the surface magnetic flux. Open symbols represent the surface magnetic flux calculated directly from HMI synoptic maps. Filled symbols represent the Sun-as-a-star, where only harmonics up to order $\ell_{\max}=10$ were used. Colours show temporal evolution with darkest colour corresponding to year 2010.5  and lightest colour to year 2019.5 (solar cycle 24). The solid lines show power-law fits to the data. Further discussion is shown in \citet{2020MNRAS.496.4017H}.}
	\label{fig.sunXUV}
\end{figure}

In the Sun, X/EUV flux is very well correlated with sunspot number and the solar surface magnetic flux \citep[Figs.~1 and 8 in][]{2020MNRAS.496.4017H}. Figure~\ref{fig.sunXUV} shows how the solar XUV flux  varies with surface magnetic flux. The XUV flux is calculated over the range 5 to 912~\AA, using data from the Solar Extreme Ultraviolet Experiment (SEE) instrument on the NASA Thermosphere Ionosphere Mesosphere Energetics Dynamics (TIMED) spacecraft \citep{2005JGRA..110.1312W}. The solar synoptic maps used to compute the magnetic fluxes are from the Helioseismic Magnetic Imager (HMI) on board  the Solar Dynamics Observatory \citep{2012SoPh..275..207S}. The empty symbols consider the HMI data in its full resolution, while the filled symbols only considers magnetic field components with  spherical harmonics orders below $\ell_{\max}=10$. This represents the Sun-as-a-star case.

Figure \ref{fig.sunXUV}  shows the relatively tight correlation between surface XUV flux and magnetic fluxes considering both the full resolution maps as well as the low resolution maps  (large-scale fields). This tight correlation led \citet{2020MNRAS.496.4017H} to suggest that the large-scale fields observed in other stars (e.g., Fig.~\ref{fig.zdi_output}) could be used to infer the stellar X/EUV fluxes of stars through the relation 
\begin{equation}
F_{\rm XUV} \propto \phi_V ^{1.04\pm 0.026}, 
\end{equation}
with $\phi_V$ defined in Eq.~(\ref{eq.fluxesB}). Note, however, that if one assumes a lower harmonics order  (e.g., $\ell_{\max}\sim 5$) for the `Sun as a star' case, the scatter in the relation increases substantially. The reason why XUV emission is better related with higher harmonics order is that the solar XUV emission mostly comes from compact active regions, which are characterised by smaller-scale fields, i.e., with higher harmonics orders (see discussion at the beginning of Section \ref{sec.activity}).
 
\section{Where are we in terms of modelling?}\label{sec.models}
The three observed ingredients discussed in Sect.~\ref{sec.ingredients} are often used to inform stellar wind models, which developed from models of the present-day solar wind. Here, I do not go into details of the present-day solar wind models. I refer the reader to  some recent reviews in the area, such as \citet{2018LRSP...15....4G} for a review using fluid-based description and \citet{2006LRSP....3....1M, 2011SGeo...32....1E} for reviews adopting kinetic theory. Other comprehensive reviews in the area, discussing the heating and acceleration of the present-day solar wind, include \citet{2009LRSP....6....3C, 2019ARA&A..57..157C, 2018LRSP...15....1R}.

This section is focused on models of winds of solar-type stars and, in particular, on creating an evolutionary sequence that considers and reproduces the stellar observations available to us, namely rotation, magnetism and activity (Sect.~\ref{sec.ingredients}). In Sect.~\ref{sec.wind_types}, I provide a brief overview of the  main types of models used to describe winds of solar-type stars. One important physical process that is not frequently dealt with in solar wind models is angular momentum loss. This is mostly because solar wind studies have focused on the properties of the Sun \emph{today}: its open magnetic flux, mass flux (densities and velocities) impinging on the solar system objects, and its mass-loss rate. The solar wind is usually treated as a snapshot in evolutionary timescales, and rotation, although considered in solar wind models, has its feedback neglected on the Sun's long-term evolution. For stellar astronomers, rotation is the key for the evolution of the Sun and its wind properties. Thus, in Sect.~\ref{sec.jdot}, I focus on modelling angular momentum losses, after presenting a brief overview of magnetohydrodynamics (MHD)  wind models in Sect.~\ref{sec.wind_model_overview}. 

\subsection{Overview on the different treatments used in stellar wind models}\label{sec.wind_types}
Studies of the solar wind have provided guidance for models of winds of low-mass stars. The detailed physical mechanism(s) that provides the heating and acceleration of  solar-like winds is not known, similarly to the solar wind itself \citep[e.g.,][]{2009LRSP....6....3C}. These winds are believed to be magnetically-driven, in which coupling between stellar magnetism and convection transports free magnetic energy, which in turn is converted into thermal energy in the upper atmosphere of stars \citep{2014MNRAS.440..971M}, giving rise to a hot corona and a hot wind ($\gtrsim 10^6$K). The X-ray emitting stellar corona, set by energy input,  varies with the properties of the star \citep{2004A&A...414L...5J,2004A&ARv..12...71G}, as do the stellar wind properties \citep{2019MNRAS.483..873O}. 

To the best of my knowledge, the currently-available stellar wind models are based on fluid descriptions (magnetohydrodynamics, MHD from now on). Although used to study the solar wind itself,  I am not aware of models that have adopted kinetic theory to study the evolution (or even a snapshot) of winds of solar-type stars. In terms of the fluid description, two modelling approaches are used in the study of the hot coronal winds of low-mass stars. The first one explicitly considers atmospheric heating that is deposited at the photospheric (or sometimes the chromospheric) level, while for the second one, a million Kelvin-temperature wind is assumed (i.e., heating is assumed a priori and thus is ``implicit''). I discuss these different treatments next. 

\paragraph{(i) Explicitly including heating  from the photosphere:} 
A possible scenario to convert magnetic  into thermal energy in the atmosphere of stars involves the dissipation of waves and turbulence, based on scenarios developed for the solar wind  \citep[e.g.,][]{1983ApJ...275..808H, 2008ApJ...689..316C, 2011ApJ...741...54C, 2013PASJ...65...98S, 2014MNRAS.440..971M}. In addition to depositing energy,  the gradient of wave pressure provides a volumetric force that accelerates the wind, similar to thermal pressure gradients.  The modelling approach that explicitly considers heating from the photosphere involves a more rigorous computation of the wave energy and momentum transfer, i.e., the computations are done from ``first principles'' \citep[e.g.,][]{1973ApJ...181..547H, 1983ApJ...275..808H, 2006ApJ...639..416V, 2008ApJ...689..316C, 2013PASJ...65...98S, 2019ApJ...880L...2S}. In these models, the increase in temperature from the colder photosphere to the hotter corona arises naturally in the solution of the equations as does the wind acceleration. Most of the models that treat the stellar wind acceleration starting from the photosphere have focused on the wind dynamics along a single open magnetic flux tube, as, depending on the level of details of the physics involved in the wind acceleration/heating mechanisms, models  can become computationally intensive. In particular, a challenging numerical aspect is the large  contrast between the dense and cold photosphere and the hot and rarefied corona  \citep[e.g.,][]{2012ApJ...749....8M}. 

\begin{figure}[ht]
	\centering
	\includegraphics[height=.35\columnwidth]{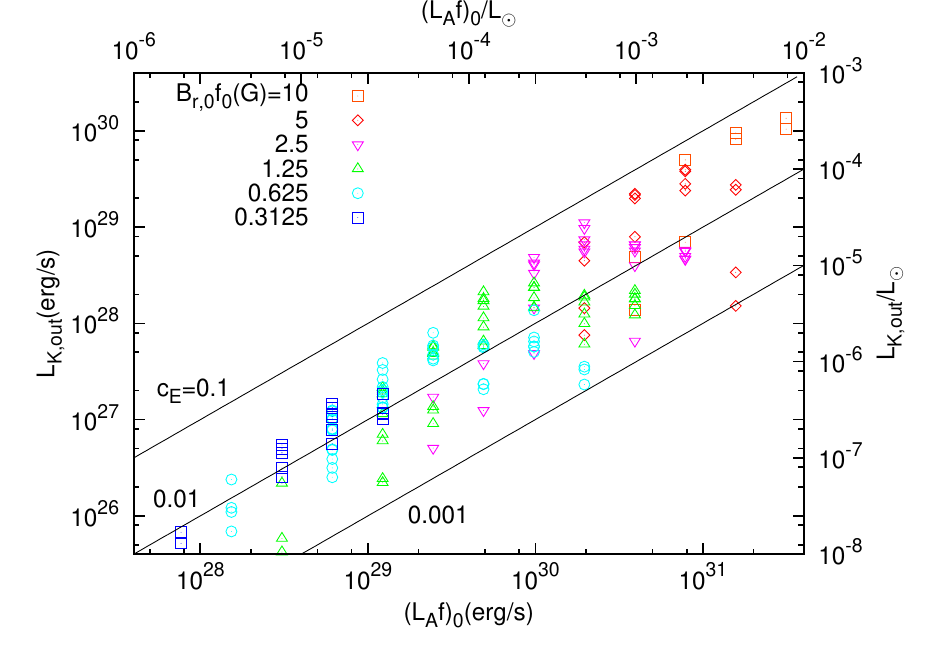}
	\includegraphics[height=.35\columnwidth]{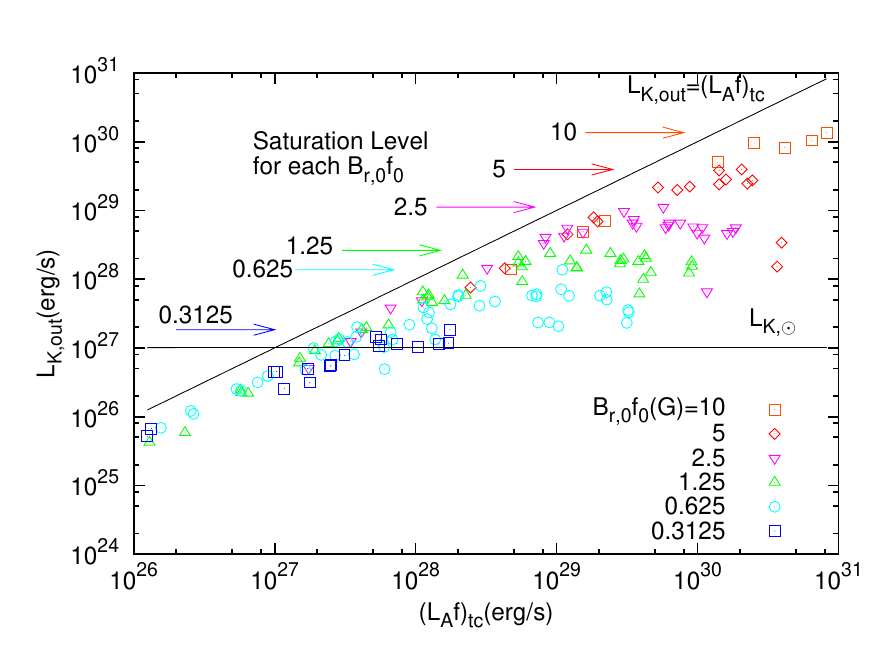}
	\caption{Left: The output kinetic luminosity ($\dot{M} u_r^2/2$, $y$-axis)  of stellar winds increases with the input wave luminosity ($x$-axis) at the photosphere (subscript `0').  Different values of magnetic field strengths  are shown by the different symbols. Curves of constant ratios between $y$-axis and $x$-axis (labeled $c_E$) are shown by the solid lines. Right: The same as the left panel, but instead of the input wave luminosity, the $x$-axis now shows the wave luminosity as measured at the transition region (top of the chromosphere, subscript `tc'). Images reproduced with permission from \citet{2013PASJ...65...98S}, copyright by ASJ. }
	\label{fig.suzuki}
\end{figure}

Models validated for the solar wind have been applied to stellar wind studies. \citet{2013PASJ...65...98S} presented a parametric study of Alfv{\'e}n-wave driven winds applied to solar-like stars.\footnote{Alfv{\'e}n waves are MHD waves that propagate along magnetic field lines and have been suggested as a possible heating mechanism for the solar corona \citep{1947MNRAS.107..211A}.} The authors simulated a  wide range of input wave fluxes, starting from the photosphere, by adopting different values of magnetic field strengths and turbulent velocities, representative of stellar values. The left panel of Fig.~\ref{fig.suzuki} shows a summary of their simulation results, where the  output kinetic luminosity of the stellar wind ($\dot{M} u_r^2/2$, $y$-axis) is plotted against the  the input wave luminosity ($x$-axis). Different values of average magnetic field strengths are shown by the different symbols. Here, $B_{r,0}$ represents the magnetic field intensity at the bottom of a magnetic flux tube and $f_0$ represents the fraction of the star that is covered in open flux tubes. Note the vertical scatter, indicating that similar input wave energies give rise to different wind kinetic energies. The right panel of Fig.~\ref{fig.suzuki} shows a similar plot, but instead of the input wave luminosity, the $x$-axis now shows the wave luminosity as measured at the transition region (top of the chromosphere). Note that these values are smaller than the input wave luminosity at the photosphere, because a fraction of the waves is lost as they reflect back downward, being mostly radiated away. \citet{2013PASJ...65...98S} found that a fraction between 1 and 30\% of the original input wave energy is able to get into the chromosphere. Of this survival wave energy, part of the energy is lost in the form of radiation and part is used to do work against the gravitational force -- thus, only a small amount of the input wave energy at the photosphere is actually used to drive the stellar winds. This study shows that waves cannot easily penetrate in the corona/wind. Overall, for a given magnetic field, an increase in the wave input generates a larger output kinetic energy, until saturation occurs. The consequence of this is that, for a given value of $B_{r,0} f_0$, the mass-loss rate of the stellar wind increases with input wave energy up to a saturation value \mdot$_{\rm sat}$. Beyond this point, the increase in input wave energy seems to no longer increase mass-loss rate. The saturation is also seen to occur at higher values for  larger average fields: \mdot$_{\rm sat} =7.86\times10^{-12} (B_{r,0} f_0/[1 \rm{G}])^{1.62} \,\msano$. This result indicates that stars more magnetically active would have higher `maximum' mass-loss rates. It is interesting to note that this model links not only the magnetic field of an open flux tube, but the product between the magnetic field of a flux tube and the amount of open flux tubes in the surface.

In Sect.~\ref{sec.evolB}, I discussed   surface filling factor of active regions, $f$, mentioning that recent results suggest that the magnetic field in active regions  does not change significantly from star to star, but the filling factor $f$ increases  for fast rotators/younger stars  \citep{2011ApJ...741...54C, 2019ApJ...876..118S, 2020A&A...635A.142K}. The filling factor in active regions is likely linked to  the filling factor of open flux tubes -- if one assumes that the field lines in active regions will remain closed, at the very least we could imagine a scenario where $f \simeq 1 - f_0$, i.e., the star has only $ f_0$ open and $f$ closed field regions (there are no unmagnetised regions in the star in my speculative scenario). Then, the question is: if $f$ saturates and eventually $f\to 1$ for high rotation, does it mean that the star will have no open flux tube and thus no stellar wind if it is very active? This is unlikely. 

Unfortunately, to determine the amount of open magnetic flux, one needs models, such as potential field extrapolations \citep[e.g.,][]{2002MNRAS.333..339J} or stellar wind models \citep[e.g.][]{2014MNRAS.438.1162V, 2015ApJ...798..116R}. Nevertheless, even without models, there is an observationally-derived proxy that could tell us a bit more about how open magnetic flux, i.e., the one embedded in the wind, varies with stellar activity. A large fraction ($\sim 90$\%, \citealt{2009A&A...496..787R, 2010MNRAS.407.2269M}) of the unsigned field derived in Zeeman broadening measurements can be missed in ZDI measurements due to flux of different polarities cancelling out within an element of resolution \citep{2014MNRAS.439.2122L}. The bulk of the ``missed'' field is thus in small-scale fields (in regions that are likely smaller than active-region sizes), which are mostly connected to each other forming closed-field structures. The smaller the loop size, and thus the larger its harmonic order $\ell$, the faster is its decay with radial distance, given that the magnetic field of order $\ell$ decays with $r^{-(\ell+2)}$. These small loops, thus, decay very fast with distance and they do not thread the wind, which represents a more global-scale quantity. The magnetic field that will thread the wind is, on the contrary, a large-scale field, with \citet{2018MNRAS.474..536S} showing that the open flux is largely dominated by the dipolar field alone. ZDI measures large-scale fields, a fraction of which will open up and contribute to the open magnetic flux. Of course, one cannot tell which fraction of the ZDI field will open up and contribute to the wind without a  model, but in light of the two available observational quantities, namely the total  field  (small+large scales) derived in Zeeman broadening measurements and the large-scale field derived in ZDI measurements, the latter represents a better proxy for the wind-bearing magnetic field lines. Thorough discussions on  the fraction of open field in the present-day Sun and on other cool dwarfs can be found in \citet{2017ApJ...840..114C} and \citet{2019ApJ...876..118S}. Given that the magnetic flux in ZDI fields increases with X-ray luminosities (see Fig.~\ref{fig.mag_flux}d), this probably means that the magnetic flux in open field lines increase for more active stars. Thus, even if the closed-field region increases in coverage for fast rotators, the remaining area of open field lines, even occupying a smaller portion of the star, harbours a larger amount of magnetic flux, which ultimately pertains to the stellar wind.

Theory and numerical simulations predict that wind quantities such as mass- and angular momentum-loss rates depend on the amount of open magnetic flux \citep[e.g.,][]{1993MNRAS.262..936W,2014MNRAS.438.1162V, 2015ApJ...798..116R, 2019ApJ...886..120S}, thus estimating the fraction $f_0$ of open field lines is important for stellar wind models. This is also seen in the work of  \citet{2013PASJ...65...98S} -- Figure \ref{fig.suzuki} (left), for example, shows that  the  output kinetic luminosity of the wind depends on the fraction $f_0$ of the photosphere covered in open flux tubes, along which Alfv{\'e}n waves with  luminosity $L_{A,0}$ can propagate. (The wave luminosity itself depends on the $\sim $kG-level magnetic field intensity at the bottom of a magnetic flux tube: $L_{A,0} \propto v_{A,0} \propto B_{r,0}$.)
 
\paragraph{(ii) Implicitly assuming heating in wind models:} The models that explicitly solve for the energy input in the upper atmosphere of Sun-like stars, such as from \citet{2013PASJ...65...98S}, reproduce the rapid increase in temperature from the photosphere to the corona. One of their limitations is that, because they are computationally expensive, they are computed along a flux tube,  and usually do not consider rotation (however, see \citet{2020ApJ...896..123S} for first efforts in including rotation in these types of model). These two limitations can be overcome in the second approach for modelling winds of solar-like stars that I discuss now (albeit other limitations appear).  This approach adopts a simplified energy equation, usually assuming that the thermal variables are connected by a polytropic relation. Typically, a power-law between the thermal pressure $P$ and the density $\rho$  or temperature $T$ are used:  $P \propto \rho^\Gamma$, or $T \propto \rho^{\Gamma -1}$ where $\Gamma$ is known as the polytropic index. A value of $\Gamma=1$, for example, indicates that the temperature is independent of the density, i.e., mimicking an isothermal relation. In the solar wind, the measured effective polytropic index in the corona is around $1.1$ \citep{2011ApJ...727L..32V}, increasing  to $1.46$ near Earth \citep{1995JGR...100...13T} and decreasing with distance in the outer heliosphere \citep{2019ApJ...885..156E}. Usually, polytropic wind models assume constant values of $\Gamma$ ranging from $1$ to $1.15$, although a few models treating a distance-dependent  $\Gamma$ exist for solar wind models \citep{2003ApJ...595L..57R,2007ApJ...654L.163C} and stellar wind models \citep{2009PhDT........94V, 2015A&A...577A..27J, 2015A&A...577A..28J}. 

 In polytropic wind models (which can be magnetised or not), the computation often starts at the point where the temperature has already reached coronal values $\sim 10^6$~K \citep[e.g.,][]{1971SoPh...18..258P,  1993MNRAS.262..936W,  2000ApJ...530.1036K, 2012ApJ...754L..26M, 2009ApJ...699..441V,  2015ApJ...798..116R}. This approach ignores the physical reason of what led temperatures to increase from photospheric to coronal values. Technically, it is as if the wind base occurred at the corona -- if we take the height of the solar corona to be at about 2000~km \citep[e.g.,][]{2001SoPh..203...71G, 2009A&A...501..745Y}, this means that the base of the corona, and of these wind models, occurs at $\sim 1.003\,R_\odot$. In practice, models usually start at $1\,R_\odot$ and this small difference in neglected in simulations.

By assuming a polytropic wind, the energy equation is simplified and no additional equation (e.g., describing wave propagation) is computed. This allows us to not only perform three-dimensional numerical simulations, but also to extend the simulation box to much larger distances from the star. For this reason, I call these ``global'' wind models, or large-scale wind models. These models usually consider rotation, allowing us to calculate angular momentum losses (cf.~Sect.~\ref{sec.jdot}). The obvious drawback of such models is that they do not consider the physics of what happens within $0.003\,R_\odot$ above the photosphere, which is where the bulk of the heating takes place. \label{page.freeparam} Therefore, the temperature, and also the density, at the wind base are free parameters of  polytropic wind models. These free parameters are usually determined from complementary observations, such as X-ray observations, radio observations and observationally-derived mass-loss rates \citep[e.g.,][]{2007A&A...463...11H, 2015A&A...577A..27J, 2015A&A...577A..28J, 2016ApJ...832..145R,2018MNRAS.476.2465O, 2018MNRAS.481.5296V, lambda_and}.

To reduce the uncertainty in these free parameters, some wind models have used X-ray properties to constrain, for example, the density and temperature at the wind base \citep[e.g.,][]{2007A&A...463...11H, 2016ApJ...832..145R, 2019MNRAS.483..873O}. In the Sun, coronal holes are X-ray dark, and the closed field line region is X-ray bright, with the more active/denser region showing higher temperatures \citep{2004ApJ...612..472P}. 
Given that the wind flows through coronal holes, relating wind base properties with X-ray properties is at the end an assumption that the thermal properties of coronal holes are \emph{proportional} to the thermal properties of the closed corona. Such an assumption  implies that, given that the corona of a fast rotating star is hotter \citep{1997ApJ...483..947G, 2005ApJ...622..653T}, their wind should be proportionally hot (corona and wind temperatures are not necessarily the same though). Likewise, stars with denser corona would have denser winds in these models. These wind models thus predict that mass-loss rates are higher for young and fast rotating stars.

An alternative approach that has been used to constrain the wind base temperature in other stars is by assuming that the thermal speed (and thus the temperature) at the base of the wind is a fixed fraction (about 22\%) of the surface escape velocity  \citep{2008ApJ...678.1109M, 2012ApJ...754L..26M}. Because the activity level is not considered in this approach, two stars with same mass and radius, but with different magnetic activity, would have winds with the same temperature.  One uncertainty is that, because  the magnetic field flux at the base of open-field regions increases with magnetic activity,  the heating efficiency may not be the same during the spin evolution. Given that the surface escape velocity does not change considerably during the main sequence, these wind temperatures change by a smaller amount during stellar spin evolution than the approach using X-ray observations. 

\begin{figure}
\centering
\includegraphics[width=0.99\textwidth]{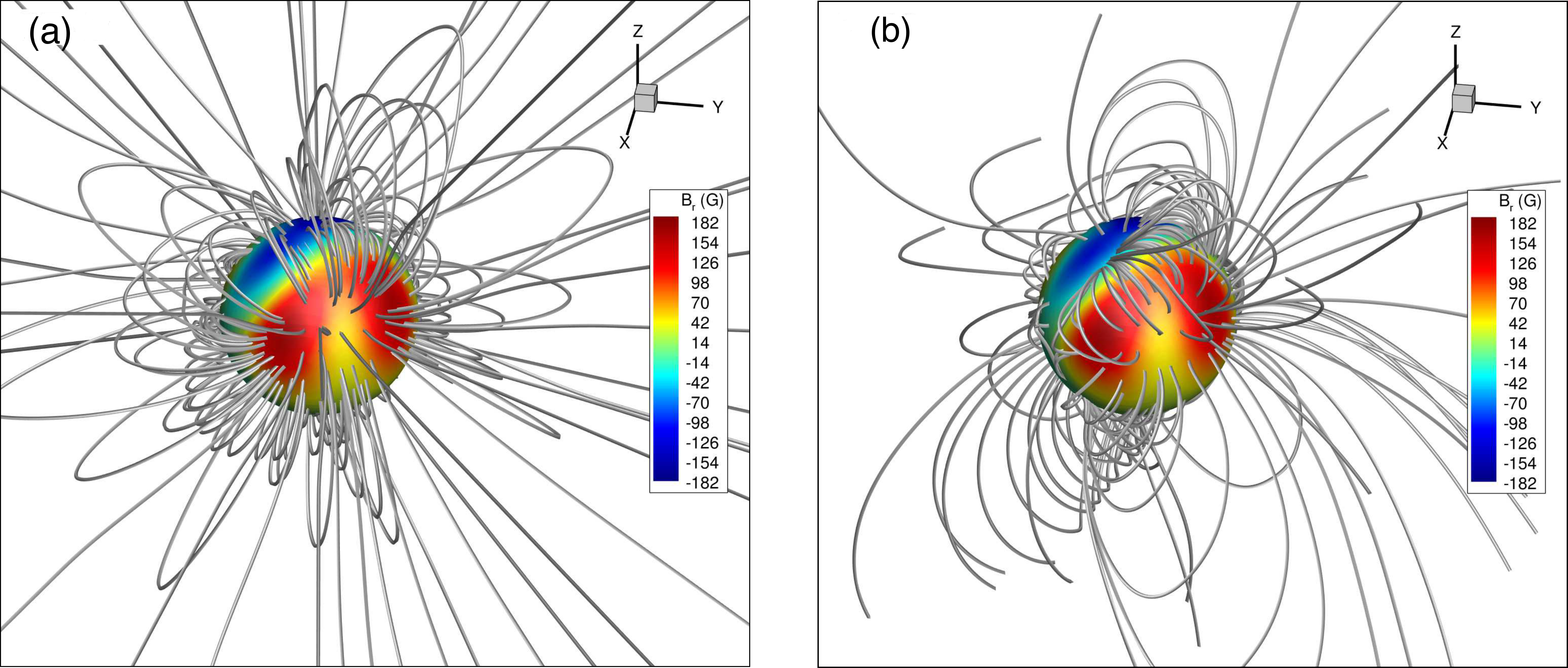}
\caption{(a) Magnetic field line extrapolations, assuming a potential field source surface model. The surface magnetic field is derived from ZDI observations \citep{2008MNRAS.390..545D}. The potential field model assumes the stellar magnetic field is in its minimum energy state. (b) The field lines are stressed after interaction with stellar wind flow. Based on the simulations presented in \citet{2014MNRAS.438.1162V}.}
\label{fig.potential_field}       
\end{figure}

There are two advantages of global wind models that I would like to highlight here. Firstly,  because the numerical grid can extend out to large distances, it is possible to characterise the {stellar wind} conditions around exoplanets \citep{2009ApJ...703.1734V, 2010ApJ...720.1262V, 2012MNRAS.423.3285V,  2015MNRAS.449.4117V, 2011ApJ...733...67C, 2011ApJ...738..166C, 2013MNRAS.436.2179L, 2016MNRAS.459.1907N, 2017AA...602A..39V, 2019ApJ...881..136S}. The characterisation of the local conditions of the stellar wind  is important to quantify the wind (magnetic and particles) effects on exoplanets, as I will discuss later on in this review. Secondly, the global wind models can also incorporate more complex magnetic field topologies, including synoptic maps derived from ZDI studies. The magnetic map is imposed as boundary condition at  the stellar wind base. Usually, the process involves extrapolating the surface field into the computational domain initially assuming the field is in its lowest energy state (e.g., a potential field). After the interaction with the stellar wind flow, the magnetic field becomes stressed. The self-consistent interaction between magnetic field lines and stellar wind  are allowed to evolve, until a relaxed solution is found \citep[for more details, see e.g.,][]{2014MNRAS.438.1162V}. Figure~\ref{fig.potential_field} illustrates the initial (left panel) and final (right panel) conditions of a  3D magnetohydrodynamics (MHD) stellar wind simulation, that adopts a polytropic heating and incorporates observed surface magnetic fields at its inner boundary. 

\begin{figure}
\includegraphics[height=5.5cm]{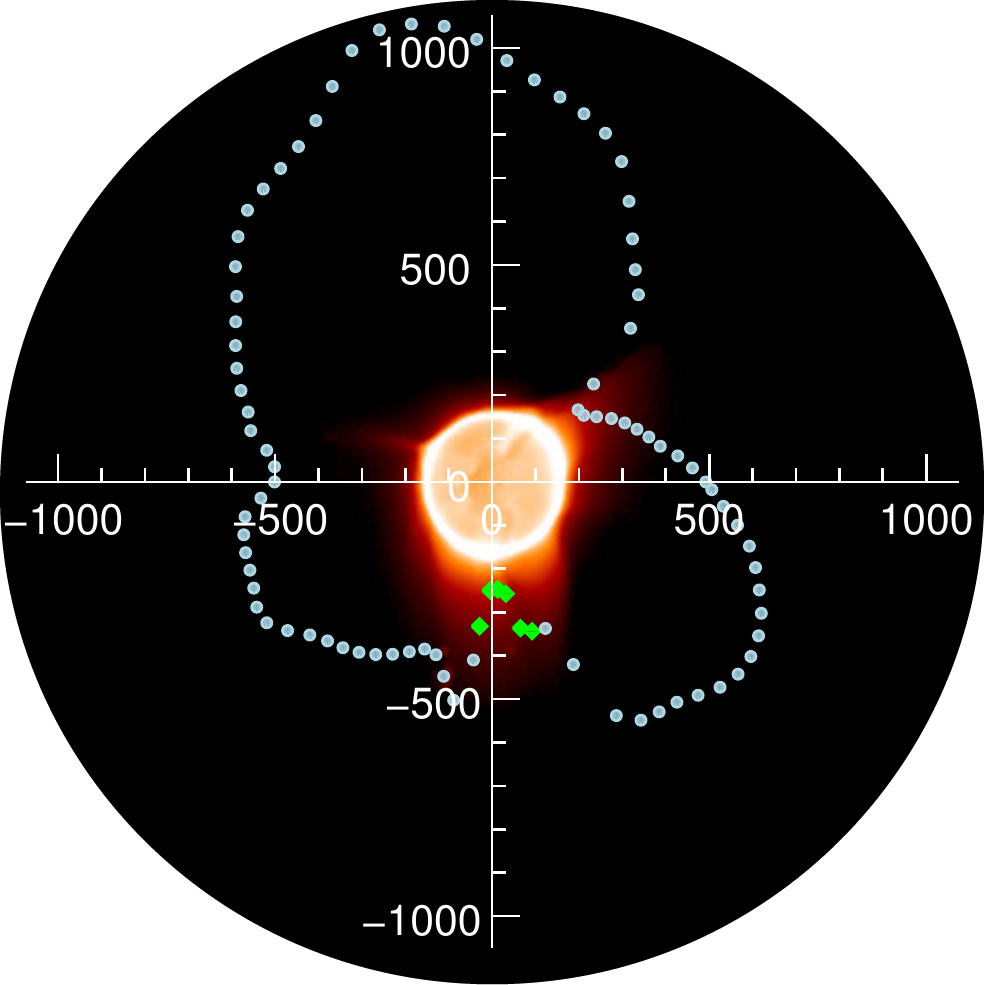} 
\includegraphics[height=5cm]{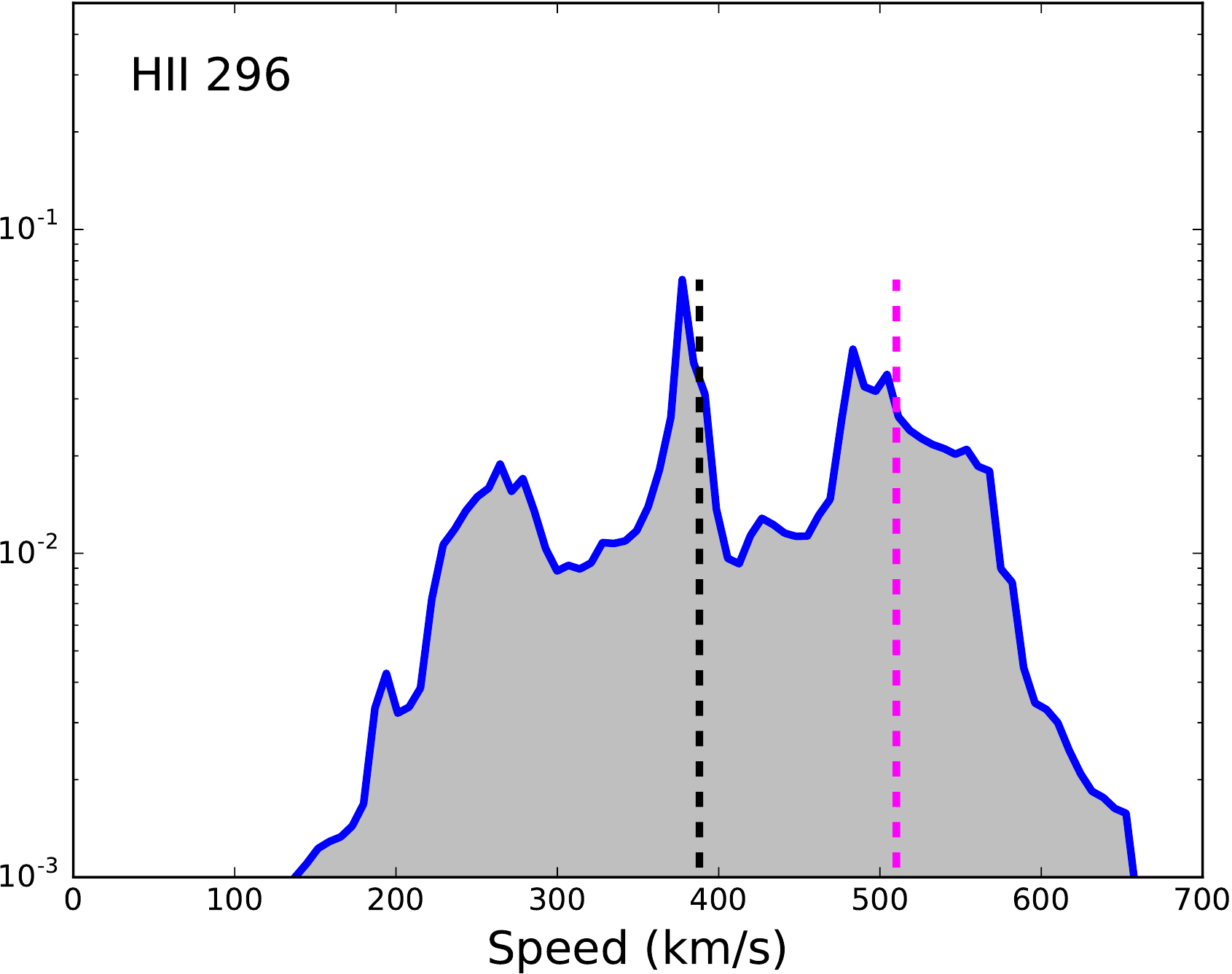}
\caption{Left: Velocity of the simulated stellar wind (in km/s) of the planet-hosting star HD\,189733, at the position of the orbit of the exoplanet HD\,189733b. The blue circles denote an inward component of the magnetic field and the green diamonds denote an outward component. The computed free-free X-ray emission of the coronal wind is shown in the background. The structure that the stellar magnetic field imposes on the X-ray corona is correlated with the structure of the stellar wind. Figure from \citet{2013MNRAS.436.2179L}. Right: Histogram of wind velocities for 3D MHD simulations of the wind of the young solar-like star HII 296. Three velocity components related to magnetic field geometry can be identified: a slow one at 250 km/s emanating from helmet streamers, a fast component at 500 km/s (dashed-magenta line) from expanded flux tubes and an intermediate velocity of 400 km/s. Figure from \citet{2016ApJ...832..145R}.}
\label{fig.av_wind}       
\end{figure}

Being able to include the diversity of observed magnetic field geometries is important. In the case of the Sun, observations of the solar wind with the Ulysses spacecraft  revealed that the geometry of the solar magnetic field affects the velocity distribution of the solar wind \citep{2008GeoRL..3518103M}. When the Sun is in its activity minimum and has a simpler magnetic field topology, close to a dipole, the wind structure is bi-modal, with larger wind velocities in the (high-latitude) coronal holes than in the low-latitude region. On the other hand, when the Sun is at its activity maximum, its magnetic field geometry becomes more complex, which is then reflected in the wind structure. Numerical simulations indeed show that the geometry of the stellar magnetic field affects the wind velocity \citep[e.g.,][]{2009ApJ...699..441V, 2013MNRAS.436.2179L, 2016ApJ...832..145R, 2019MNRAS.483..873O}. Figure \ref{fig.av_wind} shows results of two 3D MHD  simulations of winds of solar-like stars that include the observed surface magnetic fields. 
The left panel of Fig.~\ref{fig.av_wind} shows the solution of the stellar wind model of the planet-hosting star HD\,189733 from  \citet{2013MNRAS.436.2179L} that was computed using the observationally-derived ZDI magnetic map from \citet{2010MNRAS.406..409F}. This panel  has a similar format as Fig.~\ref{fig.ulysses}. The background shows the computed X-ray emission of the hot, quiescent corona of the star due to thermal free-free radiation. Note that coronal {X-ray emission} should come mainly from smaller scale active regions. 
As the small-scale magnetic structure is not resolved in ZDI observations, the X-ray emission computed in \citet{2013MNRAS.436.2179L}  captures only the emission of the background coronal wind and, as such, provide a lower limit for the emission. Overlaid to the computed X-ray image is the velocity of the stellar wind in km/s at the position of the orbit of the exoplanet HD\,189733b. Regions of fast wind correspond to X-ray dark regions in the corona. Slow wind regions tend to lie over the largest helmet streamers in the coronal field. This is a result of the nature of the magnetic force, which generates meridional flows that bring wind from open-line regions (coronal holes) to the top of the streamers \citep{2009ApJ...699..441V}. Therefore the structure that the stellar magnetic field imposes on the X-ray corona is correlated with the structure of the stellar wind. The right panel of Fig.~\ref{fig.av_wind} shows a histogram of wind velocities derived in another  3D MHD simulation of the wind of the young solar-like star HII 296,  where three velocity components can be identified \citep{2016ApJ...832..145R}. A slow component around 250 km/s originates from wind emanating from helmet-streamer structures of the magnetic field, a fast component around 500 km/s originates due to magneto-centrifugal effects and flux tube expansion (dashed magenta line), and an intermediate component at 400 km/s, which coincides with the non-magnetised polytropic wind solution (dashed black line). \citet{2016ApJ...832..145R} showed that the result of the interaction of the flow with strong, non-axisymmetric field generates a complex wind-speed distribution, which is wider in the case of stars with higher magnetisation and faster rotation \citep[see also Figure 5 in][]{2019MNRAS.483..873O}. The intermediate component, which is similar to the non-magnetised polytropic solution,  originates from regions of weak (and likely radial) field, where the interaction of the wind flow with the magnetic field is the weakest \citep{2016ApJ...832..145R}.

\paragraph{(iii) Hybrid approach:} Recently, there have been efforts in developing a hybrid model that combines the two approaches described above to study {winds of low-mass stars}. In these hybrid models, a phenomenological approach of the (solar-based) wave heating mechanism is implemented in three-dimensional simulations of solar/stellar winds, starting from the upper chromosphere \citep{2010ApJ...725.1373V,2014ApJ...782...81V, 2013ApJ...764...23S,2016A&A...595A.110G, 2016A&A...588A..28A, 2016A&A...594A..95A, 2017ApJ...835..220C, 2020A&A...635A.178B, lambda_and}. These models present a step forward in the 3D modelling of winds of low-mass stars, as they, for example, do not need to impose a polytropic index to mimic energy deposition in the wind. However, the full computation of the heating process from the photosphere is still not available. 

In the hybrid models, there are also free parameters that need to be considered, such as the energy flux of waves at the inner boundary and its dissipation length scale  \citep{2013ApJ...764...23S, lambda_and}. In the case of the solar wind, these free parameters can be constrained from solar observations \citep{2013ApJ...764...23S, 2014ApJ...782...81V}. In the case of winds of other stars, it is less clear how to constrain these free parameters. Of particular relevance is the wave energy flux, or Poynting flux, at the inner boundary of these models (chromosphere). \citet{2016A&A...594A..95A} and \citet{2017ApJ...835..220C}, for example, followed the approach of \citet{2013ApJ...764...23S} who used empirical relation between X-ray and magnetic fluxes to constrain the wave input Poynting flux. While in a polytropic formalism the heating is not directly related to the surface magnetic field, in their hybrid approach, the input wave flux at the chromosphere is taken to be proportional to the local surface magnetic field strength \citep{2013ApJ...764...23S}. By assuming that the Alfv{\'e}n wave turbulent dissipation is the main source of heating of the solar wind/corona, \citeauthor{2013ApJ...764...23S}'s approach implies that the total heating power scales with the magnetic flux (Equation 1 in \citealt{2013ApJ...764...23S}), with  a  fraction of this total power generating the observed X-ray luminosity seen in scalings from \citet{2003ApJ...598.1387P}. \citet{2013ApJ...764...23S} derived the constant of proportionality between the Poynting flux and the local surface magnetic field as $F_A / B \simeq 44.7 (4\pi R_\star^2)^{0.1488} \simeq 1.1\times 10^5 (R_\star/R_\odot)^{0.2976}$~erg~cm$^{-2}$~s$^{-1}$G$^{-1}$. Note the stellar size dependence, which should be considered when this implementation is applied to other stars \citep{2017ApJ...835..220C}. 

This is not the only approach to  determine the input wave energy flux. An alternate one was discussed in \citet{2020A&A...635A.178B}, who constrained the wave flux using far-ultraviolet spectral lines that are formed in the upper chromosphere or transition region of stars. For completeness, I also highlight alternative approaches for determining wave energy fluxes used in 1D models (which start from the photosphere, contrary to the hybrid studies discussed in this section, that start in the chromosphere). \citet{2011ApJ...741...54C} and \citet{2017ApJ...840..114C} constrained the Poynting flux used in their models from the magneto-convection models of \citet{2002A&A...386..606M}, who studied the dependence on the emerging wave energy flux with stellar gravity under the mixing-length theory. The gravity dependence can be particularly relevant for stars that evolve off the main sequence  \citep{lambda_and}. Dependence with mass and metallicity of the emerging wave energy flux has also been considered in the work of \citet{2018PASJ...70...34S}, when investigating winds of population III stars. In this case, the amplitude of the wave velocity fluctuations $\delta v$ (note that $F_A \propto \rho \langle \delta v^2\rangle$) at the photosphere is determined from the surface convective flux, which is proportional the stellar luminosity: $\rho  \delta v^3 \propto L_\star$ (see their Eq.~15). 

By considering a Poynting flux dependent on the local magnetic field strength, the Poynting flux varies across the stellar surface. As a result,  hybrid 3D models show stronger contrast between the wind properties  and the large-scale magnetic field geometry than those based on a purely polytropic formalism. This technique is recent in comparison to those discussed in (i) and (ii) and I expect it will be further developed in the years to come.

\subsection{Brief overview of MHD stellar wind theory}\label{sec.wind_model_overview}
Each of the different treatments adopted in wind models have their advantages and their drawbacks. When it comes to modelling rotational evolution, models that explicitly treat heat deposition in the wind are less suitable, as most of these models neglect rotation. To study angular momentum losses, it is more common to adopt polytropic models (i.e., with implicit energy deposition). The first polytropic model for the solar wind was developed by \citet{1958ApJ...128..664P}. This model is, in fact, isothermal, which is a special case of a polytropic wind with $\Gamma=1$. Here, I do not present the derivation of the MHD equations that are often used in stellar wind theory -- those can be found in textbooks \citep[e.g.,][]{1999stma.book.....M, 2004prma.book.....G}. Here, I only present some  key MHD equations that I will use to discuss stellar wind models and angular momentum losses. 

\subsubsection{Thermally-driven winds}\label{sec.thermalwinds}
The first important equation in wind models is that of mass continuity 
\begin{equation}\label{eq.cc}
 \frac{\partial \rho}{\partial t} + \pmb{\nabla} \cdot(\rho \vec{u}) =0 \, ,
\end{equation}
where $\rho$ is the mass density, and $\vec{u}$ is the wind velocity. In the absence of sinks or sources,  mass is conserved in a stellar wind. 

For the momentum equation, it is useful to use the Lagrangian formalism. The Lagrangian time derivative $D/Dt$, sometimes also known as the total derivative, is given by
\begin{equation}
\frac{D Q}{Dt} = \frac{\partial Q}{\partial t} + \vec{u} \cdot \pmb{\nabla}  Q . 
\end{equation}
The $D/Dt$ operator can be applied to scalar $Q$ or vector $\vec{Q}$ quantities. The first term on the right hand side is the rate of change of $Q$ computed at a fixed location (i.e., the Eulerian time derivative) and the second term arises because the fluid element with velocity $\vec{u}$ has moved to a new location where the quantity $Q$ has a different value. This  term is also known as the `convective derivative'. In the Lagrangian description, Newton's second law can be simply written as
\begin{equation}
\rho \frac{ D\vec{u}}{Dt} = \sum \frac{\vec{F}}{\rm volume} \, .
\end{equation}
 The  term of the right hand side represents the sum of all volumetric forces acting on the wind. Once derived, however, the momentum equation is more convenient to use from the Eulerian viewpoint. For a polytropic  (or an isothermal) stellar wind with only gravitational forces and pressure gradients acting on the wind, the momentum equation, using the two previous equations, is 
\begin{equation}
\rho \left(  \frac{\partial \vec{u}}{\partial t} + \vec{u} \cdot \pmb{\nabla}  \vec{u} \right) = -\frac{\rho G M_\star}{r^2} \hat{\bf{r}} - \pmb{\nabla} P \, ,
\end{equation}
 where $\hat{\bf{r}}$ is the unit vector along the radial direction. This equation can be further simplified by assuming steady state (${\partial /\partial t} =0$) and spherical symmetry (${\partial /\partial \theta} =0$, ${\partial /\partial \varphi} =0$):
\begin{equation}\label{eq.cc2}
\rho{u_r} \frac{d u_r}{d r} = -\frac{\rho G M_\star}{r^2}  - \frac{dP}{dr} \, .
\end{equation}
Under these conditions, the continuity equation (\ref{eq.cc}) can be written as
\begin{equation}\label{eq.cc3}
\frac{d}{dr} (\rho r^2 u_r) =0 \,\,\,\,\,\,\,\,\,\,\,\,\,\,\,\, \to   \,\,\,\,\,\,\,\,\,\,\,\,\,\,\,\, \rho r^2 u_r = {\rm constant.}
\end{equation}
Taking into account that the stellar wind emerges from the $4\pi$ steradians of stellar surface, we then define the mass-loss rate as 
\begin{equation}\label{eq.mdot}
\dot{M} = 4 \pi \rho r^2 u_r \, ,
\end{equation}
which is constant with radial distance in a 1D stellar wind. 

Equations  (\ref{eq.cc2}) and (\ref{eq.mdot}) form the basis of the thermally-driven wind model derived by \citet{1958ApJ...128..664P}. The thermal pressure can be written using the ideal gas law
\begin{equation}
P = \frac{\rho k_B T}{\mu m_p} = \rho a^2 \, ,
\end{equation}
where $k_B$ is the Boltzmann constant, $m_p$ the proton mass, $T$ the wind temperature and $\mu$ is the mean ``molecular'' weight. In the case of hot stellar winds, no molecules exist, and the mean molecular weight is simply the average mass of the particles. For a purely hydrogen plasma, fully ionised (50\% protons, 50\% electrons), the mean mass of the particles are $\mu m_p= 0.5 m_p + 0.5 m_e \simeq 0.5 m_p$, where $m_e$ is the electron mass, so $\mu=0.5$. In the equation above, I wrote the isothermal sound speed as $a = \sqrt{k_B T/(\mu m_p)}$. Thus, differentiating the ideal gas law along the radial coordinate, we get
\begin{equation}\label{eq.cc1}
\frac{dP}{dr} =a^2 \frac{d\rho}{dr}   + \frac{\rho a^2}{T} \frac{d T}{dr} \, .
\end{equation}
In Parker's model, the wind is considered isothermal, thus the terms with $dT/dr$ are null in the previous equation. I am going to assume here that the wind is polytropic, which implies that
\begin{equation}
P \propto \rho^\Gamma  \,\,\,\,\,\,\,\,\,\,\,\,\,\,\,\, {\rm or}    \,\,\,\,\,\,\,\,\,\,\,\,\,\,\,\, T \propto \rho^{\Gamma-1}\, ,
\end{equation}
where I used the ideal gas law to convert between the two equations. From this, the temperature gradient is
\begin{equation}
\frac{1}{T}\frac{d T}{dr} =  (\Gamma-1) \frac{1}{\rho} \frac{d\rho}{dr} 
\end{equation}
and Eq.~(\ref{eq.cc1}) is written as
\begin{equation}
\frac{dP}{dr} =a^2 \frac{d\rho}{dr}  + a^2  (\Gamma-1)\frac{d\rho}{dr} = \Gamma a^2 \frac{d\rho}{dr} = c_s ^2  \left(  -\frac{\rho}{u_r}\frac{du_r}{dr}  - \frac{2\rho }{r}\right)  \, ,
\end{equation}
where I used the mass conservation Eq.~(\ref{eq.cc3}) to write ${d\rho}/{dr}$ and wrote the sound speed for a polytropic gas as
\begin{equation}
c_s ^2 =   {\Gamma a^2}   \, .
\end{equation}

Substituting the previous equation in the momentum Eq.~(\ref{eq.cc2}) and rearranging, we arrive at the momentum equation of a thermally-driven wind:
\begin{equation}
\rho{u_r} \frac{d u_r}{d r} = -\frac{\rho G M_\star}{r^2}  + c_s^2 \left(  \frac{\rho}{u_r}\frac{du_r}{dr}  + \frac{2\rho }{r}\right) \, .
\end{equation}
\begin{equation}\label{eq.poly_mom}
\frac{1}{u_r}\frac{d u_r}{d r} = \left( -\frac{ G M_\star}{r^2}  +  \frac{2 c_s^2}{r}\right) / ( {u_r^2}  - {c_s^2} ) \, .
\end{equation}
The equation above reduces to Parker's solar wind equation if one assumes the wind is isothermal and thus $\Gamma=1$ and hence $c_s = a$. 

\begin{figure}
\includegraphics[width=0.49\textwidth]{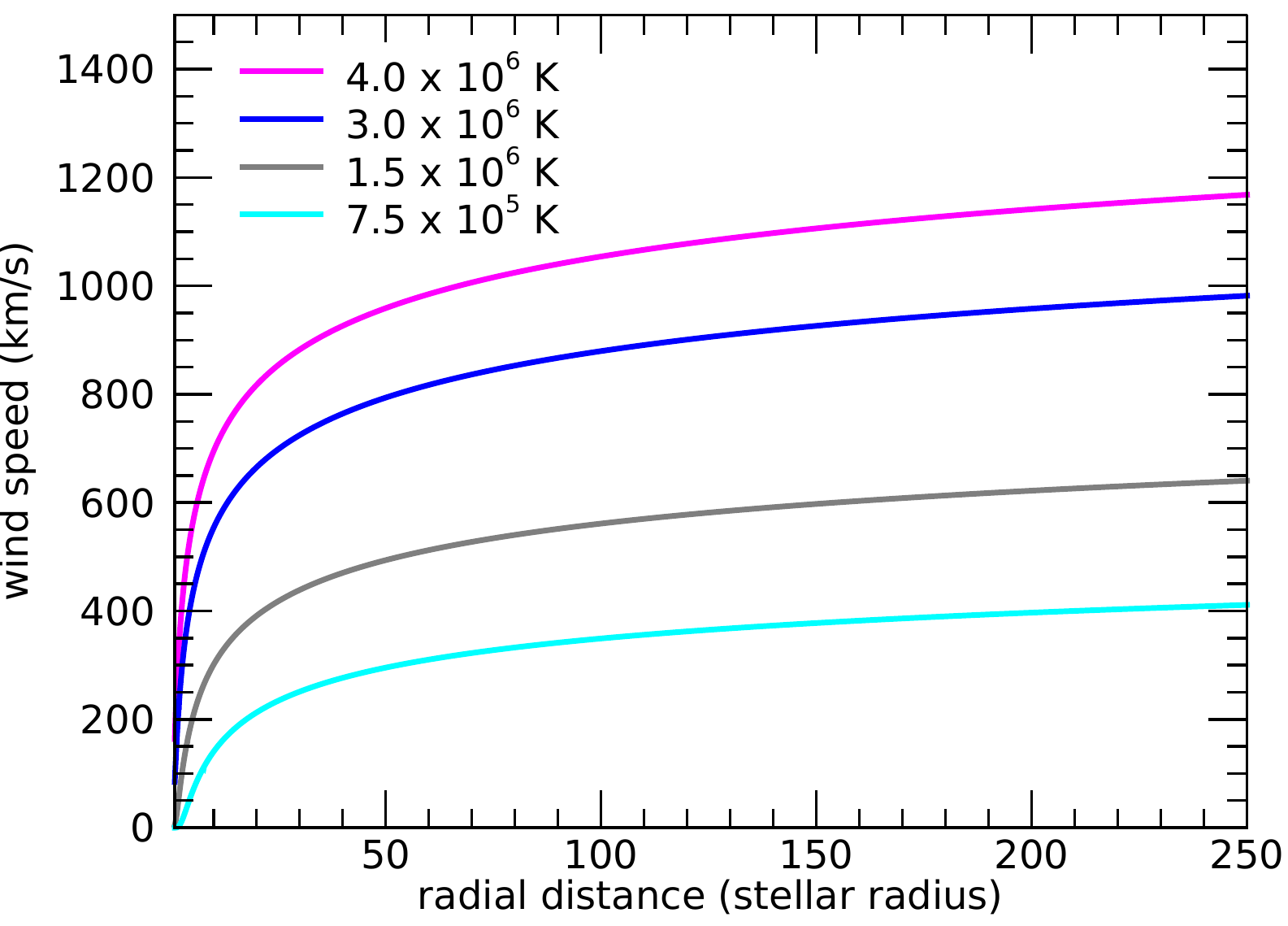}
\includegraphics[width=0.49\textwidth]{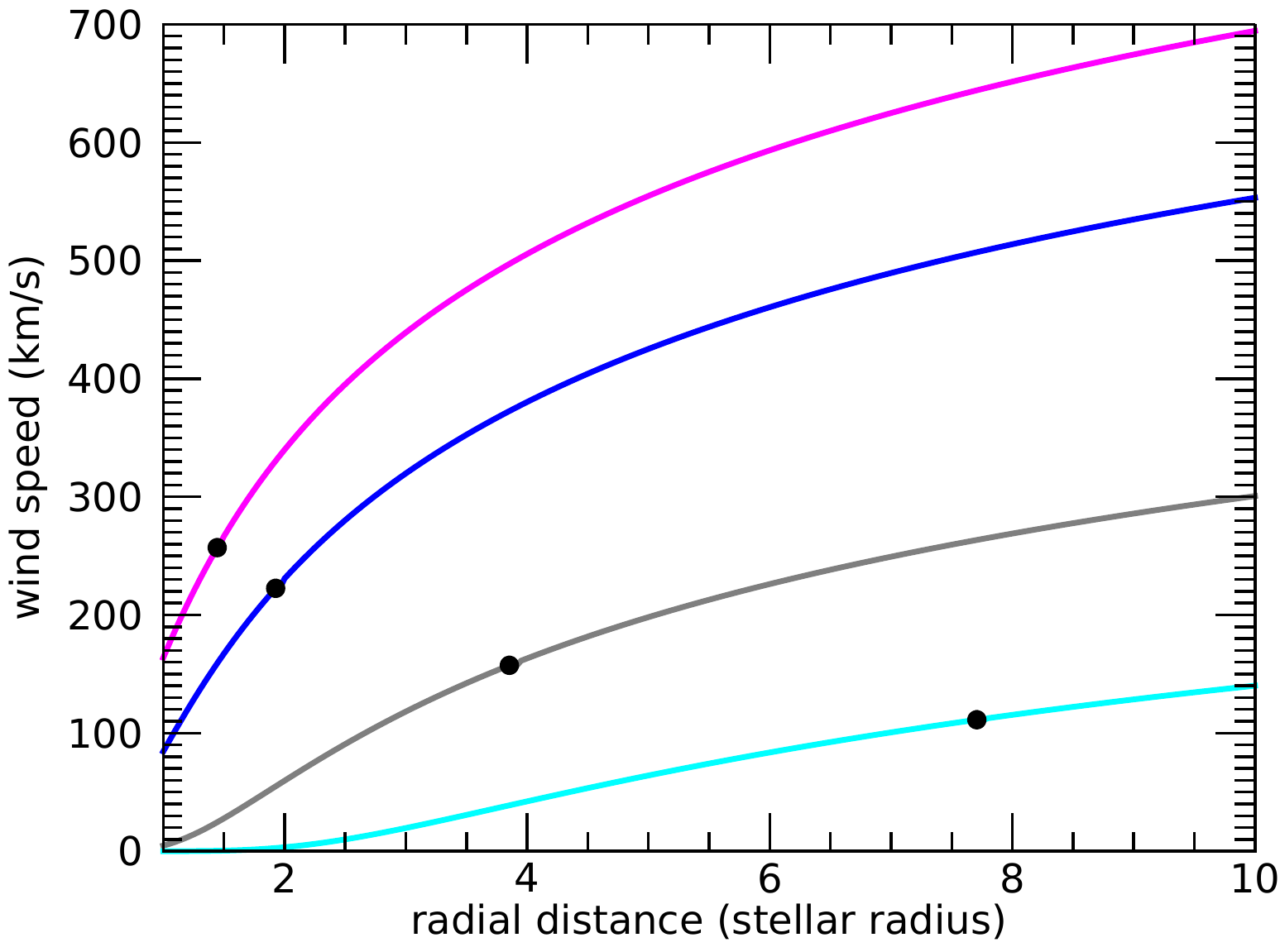}
\caption{Solutions to the momentum equation of an isothermal wind for different wind temperatures. The sonic points are marked by the filled circles, in the zoomed-in panel on the right.}
\label{fig.parker}       
\end{figure}

There are two important points to highlight here. Firstly, note that the density cancels out. Thus, in a thermally-driven wind, the acceleration of the wind does not depend on the density of the wind, which leaves the density as a scaling parameter that is usually adopted to match the required mass-loss rate (Equation \ref{eq.mdot}). Secondly, this equation has a singularity at $u_r = c_s$, i.e., when the stellar wind radial velocity reaches sound speed, the denominator goes to zero. The momentum equation for a polytropic wind is, regardless of that, continuous, as long as the numerator and the denominator go to zero simultaneously. For this to happen,  the wind speed must reach the sonic speed at $r=r_c$: $u_r (r_c) = c_s$. At this critical point, known as the sonic point, the numerator is written as
\begin{equation}\label{eq.critical_point}
 -\frac{ G M_\star}{r_c^2}  +  \frac{2 c_s^2}{r_c} = 0 \,\,\,\,\,\,\,\,\,\,\,\,\,\,\,\, \to \,\,\,\,\,\,\,\,\,\,\,\,\,\,\,\, r_c = \frac{ G M_\star}{2c_s^2} \, .
\end{equation}
This is actually a very important point in the theory of stellar winds. Equation (\ref{eq.poly_mom}) has an infinite number of mathematical solutions, but the only physical solution for a stellar wind is the transonic solution, i.e., the one that starts subsonic and becomes supersonic. Figure \ref{fig.parker} illustrates solutions for the wind radial velocity, assuming  isothermal winds ($\Gamma=1$) at four different temperatures: 0.75, 1.5, 3 and 4 MK (from bottom to top curves). I assumed here a star with $1\,M_\odot$ and $1\,R_\odot$. The curves show the typical wind radial velocity profile, which is characterised by a relatively quick increase in the velocity at inner regions and then an asymptotic solution, which tends to the terminal wind velocity, $u_\infty$. Note how sensitive $u_\infty$ is to the temperature, going from 400 km/s for a 0.75-MK wind to 1200 km/s for a 4-MK wind. 
The zoom-in panel in the right shows the position of the sonic point. Because $r_c \propto c_s^{-2} \propto 1/T$, the sonic point happens closer to the star for higher temperature winds. Likewise, because  $c_s \propto T^{1/2}$, the speed at which the sonic point is reached is increasingly larger for higher temperatures. As discussed in page \pageref{page.freeparam}, the base temperature, along with the base density, are free parameters in polytropic wind models. 

Using the mass conservation equation and the solutions for $u_r$ in Fig.~\ref{fig.parker}, the density profiles can be found: $\rho = \dot{M}/( 4\pi r^2 u_r)$. Note here that the density depends on the \emph{choice} of mass-loss rate \mdot , which is related to the discussion above on the density in thermally-driven wind being a scaling factor. For larger distances, when the wind has reached asymptotic speeds, $\rho \propto r^{-2}$. In this case, the ram pressure of the wind $P_{\rm ram} = \rho u_r^2$ also decreases with $r^{-2}$. This is relevant for calculating the size of the heliosphere, and how it evolved. I will discuss this in Sect.~\ref{sec.CR}.
  
\subsubsection{The magneto-rotator wind}
\begin{figure}
\centering
\includegraphics[width=0.8\textwidth]{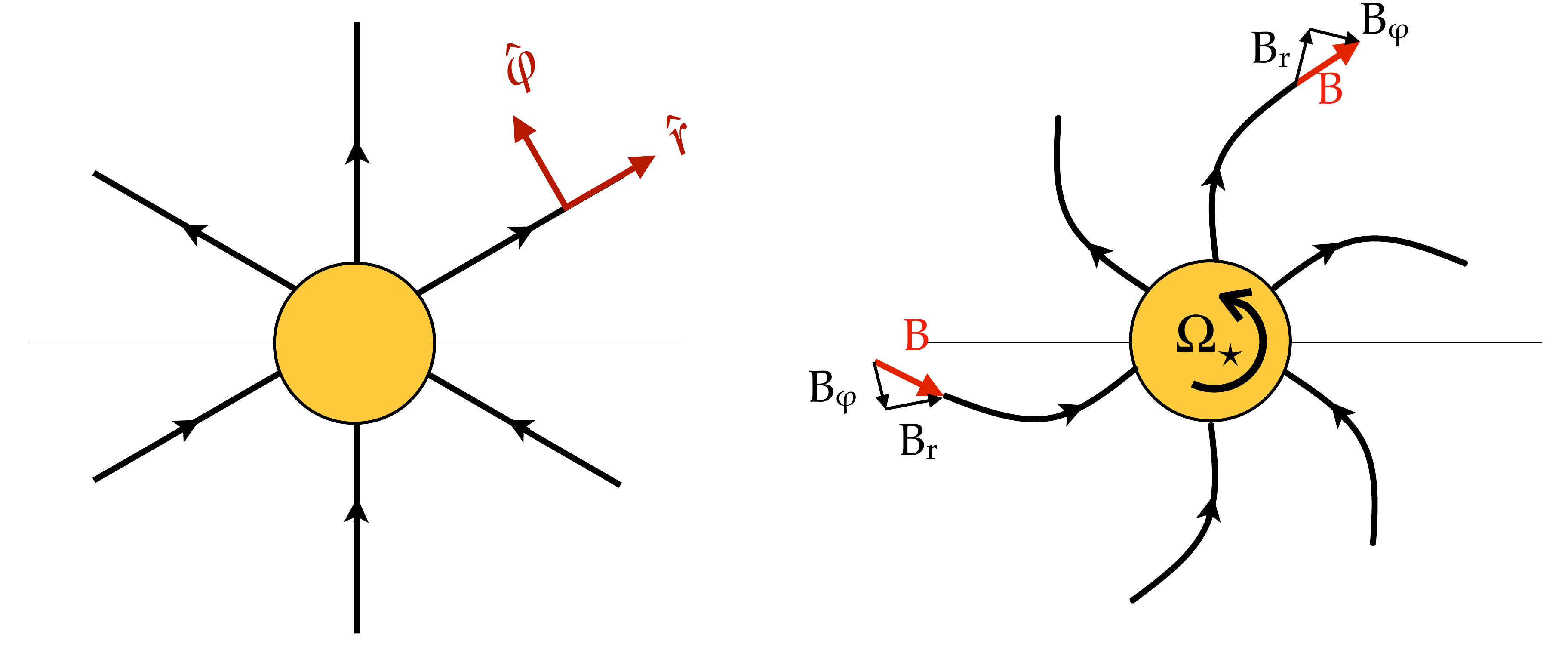}
\caption{Left: initial setup of the split monopole. Right: the trailing spiral structure that is created once stellar rotation is set. }
\label{fig.magrot}       
\end{figure}

The thermally-driven wind discussed before does not consider two physical parameters that are observed in cool stars: rotation and magnetic fields. I am going to show now an important effect when considering magnetised, rotating winds: the removal of angular momentum from the star. Consider the initial setup shown in Fig.~\ref{fig.magrot}a: a split monopole magnetic field in the equator of the star.  The magnetic field lines are anchored on the stellar surface, and thus, once stellar rotation is set, they must  co-rotate with the star (i.e., at same rotation rate \om ). I will consider the stellar rotation axis  pointing out of the page (rotation is counter-clockwise).

\paragraph{Magnetic field geometry:} I will first derive the geometry of the magnetic field lines, using the magnetic induction equation for a perfectly conducting plasma (null resistivity)
\begin{equation}
 \frac{\partial \vec{B}}{\partial t} = \pmb{\nabla}  \times (\vec{u}\times \vec{B})  \, .
\end{equation}
I will  keep assuming  steady state (${\partial /\partial t} =0$) and axi-symmetry (${\partial /\partial \varphi} =0$). Given that the equation is solved in the equatorial plane, ${\partial /\partial \theta} =0$. Thus the magnetic induction equation becomes 
\begin{equation}
\pmb{\nabla} \times (\vec{u}\times \vec{B}) =0  \, .
\end{equation}
The wind velocity vector in spherical coordinates is $\vec{u}=[u_r, 0, u_\varphi]$ and similarly the magnetic field is $\vec{B}=[B_r, 0, B_\varphi]$. Notice that even though I presented in Fig.~\ref{fig.magrot}a a split monopole ($B_\varphi=0$), the rotating star will generate an azimuthal component for the magnetic field in the wind (Fig.~\ref{fig.magrot}b). Using the curl in spherical coordinates, and the fact that $\vec{u}\times \vec{B} = [B_r u_\varphi - u_r B_\varphi] \hat{\theta}$, the last equation is written as
\begin{equation}
\frac{1}{r}\frac{\partial}{\partial r} (r[B_r u_\varphi - u_r B_\varphi] \hat{\varphi}) =0  \, .
\end{equation}
This implies that the term inside the parenthesis is constant for different radial distances: $
r(B_r u_\varphi - u_r B_\varphi )  = \rm constant $. To find this constant, I analyse the wind physical properties at the wind base, where $r=R_\star$. There, the wind radial velocity is very small $u_r(R_\star) \simeq 0$ and the azimuthal velocity is the rotational velocity of the star, $u_\varphi (R_\star) = \Omega_\star R_\star$. The  magnetic field anchored on the surface of the star is still mostly radial, thus $B_\varphi (R_\star) \simeq 0$. Hence, the constant computed at $R_\star$ is: $B_r(R_\star)  \Omega_\star R_\star^2$. This can be further generalised by using magnetic flux conservation: $B_r r^2 = B_r(R_\star) R_\star^2$. The magnetic induction equation then becomes
\begin{equation}
r(B_r u_\varphi - u_r B_\varphi ) = \Omega_\star B_r r^2 \, .
\end{equation}
Rewriting this equation, I arrive at the equation describing the magnetic field geometry of the wind of a rotating magnetised star 
\begin{equation}\label{eq.spiral}
\frac{B_\varphi}{B_r} = \frac{u_\varphi -\Omega_\star r}{u_r} \, .
\end{equation}
Close to the star (small $r$), the rotational velocity is $u_\varphi (R_\star) = \Omega_\star R_\star$ and thus $B_\varphi \simeq 0$. Far from the star (large $r$), $u_\varphi \ll \Omega_\star R_\star$ and thus $B_\varphi/B_r \simeq - \Omega_\star R_\star/u_r$. This implies that not only an azimuthal component of the magnetic field is generated at large distances, but that this component has opposite sign to the radial field: $B_\varphi/B_r  <0$.  This is illustrated in the sketch on the right panel of Fig.~\ref{fig.magrot}. As a result, the rotating wind generates a trailing spiral: if $B_r>0$ (pointing outward), then $B_\varphi <0$ (clockwise). On the contrary, if $B_r<0$ (pointing inward), then $B_\varphi >0$ (counter-clockwise).

\paragraph{The momentum equation of a magneto rotator wind:} The momentum equation of a magnetic rotator is similar to the equation used before, except that now the Lorentz force must also be considered
\begin{equation}\label{eq.wd1}
\rho \vec{u} \cdot \pmb{\nabla}  \vec{u} = -\frac{\rho G M_\star}{r^2} \hat{\bf{r}} - \pmb{\nabla} P  + \vec{F}_B\, ,
\end{equation}
where I assumed steady state and $ \vec{F}_B$ is the Lorentz force per unit volume
\begin{equation}
\vec{F}_B = \frac{1}{c} \vec{J} \times \vec{B}=  \frac{1}{4\pi} (\pmb{\nabla} \times \vec{B})\times \vec{B} \, ,
\end{equation}
where $\vec{J}$ is the current density and I used the Amp\`ere's law (${\bf J}= {c}/({4 \pi})({\bf \nabla} \times {\bf B})$) in the second equality. Using the curl in spherical coordinates, it is straightforward to demonstrate that $\pmb{\nabla} \times \vec{B} =- r^{-1} \partial (r B_\varphi) /\partial r  \hat{\theta}$ and thus
\begin{equation}\label{eq.Fb}
\vec{F}_B = \frac{1}{4\pi r} \left[ -B_\varphi \frac{\partial (rB_\varphi)}{\partial r}\hat{r}  + B_r \frac{\partial (rB_\varphi)}{\partial r}\hat{\varphi}\right] \, .
\end{equation}

Before substituting Eq.~(\ref{eq.Fb}) in the momentum equation (\ref{eq.wd1}), it is necessary to write the convective derivative of $\vec{u}$ in spherical coordinates: 
\begin{equation}\label{eq.conv_der}
\vec{u} \cdot \pmb{\nabla}  \vec{u}   = \left[ u_r\frac{\partial u_r}{\partial r}  - \frac{u_\varphi^2}{r} \right] \hat{r} + \left[ u_r\frac{\partial u_\varphi}{\partial r}  + \frac{u_r u_\varphi}{r} \right] \hat{\varphi} \, .
\end{equation}

Substituting Eqs.~(\ref{eq.conv_der}) and (\ref{eq.Fb}) into (\ref{eq.wd1}), I finally arrive at the momentum equation of a magneto-rotator wind. It is convenient to split it into the radial and azimuthal components, which are respectively 
\begin{equation}\label{eq.wbr}
\rho \left( u_r\frac{\partial u_r}{\partial r}  - \frac{u_\varphi^2}{r} \right)= -\frac{\rho G M_\star}{r^2}  - \frac{\partial P}{\partial r} -\frac{B_\varphi}{4\pi r}  \frac{\partial (rB_\varphi)}{\partial r} 
\end{equation}
and
\begin{equation}\label{eq.wbphi}
 \rho \left( u_r\frac{\partial u_\varphi}{\partial r}  + \frac{u_r u_\varphi}{r}\right) =  \frac{B_r}{4\pi r}  \frac{\partial (rB_\varphi)}{\partial r}\ \, .
\end{equation}
These are the equations derived by \citet{1967ApJ...148..217W}.

The radial component of the momentum equation of the magneto-rotator wind (\ref{eq.wbr}), compared to that of the thermally-driven wind (Equation \ref{eq.cc2}) has two additional terms: the slinging effect of the rotating magnetic field  ($-{B_\varphi}/[{4\pi r}] \, \partial [rB_\varphi]/{\partial r}$) and the centrifugal force ($\rho{u_\varphi^2}/{r}$). These two terms originate from the rotation of the star. Similar to the thermally-driven wind, the magneto-rotator wind also has critical points. One important critical point for angular momentum losses is the Alfv{\'e}n radius\footnote{Note that I discussed about the Alfv{\'e}n surface, which is the generalisation of the Alfv{\'e}n point in three dimensions, when I discussed about prominences in winds in Sect.~\ref{sec.promi}.}, $r_A$, which is defined as the position where the wind radial velocity reaches Alfv{\'e}n velocity, $v_A$: $u_r(r_A) = v_A$. The Alfv{\'e}n velocity is given by $v_A= B/\sqrt{4\pi \rho}$.

This means that the wind of a magnetised, rotating star is trans-Alfv{\'e}nic. In the sub-Alfv{\'e}nic part of the stellar wind (closer to the star), the kinetic energy is smaller than the magnetic energy: $\rho u_r^2/2 \ll B^2/(8\pi)$. In this case, the wind is too weak to modify the structure of magnetic field and the wind is forced to flow along the magnetic field lines.  In the super-Alfv{\'e}nic part of the wind, the opposite is true and  $\rho u_r^2/2 \gg B^2/(8\pi)$. In this case, the magnetic field lines are dragged with the flow. 

\begin{figure}
	\centering
	\includegraphics[width=.65\columnwidth]{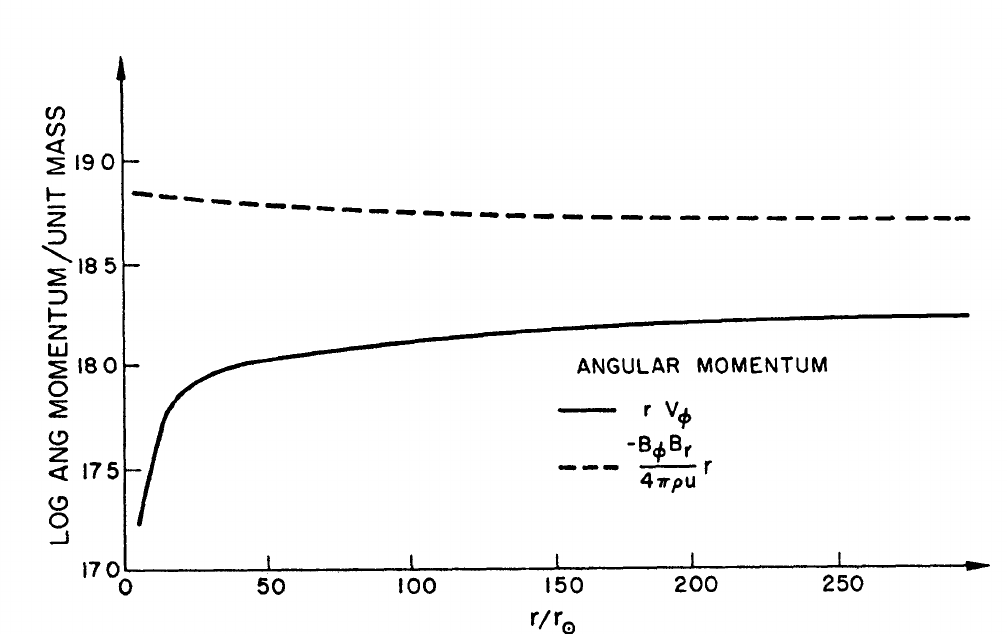}
	\caption{The specific angular momentum of the solar wind calculated by \citet{1967ApJ...148..217W}. The sum of these two terms is constant with distance and simplifies to ${\cal L} =\Omega_\star r_A^2$ (see Equations \ref{eq.wbphi3} and \ref{eq.angmom}), where $r_A$ is the distance to the Alfv{\'e}n point.}
	\label{fig.weber}
\end{figure}

\paragraph{Angular momentum losses:} The angular momentum carried away by the stellar wind can be found from the azimuthal component of the momentum equation of the magneto-rotator  (\ref{eq.wbphi}), which can be written as
\begin{equation}\label{}
\rho u_r\frac{\partial (r u_\varphi)}{\partial r}   =  \frac{B_r}{4\pi }  \frac{\partial (rB_\varphi)}{\partial r}\ \, .
\end{equation}
Rearranging this equation, one has
\begin{equation}\label{}
\frac{\partial (r u_\varphi)}{\partial r}   =  \frac{B_r}{4\pi \rho u_r}  \frac{\partial (rB_\varphi)}{\partial r} =  \frac{B_r r^2 }{\dot{M} }  \frac{\partial (rB_\varphi)}{\partial r} =   \frac{\partial}{\partial r} \left( \frac{B_r r^2 rB_\varphi} { \dot{M}} \right) =   \frac{\partial }{\partial r}  \left(  \frac{r B_r B_\varphi}{4\pi \rho u_r} \right)  \, ,
\end{equation}
where I use the facts that the mass-loss rate ($\dot{M}=4\pi r^2 \rho u_r$, Equation \ref{eq.mdot}) and the magnetic flux ($B_r r^2$) do not depend on $r$. Thus,
\begin{equation}\label{eq.wbphi2}
\frac{\partial }{\partial r}   \left(  r u_\varphi - \frac{r B_r B_\varphi}{4\pi \rho u_r} \right)  =0 \, ,
\end{equation}
which implies that the term in parenthesis is constant with radial distance
\begin{equation}\label{eq.wbphi3}
 r u_\varphi - \frac{r B_r B_\varphi}{4\pi \rho u_r} = {\rm constant} = {\cal L} \, ,
\end{equation}
where ${\cal L}$ is the specific angular momentum (angular momentum per unit mass). Fig.~\ref{fig.weber} shows the two terms as computed for the present-day solar wind. The first term on the left hand side is the specific angular momentum carried by the gas (the one that we are most familiar with) and the second term is the angular momentum carried  by magnetic stresses. We notice from  Fig.~\ref{fig.weber} that even a star that has a modest magnetic field like the Sun and a relatively weak wind, the magnetic term dominates the angular momentum losses. This can be understood in light of the size of the Alfv{\'e}n surface.  Rewriting Eq.~(\ref{eq.wbphi3}) with the help of Eq.~(\ref{eq.spiral}), one finds
\begin{equation}\label{}
{\cal L}  = r u_\varphi - \frac{r B_r^2 }{4\pi \rho u_r^2} (u_\varphi -\Omega_\star r) =  r u_\varphi  - \frac{v_A^2}{u_r^2}(r  u_\varphi -\Omega_\star r^2) \, .
\end{equation}
Given that $\cal{L}$ is a constant, we can calculate the previous equation at $r=r_A$, in which case $u_r = v_A$ and the equation simplifies to
\begin{equation}\label{eq.angmom}
{\cal L}  =  \Omega_\star r_A^2 \, .
\end{equation}
This means that, if we know where the Alfv{\'e}n radius is, then we can calculate the specific angular momentum of the wind by using the previous equation. I will show later on that this is however not an easy task. 
It is curious to see that Eq.~(\ref{eq.angmom}) resembles so much the angular momentum of a solid body with a radius $r_A$ rotating at the angular speed of the star. Because of this,  sometimes this equation is incorrectly interpreted as a wind rotating as a solid body out to $r_A$. In fact, the wind is forced to co-rotate with the magnetic field (which is anchored on the star) out to a distance that is much smaller than $r_A$.

Finally, I can now calculate the angular momentum loss rate. The flux of angular momentum is the product of the specific angular momentum ${\cal L} $ by the mass flux $\rho \vec{u}$. Integrating the flux of angular momentum over a surface area $S$, one finds the angular momentum loss rate 
\begin{equation}
\dot{J} = \oint {\cal L} \rho \vec{u} \cdot d\vec{S}  = \oint \Omega_\star r_A^2 \rho_A  v_A  d{S_A} \, ,
\end{equation}
where the subscript `A' indicates values computed at the Alfv{\'e}n radius. Note that all this calculation was done for the equatorial plane, but in reality, the wind is permeating the whole volume surrounding the star. To consider this, one needs to take into account a $\sin^2\theta$ in the equation above. This comes from the fact that the `lever arm' $r_A$ should be the distance from a point in the wind where the torque is applied to the spin axis, i.e., it should be a cylindrical radius $r_A \sin \theta$. In the equatorial plane, $\sin \theta = 1$, returning to the equation above. Inserting the $\sin^2\theta$ in the equation above and integrating it, one arrives at the angular momentum loss rate (or torque) of a stellar wind 
\begin{equation}
\dot{J} = \oint \Omega_\star r_A^2 \rho_A  v_A  \sin^2 \theta d{S_A}  = \Omega_\star \frac{\dot{M}}{4\pi} r_A^2 \oint  \sin^3 \theta d\theta d\varphi \nonumber \, ,
\end{equation}
\begin{equation}\label{eq.jdot}
\dot{J} = \frac{2}{3} \dot{M} \Omega_\star r_A^2  \, .
\end{equation}

\subsection{Modelling the evolution of the Sun's rotation}\label{sec.jdot}
\subsubsection{A simple model for the evolution of stellar rotation and its relation to gyrochronology} 
Here I show that using the magneto-rotator wind model, it is possible to explain the relation $\Omega_\star \propto t^{-1/2}$, which forms the basis of gyrochronology for stars older than after $\sim 600$~Myr. 

Using magnetic flux conservation, one can write $B_{r, \star} R_\star^2  = B_{r, A} r_A^2  $, where  the subscripts `$\star$' and `$A$' relate to the stellar surface and Alfv{\'e}n surface, respectively. With this,  Eq.~(\ref{eq.jdot}) becomes
\begin{equation}\label{eq.j1}
\dot{J} = \frac{2}{3} (4\pi \rho_A  v_A) \Omega_\star (B_{r, \star} R_\star^2/B_{r, A})^2   = \frac{2}{3}  \frac{\Omega_\star }{v_A} (B_{r, \star} R_\star^2)^2 \, ,
\end{equation}
where I substituted for the Alfv{\'e}n velocity in the second equality. 

The angular momentum of the star (assuming the moment of inertia $\cal{I}$ of a spherical solid body) is
\begin{equation}
J = {\cal I} \Omega_\star  = \frac25 M_\star R_\star^2 \Omega_\star\, .
\end{equation}
Differentiating this equation with respect to time, a second way to express \jdot\ becomes
\begin{equation}\label{eq.j2}
\dot{J} = \frac25 M_\star R_\star^2 \frac{d \Omega_\star}{dt}\, .
\end{equation}
Now here I am making the assumption that $M_\star$ and  $R_\star$ remains constant after $\sim 600$~Myr. Equating Equations (\ref{eq.j1}) and  (\ref{eq.j2}), I then find:
\begin{equation}\label{}
\frac{2}{3}  \frac{\Omega_\star }{v_A} (B_{r, \star} R_\star^2)^2  = \frac25 M_\star R_\star^2 \frac{d \Omega_\star}{dt}\, ,
\end{equation}
which can be rewritten as
\begin{equation}\label{}
  dt = \frac35 \frac{M_\star }{R_\star^2  }\frac{v_A}{B_{r, \star}^2} \frac{d \Omega_\star}{\Omega_\star } \, .
\end{equation}
The previous equation has to be integrated to derive $\Omega_\star(t)$, i.e., the rotational evolution of the star. I already assumed that $M_\star$ and  $R_\star$ are independent of age, and thus, independent of rotation. Two further assumptions need to be made before integrating the equation above. The first one is related on how  $B_\star$ depends on  $\Omega_\star$. I will assume they are related to each other in a linear way (i.e., a linear-type dynamo). This is not too far from what is observed, as I showed in Fig.~\ref{fig.mag_flux}b. The second assumption, and the most tricky one, is on how  $v_A$ depends on $\Omega_\star$. The reason why this is tricky is that $v_A= B/\sqrt{4\pi \rho}$ depends on the magnetic field and density at the Alfv{\'e}n point, which in turn depends on the thermal and magnetic properties of the wind. Thus, there is no way to derive the location of the Alfv{\'e}n point, and its velocity, before computing a wind model. I will assume, for the sake of simplicity, that $v_A$ is independent of $\Omega_\star$ in the main sequence for $t\gtrsim 600$~Myr, but I will come back to this in the next section. With these assumptions, the previous equation can be written as
\begin{equation}\label{}
\int  dt = \frac35 \frac{M_\star v_A}{R_\star^2  } \int \frac{1}{B_{r, \star}^2} \frac{d \Omega_\star}{\Omega_\star } \propto  \int \frac{1}{\Omega_\star^2} \frac{d \Omega_\star}{\Omega_\star }\, .
\end{equation}
Integrating the previous expression results in $t \propto \Omega_\star ^{-2} $, arriving at the following  relation
$$\Omega_\star \propto t^{-1/2} \, ,$$
which surprisingly agrees with observations! 

In summary, two main conclusions can be extracted from this simple evolutionary model. First, it demonstrates that rotating, magnetised  stellar winds carry away angular momentum from the star. As angular momentum leaves the star, the star spins down, explaining the observed trend that stars spin down as the square-root of age. Second, angular momentum extraction is enhanced by the magnetic field: even for a star with a $\sim$ weak magnetic field as the Sun, the angular momentum carried away by magnetic stresses is the dominant contributor to the loss rate. 

In a more sophisticated treatment, one should not use the moment of inertia of a spherical solid body nor the assumption of constant mass and radius through the main sequence. These quantities can be extracted from stellar evolution models and thus coupling the wind angular momentum loss prescription to stellar evolution models would provide a more sophisticated modelling for $\Omega_\star(t)$. The other naive assumption made above refers to the  velocity at  the Alfv{\'e}n point (or surface). I assumed that the Alfv{\'e}n velocity  is independent of stellar rotation. However, this is a very rough oversimplification of the problem, as  $v_A$ depends on the structure of the wind, which depends on the magnetic, thermal and rotational properties of the wind. Thus, there is no way to derive the location of the Alfv{\'e}n surface, and its velocity, without computing a wind model. I will explore in the next Section how this can be achieved.

\subsubsection{Magnetic braking laws:  prescriptions for angular momentum-loss rate}\label{sec.braking}
\begin{figure}
	\centering
	\includegraphics[width=.65\columnwidth]{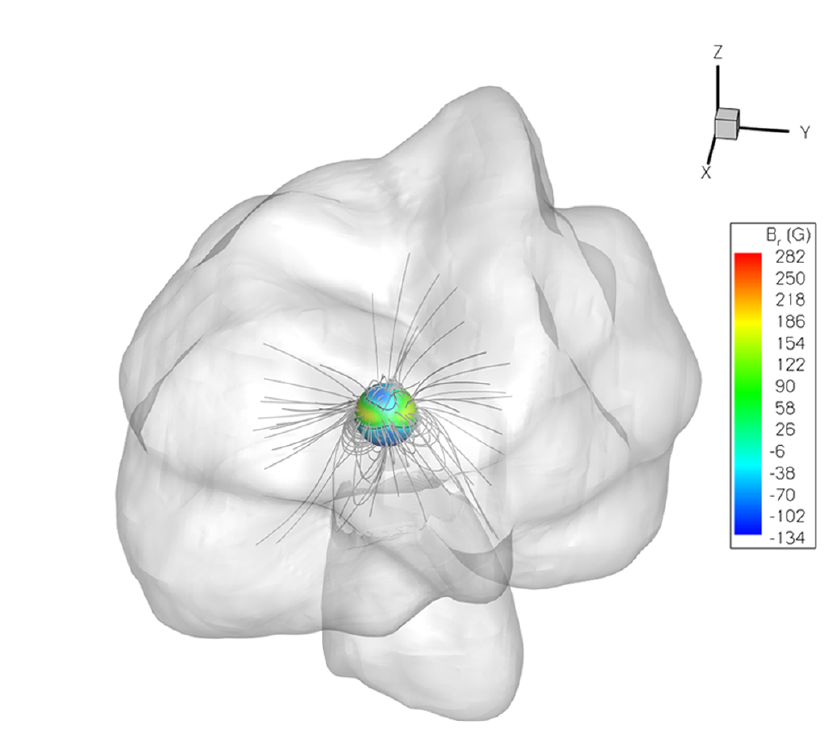}
	\caption{In 3D, the Alfv{\'e}n radius takes the form of an Alfv{\'e}n surface (grey surface). Because of the complex stellar surface magnetic field geometry (shown by the colour bar),  the Alfv{\'e}n surface can be highly asymmetric. Figure from \citet{2014MNRAS.438.1162V}.}
	\label{fig.weber2}
\end{figure}

One of the biggest challenge in stellar rotational evolution studies  is on determining the size of the Alfv{\'e}n surface, so that the angular momentum loss rate \jdot\  can be calculated. Equation (\ref{eq.jdot})  looks deceptively simple! The Alfv{\'e}n radius, in a 1D geometry, or surface, in 3D, can be interpreted as the lever arm of the wind torque -- the ``position'' where the torque is applied in order to change the angular rotation of the star. In fact, there is not a unique radius where the torque is applied in a stellar wind, as  magnetic field stresses, particle motion and even thermal pressure (not discussed here, see Appendix in \citealt{2014MNRAS.438.1162V}) contribute to the stress tensor throughout the wind.  Figure~\ref{fig.weber2} shows an example of an Alfv{\'e}n surface, computed in a 3D simulation. The surface is far from being a smooth, spherical surface as the 1D model that I derived before would suggest. To makes matter worse, both the thermally driven wind or the magneto-rotator winds depends on free parameters that directly affect the size of the Alfv{\'e}n surface. These free parameters are related to the thermal properties of the wind (density and temperature), which are not directly observed (though they can be inferred to some level).

Provided that one can determine the size of the Alfv{\'e}n surface and thus calculate \jdot , then \jdot\ can be implemented in stellar evolution codes so that the stellar structure and its rotation can be evolved forward in time. This has been done with different levels of complexity in stellar evolution models  \citep[e.g.,][]{1988ApJ...333..236K, 1989ApJ...338..424P, 1997A&A...326.1023B, 2011MNRAS.416..447S, 2015A&A...577A..28J,2015ApJ...799L..23M, 2019A&A...631A..77A}. What these models all have in common is that they rely on a parameterisation for \jdot , i.e., an analytical relation that summarises how \jdot\ varies with different physical properties of the star, such as its mass, radius, rotation, surface magnetic fields, etc. Such analytical relation can be derived from analytical considerations  \citep{1988ApJ...333..236K} or numerical simulations \citep{2012ApJ...754L..26M}. \citet{1988ApJ...333..236K}, for example, prescribed two power-laws, one relating surface magnetism and rotation ($B_\star \propto \Omega_\star^a$, where $a=1$ represents a linear-type dynamo) and another relating how \jdot\ varies with magnetic field geometry
\begin{equation}\label{eq.jdot_kawaler}
\dot{J} =\frac23 \dot{M} \Omega_\star R_\star^2 \left(\frac{r_A}{R_\star} \right)^n  ,
\end{equation}
in which $n=2$ corresponds to a radial field geometry (as I derived before) and $n=3/7$ to a dipolar field geometry. Combining these two properties, \citet{1988ApJ...333..236K} obtained an angular-momentum loss prescription of the form
\begin{equation}\label{eq.jdot_kawaler2}
\dot{J} \propto \Omega_\star ^{1+4an/3} R_\star^{2-n} \dot{M}^{1-2n/3} M_\star^{-n/3}  \, .
\end{equation}
To  find the proportionality constant (omitted) in the previous equation, models are usually calibrated to the solar values, i.e., for a given $n$ (field geometry) such a constant is found when the solar rotational velocity is obtained at the solar age and radius.  Similarly, $n$ is adjusted to find the best fit to observational data.  \citet{1988ApJ...333..236K} found that $n=1.5$ reproduces the Skumanich relation (\om $\propto t^{-1/2}$). Also very illuminating, but not discussed in this review, is that once the solar calibration and calibration for solar-mass stars in clusters are done, the model can be applied to low-mass stars with a range of  masses.

The power-law indices and constants in the analytical description of \jdot\  (e.g., from Eq.~(\ref{eq.jdot_kawaler2})) hide our ignorance of the  physics that is driving winds of Sun-like stars \citep{1988ApJ...333..236K}. For example, different surface magnetic field geometries and thermal properties of the wind  change the size of the Alfv{\'e}n surface. By sweeping a large parameter space in, for example, magnetic field strengths, geometry, and wind thermal properties, numerical simulations can provide a more general view of \jdot , and thus constants that were previously assumed (e.g., $a$, $n$, etc) can be derived from numerical studies. \citet{2012ApJ...754L..26M} performed a range of 2D simulations for solar-type stars with different rotation rates \om\ and dipolar magnetic field strengths $B_{{\rm dip},\star}$. The authors found a generalised fit for their simulations that can be written in terms of constants $k_1$, $k_2$ and $m$ as
\begin{equation}\label{eq.matt_fit}
\dot{J} =k_1^2  B_{{\rm dip},\star}^{4m} \dot{M}^{1-2m} R_\star^{2+4m} \frac {\Omega_\star}{(k_2^2 v_{{\rm esc}, \star}^2 + \Omega_\star^2 R_\star^2)^m } ,
\end{equation}
with $k_1 = 1.30$, $k_2 = 0.0506$, and $m = 0.2177$. The surface escape velocity is $v_{{\rm esc}, \star} = (2GM_\star/R_\star)^{1/2}$. With all values given in cgs units, then the angular momentum loss rate has units of g~cm$^2$~s$^{-2}$, which is an energy unit and thus \jdot\ is frequently expressed in erg. 

The angular-momentum loss prescriptions shown, for example, in Eqs.~(\ref{eq.jdot_kawaler2}) and (\ref{eq.matt_fit}) can then be implemented in stellar evolution codes. Using Eq.~(\ref{eq.matt_fit})  is an advantage over the use of Eq.~(\ref{eq.jdot_kawaler2}), as discussed in \citet{2015A&A...577A..28J}. However, stellar evolution models still need to adjust the value of $k_1$ so that they can be properly calibrated for the Sun \citep{2013A&A...556A..36G,2015A&A...577A..28J,2019ApJ...886..120S}. One challenge in such prescriptions for \jdot\ is that they require a value of \mdot , which, as discussed in Sect.~\ref{sec.obs}, is difficult to observationally derive.  In particular, the mass-loss rates are dependent of the thermal properties of the wind. In wind models that implicitly assume heating, the thermal properties at the wind base are free parameters (see page \pageref{page.freeparam}). Models that assume, for example, that wind temperatures increase with stellar activity \citep[e.g.,][]{2018MNRAS.476.2465O} predict that mass-loss rates are higher for young and fast rotating stars.

Would then mass-loss rates increase indefinitely with rotation? As I presented in Sect.~\ref{ref.Mdotevol}, although it is believed that younger stars should have higher \mdot , there is still no clear picture from observations (see Fig.~\ref{fig.sequence}). 
Recent observations, using prominences to probe stellar winds, suggest that  mass-loss rates seem to keep increasing towards  young ages (Figs.~\ref{fig.detection_summary} and \ref{fig.sequence}) -- AB~Dor, for instance, is a $\sim 120$ Myr-old star, with 350 times the present-day solar mass-loss rate, or the even younger LQ Lup ($\sim 25$~Myr), with \mdot $\sim$ 4500 times the solar value \citep{2019MNRAS.482.2853J}. These findings challenge the results from the astrosphere method, which indicated that mass-loss rate might actually reduce at  ages below $\sim 600$~Myr, with $\pi^1$ UMa ($\sim 500$~Myr) showing half the present-day solar mass-loss rate \citep{2014ApJ...781L..33W}. I believe that the answer to this should come from a combined approach between modelling and observations.

\section{Implications for the evolution of the solar system planets and exoplanets}\label{sec.implications}
In this Section, I briefly discuss possible implications of the solar wind evolution on planets in the solar system and on exoplanets orbiting Sun-like stars.

\subsection{The evolution of Earth's magnetosphere}
One frequent argument in favour of the larger mass-loss rates of the young Sun is the lack of a substantial atmosphere in Mars. Given that Mars lacks a large-scale protective magnetic field similar to that of Earth, the interaction between the planet and the solar wind is not shielded by the planet's magnetic field. As a result, the young Martian atmosphere is believed to have been eroded by a stronger solar wind \citep[e.g.,][]{2007SSRv..129..207K}. This reasoning has been recently challenged in light of ion escape measurements in terrestrial planets \citep{2013cctp.book..487B}. Escape rates of O$^+$ in Earth, Mars and Venus are reasonably similar, yet these terrestrial planets have very different properties, such as gravity, orbital distances and the presence of a large-scale magnetic field. The fact that there are not two planets in the solar system that are identical, except for their magnetic fields, makes it very difficult to isolate how, or if, a large-scale magnetic field can be an effective atmospheric shield \citep{2013cctp.book..487B}. 

There are two arguments relating to whether a planetary magnetic field might, or not, protect planetary atmospheres.

\begin{description}
\item In favour: The magnetosphere of a planet acts to deflect the stellar wind particles, preventing its direct interaction with planetary atmospheres and, therefore, potential atmospheric erosion.  If the {magnetosphere} is too small, part of the atmosphere of the planet becomes exposed to the interaction with stellar wind particles. \citet{2007AsBio...7..185L} suggested that, for a magnetosphere to protect the atmosphere of planets, planets are required to have magnetospheric sizes $\gtrsim 2$ times the planet's radius. This can have important effects on the habitability of exoplanets, including terrestrial type planets orbiting inside the habitable zones of their host stars \citep[e.g.,][]{2007AsBio...7..185L, 2013A&A...557A..67V, 2014A&A...570A..99S, 2016A&A...596A.111R}. 

\item Against: However, it is not only the size of the magnetosphere that matters. Depending on the orientation of the stellar wind (i.e., interplanetary) magnetic field relative to the planetary one, magnetic reconnection can occur and the interplanetary magnetic field lines can connect to auroral zones around the planet's magnetic pole. Because these field lines are open, stellar wind plasma can be channeled towards the polar caps, causing local heating that could ultimately increase atmospheric losses through polar/auroral flows (\citealt{2007RvGeo..45.3002M}. This is potentially more important for heavier species like O$^+$  than for lighter ones like H$^+$, \citealt{2013cctp.book..487B}.) The amount of stellar wind plasma that is channeled towards the polar caps depends on the collecting area of the magnetosphere (i.e., its cross section $\sim \pi r_M^2$), which is larger for higher magnetic fields (equations will be derived below). 
\end{description}

Thus, while a magnetosphere larger than the extent of the upper atmosphere can protect atmospheric losses, larger magnetospheres also have greater `collecting area' for stellar wind plasma, which could lead to enhanced atmospheric losses.  It is likely that these two processes coexist and that their contributions differ throughout the life of the planet. \citet{2018MNRAS.481.5146B} suggested that the competition between the strength of the stellar wind  and the large collecting areas of magnetospheres indicates whether a planetary magnetic field has a protective effect on the planetary atmosphere or not. In the case of Earth, they suggested that our planet's magnetic field has played a crucial role in protecting our atmosphere. However, different solar wind characteristic in the future could reverse this protective effect, where a reduced inflow speed (related to the speed of magnetic reconnection) and larger planetary magnetospheres would conspire to generate more atmospheric losses than what an unmagnetised Earth would have presented.
 
In a first order approximation, we can estimate the magnetospheric stand-off distance by setting pressure balance between the incoming stellar wind and the planetary magnetic pressures. At the planet--stellar wind interaction zone, pressure balance between the stellar wind (left-hand side) and planetary magnetosphere (right-hand side) can be written as
\begin{equation}\label{eq.equilibrium}
\rho_w  u_w^2 \simeq \frac{B_{{p},r_M}^2}{8\pi} ,
\end{equation}
where $\rho_w$ and $u_w$ are the density and speed of the stellar wind at the planetary orbit, and  $B_{{p},r_M}$ is the planetary magnetic field intensity at a distance $r_M$ from the planet centre. 
Assuming that the planet's magnetic field is dipolar, with  $B_{p, {\rm eq}}$ its surface magnetic field at the equator, the previous equation can be rewritten so that magnetospheric size of the planet is given approximately by
\begin{equation}\label{eq.r_M}
 \frac{r_M}{R_p} \simeq \left[ \frac{B_{p, {\rm eq}}^2}{8 \pi \rho_w  u_w^2} \right]^{1/6}.
\end{equation}
Equation (\ref{eq.r_M}) shows that a large stellar wind  pressure acts to reduce the size of {planetary magnetospheres} for a given planetary magnetisation.  This is the Chapman--Ferraro equation firstly derived for Earth's magnetosphere. Note that this equation needs to be modified for exoplanets in closer orbits, as in those cases, the magnetic pressure and thermal pressure of the stellar wind, as well as ram pressure of the planetary atmosphere, might have to be incorporated in Equations (\ref{eq.equilibrium}) and (\ref{eq.r_M}) \citep{2013A&A...557A..67V,2020MNRAS.494.2417V}. 

Recently,  \citet{2019MNRAS.489.5784C} studied the evolution of Earth's magnetosphere during the solar main sequence. For that, they first performed simulations of the evolution of the solar wind using a magneto-rotator wind model (1.5D MHD). With that, they were able to model the conditions of the evolving solar wind at the orbit of the Earth. These conditions were then implemented as external boundary conditions into 3D MHD simulations of the Earth's magnetosphere. \citet{2019MNRAS.489.5784C} showed that, as the Sun spun down and the solar wind density and velocity values at Earth's orbit evolved, the Earth's magnetosphere gradually increased following a power-law $r_M \propto \Omega_\star^{-0.27}$ up until the Sun reached a rotation of $\Omega_\star \sim 1.4\Omega_\odot$. After that (i.e., for lower rotations $\Omega_\star \lesssim 1.4\Omega_\odot$), their model predicted a more rapid increase of the Earth's magnetosphere with $r_M \propto \Omega_\star^{-2}$. The piece-wise evolution they predict for the Earth's magnetosphere lies on  assumptions for the temperature of their stellar wind  models. 

 \citet{2019MNRAS.489.5784C} used a simplified stellar wind model, which allowed them to predict the evolution of the solar wind for a large interval of time. More sophisticated wind models have opted for a more focused approach. For example, \citet{2011JGRA..116.1217S} used modified solar magnetograms as input for their 3D MHD simulations for how the solar wind was 3.5~Gyr ago. Their wind results were  then implemented in simulations of Earth's ``paleomagnetosphere'', in which the strength and orientation of Earth's  magnetic field  were varied. Depending on the specific conditions of the Earth's magnetic field, \citet{2011JGRA..116.1217S}  showed that the paleomagnetosphere stand-off distance could have been as small as 4.25 Earth radii. 
 
 As the solar wind interacts with Earth's magnetosphere, not only the magnetospheric stand-off distance changes, but also does the area of Earth's polar cap containing open magnetic field lines. \citet{2011JGRA..116.1217S}   showed that at 3.5Ga the Earth's open-field polar cap area was larger for several conditions they explored. The only exception was for when Earth's dipolar field had a substantial tilt, or when the solar wind magnetic field was parallel to the Earth's magnetic field at the interaction zone. In this case, the Earth's magnetosphere was closed, without any open-field polar cap area. Appendix~B in \citet{2019MNRAS.489.5784C} discusses the differences between open and closed planetary magnetospheres, which can more clearly be seen in their Fig.~B1.

Another focused work on the young Sun-Earth study was presented in \citet{2016ApJ...820L..15D}, who developed a 3D study of the magnetohydrodynamic environment surrounding $\kappa^1$ Cet, which has been recognised as a good proxy for the young Sun at the age when life arose on Earth.  According to the authors, the mass-loss rate of $\kappa^1$ Cet is about 50 times larger than the present-day Sun \citep[see also][]{2016ApJ...817L..24A}. Due to this larger mass-loss rate, the ram pressure of the young solar wind impacting on the magnetosphere of the young Earth was larger than the present-day values. \citet{2016ApJ...820L..15D} showed that the magnetospheric sizes of the young Earth should have been reduced to approximately half to a third of the value it is today ($\sim 11$ Earth radii). The relatively large size of the early Earth's magnetosphere  may  have been the reason that prevented the volatile losses from Earth and created conditions to support life \citep{2010Sci...327.1238T, 2016ApJ...820L..15D}. This is opposed to the fate of a young, non-magnetised Mars, which could have lost its  atmosphere due to a stronger younger solar wind \citep{2007SSRv..129..207K}.

\subsection{Implication for exoplanets and new lessons for the solar system}
There are several ways in which the solar wind evolution, and the linked activity evolution, can affect planets. One way, already mentioned above, refers to the direct interaction between the evolving solar wind and the planetary magnetosphere, in the case of magnetised planets. In the case of unmagnetised or weakly magnetised planets, the magnetic pressure on the right-hand side of Eq.~(\ref{eq.equilibrium}) should be replaced by the thermal and ram pressure of the planetary atmosphere. In particular, planets that receive a large amount of high-energy irradiation  can have evaporating atmospheres. This happens in either planets that orbit old stars at close distances, or planets with larger semi-major axis orbiting more active (young) stars. These two conditions lead to enhanced XUV energy flux deposition  on the top layers of planet atmospheres, which leads to an increased heating, causing atmospheres to inflate and to more easily evaporate \citep[e.g.,][]{2003ApJ...598L.121L, 2004A&A...419L..13B, 2011A&A...532A...6S}. Since the XUV emission of solar-type stars is observed to evolve (Sect.~\ref{sec.activity}), exoplanetary evaporation is also expected to evolve in time.

This implies that the history of the host star plays an essential role in the evolution of atmospheric evaporation of their planets.  For example, a young host star rotating more slowly than other stars at the same age would generate less EUV radiation and thus their planets would evaporate less and end up with a higher atmospheric mass fraction. On the contrary, a young host star that is a fast rotator would emit more EUV radiation, which could potentially lead to a planet with a reduced  atmosphere. \citet{2015A&A...577L...3T} showed that a 0.5-Earth-mass planet orbiting at 1 au around a solar-mass star will present an atmosphere with a very different hydrogen content at 4.5 Gyr, depending on whether the host star was a slow or fast rotator during its youth. Starting with an initial hydrogen atmosphere of $0.005$ Earth mass, they showed that the atmosphere of the terrestrial planet is entirely lost before an age of 100 Myr if the host star begins its main-sequence evolution as a rapid rotator, while if the host star was a slow rotator, at 4.5 Gyr, the planet would still have about 45\% of its initial atmosphere.

Because the `final' (i.e., observed) planetary atmospheric fraction depends on the EUV history of the star, \citet{2019ApJ...879...26K} suggested that detecting the atmospheres of planets today can provide a way to probe the evolutionary path of the host star. \citet{2019A&A...632A..65K} applied this model to the Kepler-11 system, formed by a star that is very similar to the Sun (in mass and age) and that has six known transiting planets with orbital distances between 0.09 and 0.5 au. Given that these six close-in planets seem to have retained  hydrogen-dominated atmospheres, these authors concluded that this Sun-like star was a slow rotator at its youth. In a similar reasoning,  models using instead the Earth and Venus atmospheric composition were conducted by \citet{2020Icar..33913551L}, who showed that one possible explanation for the noble gas composition in Earth's and Venus' atmospheres could be due to the Sun being  a slow to moderate rotator at young ages. 

Works like these demonstrate that modelling star-exoplanet interactions could open up new avenues for us to progress in our understanding of the solar system, and the Sun and solar wind evolutions. 

Another important consideration for the young solar system is that the wind of the young Sun  might have had an increased rate of CMEs and even might have been CME-dominated at an early point in time (see Sect.~\ref{sec.cme}). Although CME rates decrease with age, due to geometrical effects, close-in planets interact much more frequently with stellar CMEs than planets at larger orbits \citep{2016ApJ...826..195K}. The CME itself might also be highly affected by the presence of a close-in planet, given that the CME does not have enough time to expand before it encounters the close-in planet \citep{2011ApJ...738..166C}. CMEs cause perturbations in stellar wind conditions, which can then lead to changes in the magnetospheres and extended atmospheres of exoplanets \citep{2007AsBio...7..167K, 2017ApJ...846...31C}. CME-planet interaction might be more common in planets orbiting M dwarfs, as these stars remain active for a long part of their lives \citep{2008AJ....135..785W,2011ApJ...727...56I} and have high flare and CME rates \citep{2016A&A...590A..11V,2017ApJ...841..124V}. M dwarf stars are the prime targets for detecting terrestrial planets in their habitable zone. However, the consequence of their high rate of energetic events might affect habitability of M dwarf planets \citep{2020IJAsB..19..136A}.

Although stellar winds constantly interact with exoplanets, exoplanets do not interact with every stellar CME, because the planet may not lie along certain CME trajectories. In the Sun, CMEs are observed within $\pm 30^o$ latitude, corresponding to the active region belt \citep{2019SunGe..14...111}. However, in young, fast rotating stars, active regions are often located at high, polar-region latitudes \citep{2006MNRAS.370..468M, 2007MNRAS.375..567J}. If CMEs originate in active regions, it is therefore more likely that CMEs in young stars could emerge at high latitudes and be directed above/below the equatorial plane. Considering planet formation in the equatorial plane,  this could mean that young planets might at the end not be strongly affected by CMEs, even though their host stars are expected to have higher CME rates. Future research in this area might shed more light into the  effects of CMEs in young planetary systems.

Regardless of whether  exoplanets lie along CME trajectories, they can nevertheless be affected by the almost-instantaneous increase in irradiation caused by an eventual CME-associated flare, since this emission covers a wider solid angle. An increase in irradiation input caused by a flare can lead to  increase in evaporation in close-in gas giants \citep{2017ApJ...846...31C, 2018ApJ...869..108B, 2020MNRAS.496.4017H}. Transit observations of the hot Jupiter HD189733b showed an enhancement of  atmospheric evaporation that took place 8h after the host star flared.  \citet{2012A&A...543L...4L} suggested that the violent event and the enhanced evaporation are potentially related to each other. The remaining open question is whether it is the increase in XUV emission caused by the flare or the interaction between an unseen associated CME and the exoplanet (or even both) that caused the increase in the observed evaporation rate.

\subsection{Evolution of the heliospheric size and implications for the penetration of Galactic cosmic rays}\label{sec.CR}
The solar wind expands into the interstellar medium (ISM), giving rise to the heliosphere. Analogously to the Chapman--Ferraro equation for calculating the Earth's magnetospheric stand-off distance, one can use pressure balance between the solar wind and ISM to derive the size of the heliosphere. The ram pressure of the solar wind can be rewritten as $P_{\rm ram}  = \rho u^2 = \dot{M} u_\infty/(4\pi r^2)$, where I assumed that for larger $r$, the wind has reached terminal speed $u_\infty$. Thus, pressure balance leads to 
\begin{equation}
\frac{\dot{M} u_\infty }{4\pi r^2} = \rho_{\rm ISM} u_{\rm ISM}^2 \,\,\,\,\,\,\,\,\,\,\,\,\, \to  \,\,\,\,\,\,\,\,\,\,\,\,\,
r = \left( \frac{\dot{M} u_\infty }{4\pi \rho_{\rm ISM} u_{\rm ISM}^2}  \right)^{1/2}
\end{equation}
where the subscript `ISM' refer to ISM `wind' quantities. If one assumes that the ISM density and velocity have not changed significantly over the past 4.5~Gyr, the larger mass-loss rates of the young Sun implies that the heliosphere was larger in the past. \citet{2020MNRAS.499.2124R} estimated that heliospheric size of the solar wind extended out to 950~au when the Sun had an age of 1~Gyr and could have ranged between 1300 and 1700~au at $t=600$~Myr, i.e., more than a factor of ten larger than the present-day value of 122~au. 

The size of the heliosphere, as well as the solar wind conditions, have consequences for the propagation of cosmic rays. Galactic cosmic rays are suppressed as they travel through the solar wind all the way to the Earth -- this ``suppression'' is usually referred to as cosmic ray ``modulation''. I refer the reader to  \citet{2013LRSP...10....3P} for a comprehensive review on  the modulation of cosmic rays in the heliosphere. In the framework of a diffusive transport equation for the Galactic cosmic rays, there are three main competing processes in place: the  diffusion of cosmic rays into the solar system, and their advection, in momentum and in space, which acts to suppress the flux of Galactic cosmic rays reaching Earth. These processes depend on the level of turbulence present in the solar wind magnetic field and on the wind velocity, both of which evolve in time. 

\begin{figure}
\centering
 \includegraphics[width=0.65\textwidth]{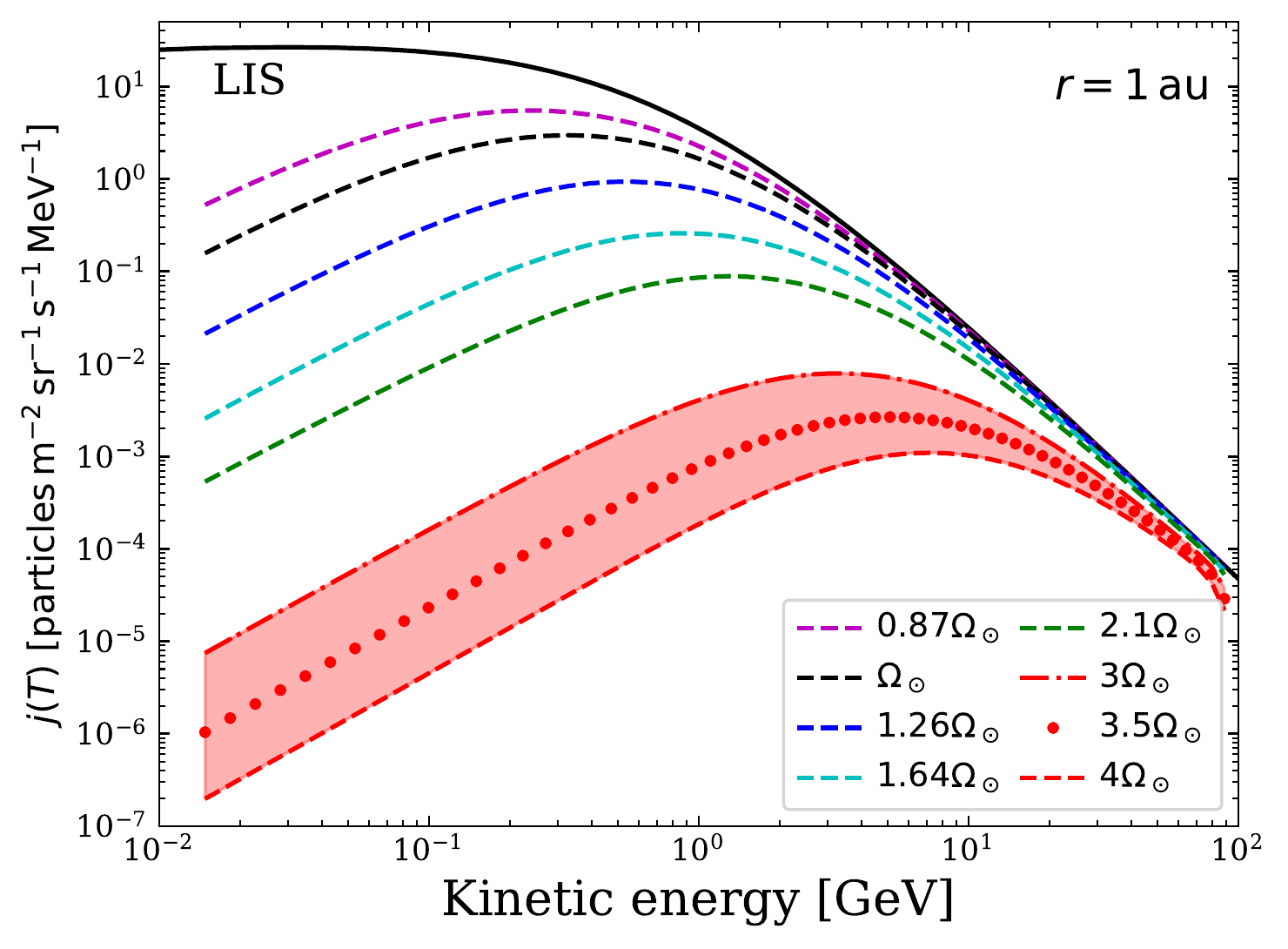}
\caption{Evolution of the Galactic cosmic ray spectrum at  Earth's orbit. The figure above shows Galactic cosmic ray differential intensity as a function of their kinetic energies. The different curves refer to different solar rotation rates normalised to present-day value (here denoted by $\Omega_0$). Each curve thus represents a different age, which ranges from about 6 to 1 Gyr from the top to the bottom dashed lines.  The black dashed line is thus for the present-day value (1$\Omega_0$) at Earth's orbit.  The black solid line is the cosmic ray spectrum at the local interstellar medium. Cosmic rays are suppressed (i.e., modulated) as they travel through the solar wind and, given the solar wind evolution, the spectrum of Galactic cosmic rays at Earth changes with time. The red lines and red shaded area represent a range of rotation values that the Sun could have had at 600~Myr. Figure from \citet{2020MNRAS.499.2124R}} \label{fig.CR}       
\end{figure}

Using the solar wind evolution model of \citet{2019MNRAS.489.5784C}, \citet{2020MNRAS.499.2124R} showed that, at the Earth's orbit, the advective processes (pushing cosmic rays out) are relatively more important than diffusive processes (allowing them in) in the Sun's past, for GeV and MeV cosmic rays. This means that  cosmic rays of these energies are more suppressed and thus fewer of those have reached Earth in the past, compared to present-day values. This is seen in Fig.~\ref{fig.CR}, which shows the modelled evolution of Galactic cosmic ray spectrum at Earth for a range of solar rotation rates. The black solid line is the cosmic ray spectrum at the local interstellar medium \citep{2015ApJ...815..119V}. The black dashed-line is the present-day value (which was calibrated to present-day measurements) at Earth's orbit, at approximately solar minimum. The red lines and shaded area represent a range of rotation values that the Sun could have had at 600~Myr, i.e., the predicted flux of Galactic cosmic rays can change by more than one order of magnitude depending on whether the Sun was a fast/slow rotator. \citet{2020MNRAS.499.2124R}  considered a 1D propagation model for cosmic rays. An interesting future study would be to use a 3D cosmic ray propagation model, in 3D wind simulations that include  ZDI observations, similar to the approach adopted in \citet{2012ApJ...760...85C}. 

With the knowledge of the spectrum of cosmic rays arriving at the top of Earth's atmosphere, further modelling can then be conducted to investigate variations of high-energy particle fluxes into the young Earth's atmosphere \citep[e.g.,][]{2007AsBio...7..208G, 2010EP&S...62..333S, 2013ApJ...774..108R, 2016A&A...587A.159G}. Interestingly, it has been suggested that the variation of cosmic ray flux over the past 3 billion years coincide with periods of glaciations at Earth  (\citealt{2006AN....327..871S}, see also \citealt{2002PhRvL..89e1102S}). Another  interesting application, not discussed in this review, is to investigate cosmic ray propagation through the winds of M dwarf stars \citep{2005AsBio...5..587G, 2015A&A...581A..44G}. These stars are currently among the main targets in search for habitable planets.

\section{Final remarks and open questions} \label{sec.open}
Over the years, we have collected a large amount of information on  the solar wind, in an incredible level of detail. All this information, however, can only tell us about how the solar wind looks like {now}. To understand the past, and future, evolution of the solar wind, we rely on observations from other suns in the Universe, i.e., stars that resemble the Sun and are at different evolutionary stages. In theory, connecting the observations of  winds of different solar-like stars under one unique evolutionary sequence would tell us how the wind of our own star has evolved. In practice, however, this is not straightforward. 

One reason for this is that detecting winds of solar-like stars is a very challenging task. Direct detection of winds of other Suns has not been possible and the Sun still remains the only star for which the wind has been probed in situ. However, over the years, a small number of clever techniques have been developed to indirectly derive stellar wind properties, such as mass-loss rates and speeds. Overall, these techniques have shown that the solar wind had a higher mass-loss rate for the past $\sim$3.5--4 billion years. Before that, from the time the Sun started its evolution in the main sequence until about 600--800~Myr, clues point in different directions. 

It has been suggested that the young solar wind had a lower mass loss rate, probed in the lower-than-expected mass-loss rate of $\pi^1$ UMa, a 500-Myr-old solar analogue, which has only half the present-day solar mass-loss rate $ \dot{M}_\odot$ \citep{2014ApJ...781L..33W}. In particular, $\pi^1$ UMa does not fit the trend of $\dot{M}$--\fx\ relation built from the astrosphere wind detection method, that older solar-like stars seem to fall on. However, a recent method for detecting winds of rapidly rotating stars, using prominences to probe the wind mass-loss rate, showed that young Suns could have much higher mass-loss rates, with the solar analogue AB~Dor ($\sim 120$ Myr) presenting a mass-loss rate of $350\,\dot{M}_\odot$ and the even younger LQ Lup ($\sim 25$~Myr), with $4500\,\dot{M}_\odot$  \citep{2019MNRAS.482.2853J}.

One way to shed light on the solar wind evolution is to look at physical characteristics known to be intimately related to winds of solar-like stars: rotation, magnetism and activity.  These three `ingredients' are related to the evolution of the solar wind in a feedback loop that I summarised in Fig.~\ref{fig.big_picture}. The stellar wind carries away angular momentum, which leads stars to spin down. A decrease in rotation rate implies that the dynamo weakens, generating weaker magnetic fields. This affects the stellar wind and the amount of angular momentum it can carry away. Then, the loop restarts. As a result, rotation, magnetism and its proxy in the form of stellar activity decrease with time. 

These ingredients, which are better observationally constrained than wind properties, can be implemented in theoretical models. Models of winds of solar-like stars have been mainly developed in the magnetohydrodynamics (MHD) framework, which treats the  wind as a magnetised fluid. Within this framework, models can be broadly divided into ones that calculate explicitly heating deposition (due to dissipation of MHD waves mostly) and those that assume the plasma has already been heated to MK temperatures, and thus energy deposition is treated implicitly. Each model set has its advantages and drawbacks. The first group has a more sophisticated treatment of the physical wind driving mechanism, but is usually computed along a magnetic flux tube, neglects rotation, and extend only out to a few solar radii. The second group has a less sophisticated treatment of the physics, which implies that they can be computed over the entire stellar surface, extend out to larger stellar distances and include rotation. 

For studying the evolution of angular momentum, the second group of models are usually adopted. In that framework, it is fascinating to see that the early model of  \citet{1967ApJ...148..217W}, which considers a simple split-monopole magnetic field embed in a thermally-driven wind of a rotating Sun already does a good job in reproducing the observed relation  $\Omega_\star \propto t^{-1/2}$ for older stars (Sect.~\ref{sec.jdot}). In this sense, stellar evolution models, calibrated to the observed rotation distribution of stars in open clusters,   have become an important aid for constraining the solar wind evolution.  
 
In this review, I focused on the solar wind evolution from the early main sequence until today. That does not mean that what lies ahead is less interesting! The Sun is about half-way through its main-sequence journey, and will have thus another $\sim 4$ Gyr worth of evolution in a relatively calm way. When the hydrogen fuel in its core is finally exhausted, the Sun will evolve off the main sequence and become a red giant. The timescales of this phase and subsequent ones become shorter and thus the Sun will pass through these phases very quickly in comparison to the long evolution during its main sequence. Eventually, the Sun will end its life as a white dwarf and cool down indefinitely. Certainly, the solar wind will change radically. In the red giant phase alone, the solar wind will likely be cooler ($\lesssim 10^4$~K), with lower velocities, but higher densities, leading to mass-loss rate that will be several magnitudes higher than today \citep{2016ApJ...829...74W}. Solar rotation will change considerably due to its radius expansion. Given that rotation  also seems to be related to magnetism in evolved stars \citep{2015A&A...574A..90A}, the Sun's magnetic field will change as well. 

The evolution of the solar wind has applications that goes beyond understanding stellar rotational evolution -- it affects planets and exoplanets, which at the end of the day are embedded in the outward-streaming plasma from their host stars. A strong solar wind in the past is often cited to explain the lack of a substantial atmosphere in Mars today -- planetary atmospheric erosion can be enhanced under strong stellar wind conditions. Such strong stellar wind conditions are also faced by close-in exoplanets, even when orbiting old and less active stars. Additionally, the Earth's magnetosphere is believed to have been smaller than it is today, as a result of a strong stellar wind. Similarly, the strong solar wind in the past was able to push further the boundary with the ISM -- the heliosphere was likely  larger in the past. A different solar wind condition and much more expanded heliosphere would then have likely reduced the amount of Galactic cosmic rays reaching Earth. Both of these (Earth's magnetosphere and cosmic ray flux) are believed to be important factors for life formation and sustainability  in this planet. 

I hope to have shown in this review that significant progress in understanding the evolution of the solar wind is being made. Nevertheless this is currently a ``living'' area of research and, as such, still many open questions remain. Understanding the physical processes heating and driving the present-day solar wind  could provide a step further in modelling the solar wind at different ages. Missions like NASA's Parker Solar Probe and ESA's Solar Orbiter will obtain in situ measurements of the solar wind at unprecedented close heliospheric distances. These regions are where the bulk of the acceleration of the solar wind takes place, and thus these missions are expected to soon shed light on the physical mechanisms that accelerate the solar wind. As an added bonus, these close heliospheric distances coincide with the regions where many close-in exoplanets are located -- these missions will thus provide information about the environment at the orbits of close-in exoplanets, which will aid in further understanding how extreme wind conditions can affect exoplanets, and potentially, link this to how the stronger solar wind in the past interacted with the young solar system planets.

\begin{acknowledgements}
I thank the editors for inviting me to write this review and the reviewers (Manuel G\"udel and anonymous) for their very thorough and constructive feedback. There are several colleagues I would like to warmly thank for fruitful discussions over the years: Moira Jardine, Jean-Francois Donati, Rim Fares, Pascal Petit, Victor See, Julien Morin, Luca Fossati, Vincent Bourrier, Sandra Jeffers, Colin Folsom, Lisa Lehmann, Joe Llama, Scott Gregory, Andrew Cameron, Stephen Marsden. I thank Gopal Hazra for providing the data used Fig.~\ref{fig.sunXUV}.
I acknowledge funding from the European Research Council (ERC) under the European Union's Horizon 2020 research and innovation programme (grant agreement No 817540, ASTROFLOW).
\end{acknowledgements}

\addcontentsline{toc}{section}{References}

\bibliographystyle{spbasic-FS}      

\end{document}